\documentclass[graybox]{svmult}

\usepackage{mathptmx}       
\usepackage{helvet}         
\usepackage{courier}        
\usepackage{type1cm}        
\usepackage{makeidx}         
\usepackage{graphicx}        
\usepackage{multicol}        
\usepackage[bottom]{footmisc}

\usepackage{amsmath}

\usepackage{amssymb}

\usepackage{enumerate}

\newcommand{\alert}[1]{{\color{red} #1}}

\newcommand\beq{\begin{equation}}
\newcommand\eeq{\end{equation}}
\newcommand\beqa{\begin{eqnarray}}
\newcommand\eeqa{\end{eqnarray}}
\newcommand\nn{\nonumber\\}

\newcommand{\id}{\text{id}}
\newcommand{\ex}{\text{ex}}
\newcommand{\XX}{\mathbf{x}}
\newcommand{\zz}{z_\Lambda}
\newcommand{\bb}{\mathfrak{b}}
\newcommand{\rr}{\mathbf{r}}
\newcommand{\pp}{\mathbf{p}}
\newcommand{\llangle}{\left\langle}
\newcommand{\rrangle}{\right\rangle}
\newcommand{\xxx}{{(\xi)}}

\newcommand{\Ss}{\text{SS}}
\newcommand{\hs}{\text{HS}}
\newcommand{\sw}{\text{SW}}

\newcommand{\cp}{\text{cp}}
\newcommand{\hnc}{\text{HNC}}
\newcommand{\py}{\text{PY}}
\newcommand{\one}{{(1)}}
\newcommand{\two}{{(2)}}
\newcommand{\three}{{(3)}}
\newcommand{\four}{{(4)}}

\definecolor{M_Beige}{rgb}{0.96 , 0.96 , 0.86}

\definecolor{M_Brown}{rgb}{0.65 , 0.16 , 0.16}

\definecolor{M_Gold}{rgb}{1.00 , 0.84 , 0.00}

\definecolor{M_LemonChiffon}{rgb}{1.00 , 0.98 , 0.80}

\definecolor{M_Orange}{rgb}{1.00 , 0.60 , 0.00}

\definecolor{M_Pink}{rgb}{1.00 , 0.75 , 0.80}

\definecolor{M_Violet}{rgb}{0.93 , 0.51 , 0.93}

\newcommand{\Cblu}[1]{\textcolor{blue}{#1}}

\newcommand{\Cred}[1]{\textcolor{red}{#1}}

\usepackage{epic}

\newcommand{\xx}{5}
\newcommand{\yy}{5}

\newcommand{\xmax}{30}
\newcommand{\ymax}{22}

\newcommand{\xmaxtiny}{20}
\newcommand{\ymaxtiny}{14}

\newcommand{\sone}{\begin{picture}(\xmax,\ymax)(-\xx,\yy)
\setlength{\unitlength}{.1mm}
\put(0,30){\circle{18}}
\end{picture}}

\newcommand{\dtwo}{\begin{picture}(\xmax,\ymax)(-\xx,\yy)
\setlength{\unitlength}{.1mm}
\put(0,30){\circle{18}}
\put(60,30){\circle{18}}
\end{picture}}

\newcommand{\stwo}{\begin{picture}(\xmax,\ymax)(-\xx,\yy)
\setlength{\unitlength}{.1mm}
\put(0,30){\circle{18}}
\put(60,30){\circle{18}}
\put(9,30){\line(1,0){42}}
\end{picture}}

\newcommand{\soneSone}{\begin{picture}(\xmax,\ymax)(-\xx,\yy)
\setlength{\unitlength}{.1mm}
\put(0,30){\circle{18}}
\put(60,30){\circle*{18}}
\put(9,30){\line(1,0){42}}
\end{picture}}

\newcommand{\dthreeA}{\begin{picture}(\xmax,\ymax)(-\xx,\yy)
\setlength{\unitlength}{.1mm}
\put(0,0){\circle{18}}
\put(60,0){\circle{18}}
\put(30,60){\circle{18}}
\end{picture}}

\newcommand{\dthreeB}{\begin{picture}(\xmax,\ymax)(-\xx,\yy)
\setlength{\unitlength}{.1mm}
\put(0,0){\circle{18}}
\put(60,0){\circle{18}}
\put(30,60){\circle{18}}
\put(9,0){\line(1,0){42}}
\end{picture}}

\newcommand{\dtwoDoneB}{\begin{picture}(\xmax,\ymax)(-\xx,\yy)
\setlength{\unitlength}{.1mm}
\put(0,0){\circle{18}}
\put(60,0){\circle*{18}}
\put(30,60){\circle{18}}
\put(9,0){\line(1,0){42}}
\end{picture}}

\newcommand{\rthree}{\begin{picture}(\xmax,\ymax)(-\xx,\yy)
\setlength{\unitlength}{.1mm}
\put(0,0){\circle{18}}
\put(60,0){\circle{18}}
\put(30,60){\circle{18}}
\put(4,8){\line(1,2){22}}
\put(56,8){\line(-1,2){22}}
\end{picture}}

\newcommand{\rthreeART}{\begin{picture}(\xmax,\ymax)(-\xx,\yy)
\setlength{\unitlength}{.1mm}
\put(0,0){\circle{18}}
\put(60,0){\circle{18}}
\put(30,60){\circle{18}} \alert{\put(30,60){\circle{30}}}
\put(4,8){\line(1,2){22}}
\put(56,8){\line(-1,2){22}}
\end{picture}}

\newcommand{\roneRtwoA}{\begin{picture}(\xmax,\ymax)(-\xx,\yy)
\setlength{\unitlength}{.1mm}
\put(0,0){\circle*{18}}
\put(60,0){\circle*{18}}
\put(30,60){\circle{18}}
\put(4,8){\line(1,2){22}}
\put(56,8){\line(-1,2){22}}
\end{picture}}

\newcommand{\roneRtwoB}{\begin{picture}(\xmax,\ymax)(-\xx,\yy)
\setlength{\unitlength}{.1mm}
\put(0,0){\circle{18}}
\put(60,0){\circle*{18}}
\put(30,60){\circle*{18}}
\put(4,8){\line(1,2){22}}
\put(56,8){\line(-1,2){22}}
\end{picture}}

\newcommand{\rtwoRoneA}{\begin{picture}(\xmax,\ymax)(-\xx,\yy)
\setlength{\unitlength}{.1mm}
\put(0,0){\circle{18}}
\put(60,0){\circle{18}}
\put(30,60){\circle*{18}}
\put(4,8){\line(1,2){22}}
\put(56,8){\line(-1,2){22}}
\end{picture}}

\newcommand{\rtwoRoneB}{\begin{picture}(\xmax,\ymax)(-\xx,\yy)
\setlength{\unitlength}{.1mm}
\put(0,0){\circle*{18}}
\put(60,0){\circle{18}}
\put(30,60){\circle{18}}
\put(4,8){\line(1,2){22}}
\put(56,8){\line(-1,2){22}}
\end{picture}}

\newcommand{\sthree}{\begin{picture}(\xmax,\ymax)(-\xx,\yy)
\setlength{\unitlength}{.1mm}
\put(0,0){\circle{18}}
\put(60,0){\circle{18}}
\put(30,60){\circle{18}}
\put(4,8){\line(1,2){22}}
\put(9,0){\line(1,0){42}}
\put(56,8){\line(-1,2){22}}
\end{picture}}

\newcommand{\soneStwo}{\begin{picture}(\xmax,\ymax)(-\xx,\yy)
\setlength{\unitlength}{.1mm}
\put(0,0){\circle*{18}}
\put(60,0){\circle*{18}}
\put(30,60){\circle{18}}
\put(4,8){\line(1,2){22}}
\put(9,0){\line(1,0){42}}
\put(56,8){\line(-1,2){22}}
\end{picture}}

\newcommand{\stwoSone}{\begin{picture}(\xmax,\ymax)(-\xx,\yy)
\setlength{\unitlength}{.1mm}
\put(0,0){\circle{18}}
\put(60,0){\circle{18}}
\put(30,60){\circle*{18}}
\put(4,8){\line(1,2){22}}
\put(9,0){\line(1,0){42}}
\put(56,8){\line(-1,2){22}}
\end{picture}}

\newcommand{\dfourA}{\begin{picture}(\xmax,\ymax)(-\xx,\yy)
\setlength{\unitlength}{.1mm}
\put(0,60){\circle{18}}
\put(60,60){\circle{18}}
\put(0,0){\circle{18}}
\put(60,0){\circle{18}}
\end{picture}}

\newcommand{\dfourB}{\begin{picture}(\xmax,\ymax)(-\xx,\yy)
\setlength{\unitlength}{.1mm}
\put(0,60){\circle{18}}
\put(60,60){\circle{18}}
\put(0,0){\circle{18}}
\put(60,0){\circle{18}}
\put(9,0){\line(1,0){42}}
\end{picture}}

\newcommand{\dfourC}{\begin{picture}(\xmax,\ymax)(-\xx,\yy)
\setlength{\unitlength}{.1mm}
\put(0,60){\circle{18}}
\put(60,60){\circle{18}}
\put(0,0){\circle{18}}
\put(60,0){\circle{18}}
\put(9,0){\line(1,0){42}}
\put(0,9){\line(0,1){42}}
\end{picture}}

\newcommand{\dtwoDtwoCA}{\begin{picture}(\xmax,\ymax)(-\xx,\yy)
\setlength{\unitlength}{.1mm}
\put(0,60){\circle*{18}}
\put(60,60){\circle{18}}
\put(0,0){\circle{18}}
\put(60,0){\circle*{18}}
\put(9,0){\line(1,0){42}}
\put(0,9){\line(0,1){42}}
\end{picture}}

\newcommand{\dtwoDtwoCB}{\begin{picture}(\xmax,\ymax)(-\xx,\yy)
\setlength{\unitlength}{.1mm}
\put(0,60){\circle{18}}
\put(60,60){\circle{18}}
\put(0,0){\circle*{18}}
\put(60,0){\circle*{18}}
\put(9,0){\line(1,0){42}}
\put(0,9){\line(0,1){42}}
\end{picture}}

\newcommand{\dfourD}{\begin{picture}(\xmax,\ymax)(-\xx,\yy)
\setlength{\unitlength}{.1mm}
\put(0,60){\circle{18}}
\put(60,60){\circle{18}}
\put(0,0){\circle{18}}
\put(60,0){\circle{18}}
\put(9,60){\line(1,0){42}}
\put(9,0){\line(1,0){42}}
\end{picture}}

\newcommand{\dtwoDtwoD}{\begin{picture}(\xmax,\ymax)(-\xx,\yy)
\setlength{\unitlength}{.1mm}
\put(0,60){\circle{18}}
\put(60,60){\circle*{18}}
\put(0,0){\circle{18}}
\put(60,0){\circle*{18}}
\put(9,60){\line(1,0){42}}
\put(9,0){\line(1,0){42}}
\end{picture}}

\newcommand{\dfourE}{\begin{picture}(\xmax,\ymax)(-\xx,\yy)
\setlength{\unitlength}{.1mm}
\put(0,60){\circle{18}}
\put(60,60){\circle{18}}
\put(0,0){\circle{18}}
\put(60,0){\circle{18}}
\put(9,0){\line(1,0){42}}
\put(0,9){\line(0,1){42}}
\put(7,53){\line(1,-1) {46.5}}
\end{picture}}

\newcommand{\dtwoDtwoE}{\begin{picture}(\xmax,\ymax)(-\xx,\yy)
\setlength{\unitlength}{.1mm}
\put(0,60){\circle{18}}
\put(60,60){\circle{18}}
\put(0,0){\circle*{18}}
\put(60,0){\circle*{18}}
\put(9,0){\line(1,0){42}}
\put(0,9){\line(0,1){42}}
\put(7,53){\line(1,-1) {46.5}}
\end{picture}}

\newcommand{\rfourA}{\begin{picture}(\xmax,\ymax)(-\xx,\yy)
\setlength{\unitlength}{.1mm}
\put(0,60){\circle{18}}
\put(60,60){\circle{18}}
\put(0,0){\circle{18}}
\put(60,0){\circle{18}}
\put(9,0){\line(1,0){42}}
\put(0,9){\line(0,1){42}}
\put(60,9){\line(0,1){42}}
\end{picture}}

\newcommand{\rfourAART}{\begin{picture}(\xmax,\ymax)(-\xx,\yy)
\setlength{\unitlength}{.1mm}
\put(0,60){\circle{18}}
\put(60,60){\circle{18}}
\put(0,0){\circle{18}} \alert{\put(0,0){\circle{30}}}
\put(60,0){\circle{18}} \alert{\put(60,0){\circle{30}}}
\put(9,0){\line(1,0){42}}
\put(0,9){\line(0,1){42}}
\put(60,9){\line(0,1){42}}
\end{picture}}

\newcommand{\roneRthreeAA}{\begin{picture}(\xmax,\ymax)(-\xx,\yy)
\setlength{\unitlength}{.1mm}
\put(0,60){\circle{18}}
\put(60,60){\circle*{18}}
\put(0,0){\circle*{18}}
\put(60,0){\circle*{18}}
\put(9,0){\line(1,0){42}}
\put(0,9){\line(0,1){42}}
\put(60,9){\line(0,1){42}}
\end{picture}}

\newcommand{\roneRthreeAB}{\begin{picture}(\xmax,\ymax)(-\xx,\yy)
\setlength{\unitlength}{.1mm}
\put(0,60){\circle*{18}}
\put(60,60){\circle*{18}}
\put(0,0){\circle{18}}
\put(60,0){\circle*{18}}
\put(9,0){\line(1,0){42}}
\put(0,9){\line(0,1){42}}
\put(60,9){\line(0,1){42}}
\end{picture}}

\newcommand{\rtwoRtwoAA}{\begin{picture}(\xmax,\ymax)(-\xx,\yy)
\setlength{\unitlength}{.1mm}
\put(0,60){\circle{18}}
\put(60,60){\circle{18}}
\put(0,0){\circle*{18}}
\put(60,0){\circle*{18}}
\put(9,0){\line(1,0){42}}
\put(0,9){\line(0,1){42}}
\put(60,9){\line(0,1){42}}
\end{picture}}

\newcommand{\ytwoAd}{\begin{picture}(\xmax,\ymax)(-\xx,\yy)
\setlength{\unitlength}{.1mm}
\put(0,60){\circle{18}}
\put(60,60){\circle*{18}}
\dashline[60]{10}(8,60)(60,60)
\put(0,0){\circle*{18}}
\put(60,0){\circle*{18}}
\put(9,0){\line(1,0){42}}
\put(0,9){\line(0,1){42}}
\put(60,9){\line(0,1){42}}
\end{picture}}

\newcommand{\rtwoRtwoAB}{\begin{picture}(\xmax,\ymax)(-\xx,\yy)
\setlength{\unitlength}{.1mm}
\put(0,60){\circle{18}}
\put(60,60){\circle*{18}}
\put(0,0){\circle{18}}
\put(60,0){\circle*{18}}
\put(9,0){\line(1,0){42}}
\put(0,9){\line(0,1){42}}
\put(60,9){\line(0,1){42}}
\end{picture}}

\newcommand{\rtwoRtwoAC}{\begin{picture}(\xmax,\ymax)(-\xx,\yy)
\setlength{\unitlength}{.1mm}
\put(0,60){\circle*{18}}
\put(60,60){\circle*{18}}
\put(0,0){\circle{18}}
\put(60,0){\circle{18}}
\put(9,0){\line(1,0){42}}
\put(0,9){\line(0,1){42}}
\put(60,9){\line(0,1){42}}
\end{picture}}

\newcommand{\rtwoRtwoAD}{\begin{picture}(\xmax,\ymax)(-\xx,\yy)
\setlength{\unitlength}{.1mm}
\put(0,60){\circle*{18}}
\put(60,60){\circle{18}}
\put(0,0){\circle{18}}
\put(60,0){\circle*{18}}
\put(9,0){\line(1,0){42}}
\put(0,9){\line(0,1){42}}
\put(60,9){\line(0,1){42}}
\end{picture}}

\newcommand{\rfourB}{\begin{picture}(\xmax,\ymax)(-\xx,\yy)
\setlength{\unitlength}{.1mm}
\put(0,60){\circle{18}}
\put(60,60){\circle{18}}
\put(0,0){\circle{18}}
\put(60,0){\circle{18}}
\put(9,0){\line(1,0){42}}
\put(0,9){\line(0,1){42}}
\put(7,7){\line(1,1) {46.5}}
\end{picture}}

\newcommand{\rfourBART}{\begin{picture}(\xmax,\ymax)(-\xx,\yy)
\setlength{\unitlength}{.1mm}
\put(0,60){\circle{18}}
\put(60,60){\circle{18}}
\put(0,0){\circle{18}} \alert{\put(0,0){\circle{30}}}
\put(60,0){\circle{18}}
\put(9,0){\line(1,0){42}}
\put(0,9){\line(0,1){42}}
\put(7,7){\line(1,1) {46.5}}
\end{picture}}

\newcommand{\roneRthreeBA}{\begin{picture}(\xmax,\ymax)(-\xx,\yy)
\setlength{\unitlength}{.1mm}
\put(0,60){\circle*{18}}
\put(60,60){\circle*{18}}
\put(0,0){\circle{18}}
\put(60,0){\circle*{18}}
\put(9,0){\line(1,0){42}}
\put(0,9){\line(0,1){42}}
\put(7,7){\line(1,1) {46.5}}
\end{picture}}

\newcommand{\roneRthreeBB}{\begin{picture}(\xmax,\ymax)(-\xx,\yy)
\setlength{\unitlength}{.1mm}
\put(0,60){\circle{18}}
\put(60,60){\circle*{18}}
\put(0,0){\circle*{18}}
\put(60,0){\circle*{18}}
\put(9,0){\line(1,0){42}}
\put(0,9){\line(0,1){42}}
\put(7,7){\line(1,1) {46.5}}
\end{picture}}

\newcommand{\rtwoRtwoBA}{\begin{picture}(\xmax,\ymax)(-\xx,\yy)
\setlength{\unitlength}{.1mm}
\put(0,60){\circle*{18}}
\put(60,60){\circle*{18}}
\put(0,0){\circle{18}}
\put(60,0){\circle{18}}
\put(9,0){\line(1,0){42}}
\put(0,9){\line(0,1){42}}
\put(7,7){\line(1,1) {46.5}}
\end{picture}}

\newcommand{\rtwoRtwoBB}{\begin{picture}(\xmax,\ymax)(-\xx,\yy)
\setlength{\unitlength}{.1mm}
\put(0,60){\circle{18}}
\put(60,60){\circle{18}}
\put(0,0){\circle*{18}}
\put(60,0){\circle*{18}}
\put(9,0){\line(1,0){42}}
\put(0,9){\line(0,1){42}}
\put(7,7){\line(1,1) {46.5}}
\end{picture}}

\newcommand{\rfourC}{\begin{picture}(\xmax,\ymax)(-\xx,\yy)
\setlength{\unitlength}{.1mm}
\put(0,60){\circle{18}}
\put(60,60){\circle{18}}
\put(0,0){\circle{18}}
\put(60,0){\circle{18}}
\put(9,0){\line(1,0){42}}
\put(0,9){\line(0,1){42}}
\put(60,9){\line(0,1){42}}
\put(7,7){\line(1,1) {46.5}}
\end{picture}}

\newcommand{\rfourCART}{\begin{picture}(\xmax,\ymax)(-\xx,\yy)
\setlength{\unitlength}{.1mm}
\put(0,60){\circle{18}}
\put(60,60){\circle{18}}
\put(0,0){\circle{18}} \alert{\put(0,0){\circle{30}}}
\put(60,0){\circle{18}}
\put(9,0){\line(1,0){42}}
\put(0,9){\line(0,1){42}}
\put(60,9){\line(0,1){42}}
\put(7,7){\line(1,1) {46.5}}
\end{picture}}

\newcommand{\roneRthreeCA}{\begin{picture}(\xmax,\ymax)(-\xx,\yy)
\setlength{\unitlength}{.1mm}
\put(0,60){\circle{18}}
\put(60,60){\circle*{18}}
\put(0,0){\circle*{18}}
\put(60,0){\circle*{18}}
\put(9,0){\line(1,0){42}}
\put(0,9){\line(0,1){42}}
\put(60,9){\line(0,1){42}}
\put(7,7){\line(1,1) {46.5}}
\end{picture}}

\newcommand{\roneRthreeCB}{\begin{picture}(\xmax,\ymax)(-\xx,\yy)
\setlength{\unitlength}{.1mm}
\put(0,60){\circle*{18}}
\put(60,60){\circle*{18}}
\put(0,0){\circle{18}}
\put(60,0){\circle*{18}}
\put(9,0){\line(1,0){42}}
\put(0,9){\line(0,1){42}}
\put(60,9){\line(0,1){42}}
\put(7,7){\line(1,1) {46.5}}
\end{picture}}

\newcommand{\roneRthreeCC}{\begin{picture}(\xmax,\ymax)(-\xx,\yy)
\setlength{\unitlength}{.1mm}
\put(0,60){\circle*{18}}
\put(60,60){\circle*{18}}
\put(0,0){\circle*{18}}
\put(60,0){\circle{18}}
\put(9,0){\line(1,0){42}}
\put(0,9){\line(0,1){42}}
\put(60,9){\line(0,1){42}}
\put(7,7){\line(1,1) {46.5}}
\end{picture}}

\newcommand{\rtwoRtwoCA}{\begin{picture}(\xmax,\ymax)(-\xx,\yy)
\setlength{\unitlength}{.1mm}
\put(0,60){\circle*{18}}
\put(60,60){\circle{18}}
\put(0,0){\circle*{18}}
\put(60,0){\circle{18}}
\put(9,0){\line(1,0){42}}
\put(0,9){\line(0,1){42}}
\put(60,9){\line(0,1){42}}
\put(7,7){\line(1,1) {46.5}}
\end{picture}}

\newcommand{\rtwoRtwoCB}{\begin{picture}(\xmax,\ymax)(-\xx,\yy)
\setlength{\unitlength}{.1mm}
\put(0,60){\circle*{18}}
\put(60,60){\circle*{18}}
\put(0,0){\circle{18}}
\put(60,0){\circle{18}}
\put(9,0){\line(1,0){42}}
\put(0,9){\line(0,1){42}}
\put(60,9){\line(0,1){42}}
\put(7,7){\line(1,1) {46.5}}
\end{picture}}

\newcommand{\rtwoRtwoCC}{\begin{picture}(\xmax,\ymax)(-\xx,\yy)
\setlength{\unitlength}{.1mm}
\put(0,60){\circle{18}}
\put(60,60){\circle{18}}
\put(0,0){\circle*{18}}
\put(60,0){\circle*{18}}
\put(9,0){\line(1,0){42}}
\put(0,9){\line(0,1){42}}
\put(60,9){\line(0,1){42}}
\put(7,7){\line(1,1) {46.5}}
\end{picture}}

\newcommand{\ytwoBd}{\begin{picture}(\xmax,\ymax)(-\xx,\yy)
\setlength{\unitlength}{.1mm}
\put(0,60){\circle{18}}
\put(60,60){\circle*{18}}
\dashline[60]{10}(8,60)(60,60)
\put(0,0){\circle*{18}}
\put(60,0){\circle*{18}}
\put(9,0){\line(1,0){42}}
\put(0,9){\line(0,1){42}}
\put(60,9){\line(0,1){42}}
\put(7,7){\line(1,1) {46.5}}
\end{picture}}

\newcommand{\rtwoRtwoCD}{\begin{picture}(\xmax,\ymax)(-\xx,\yy)
\setlength{\unitlength}{.1mm}
\put(0,60){\circle{18}}
\put(60,60){\circle*{18}}
\put(0,0){\circle{18}}
\put(60,0){\circle*{18}}
\put(9,0){\line(1,0){42}}
\put(0,9){\line(0,1){42}}
\put(60,9){\line(0,1){42}}
\put(7,7){\line(1,1) {46.5}}
\end{picture}}

\newcommand{\sfourA}{\begin{picture}(\xmax,\ymax)(-\xx,\yy)
\setlength{\unitlength}{.1mm}
\put(0,60){\circle{18}}
\put(60,60){\circle{18}}
\put(0,0){\circle{18}}
\put(60,0){\circle{18}}
\put(9,60){\line(1,0){42}}
\put(9,0){\line(1,0){42}}
\put(0,9){\line(0,1){42}}
\put(60,9){\line(0,1){42}}
\end{picture}}

\newcommand{\soneSthreeA}{\begin{picture}(\xmax,\ymax)(-\xx,\yy)
\setlength{\unitlength}{.1mm}
\put(0,60){\circle*{18}}
\put(60,60){\circle*{18}}
\put(0,0){\circle{18}}
\put(60,0){\circle*{18}}
\put(9,60){\line(1,0){42}}
\put(9,0){\line(1,0){42}}
\put(0,9){\line(0,1){42}}
\put(60,9){\line(0,1){42}}
\end{picture}}

\newcommand{\stwoStwoAA}{\begin{picture}(\xmax,\ymax)(-\xx,\yy)
\setlength{\unitlength}{.1mm}
\put(0,60){\circle*{18}}
\put(60,60){\circle*{18}}
\put(0,0){\circle{18}}
\put(60,0){\circle{18}}
\put(9,60){\line(1,0){42}}
\put(9,0){\line(1,0){42}}
\put(0,9){\line(0,1){42}}
\put(60,9){\line(0,1){42}}
\end{picture}}

\newcommand{\stwoStwoAB}{\begin{picture}(\xmax,\ymax)(-\xx,\yy)
\setlength{\unitlength}{.1mm}
\put(0,60){\circle*{18}}
\put(60,60){\circle{18}}
\put(0,0){\circle{18}}
\put(60,0){\circle*{18}}
\put(9,60){\line(1,0){42}}
\put(9,0){\line(1,0){42}}
\put(0,9){\line(0,1){42}}
\put(60,9){\line(0,1){42}}
\end{picture}}

\newcommand{\stwoStwoABtiny}{\begin{picture}(\xmaxtiny,\ymaxtiny)(-\xx,.5\yy)
\setlength{\unitlength}{.05mm}
\put(0,60){\circle*{18}}
\put(60,60){\circle{18}}
\put(0,0){\circle{18}}
\put(60,0){\circle*{18}}
\put(9,60){\line(1,0){42}}
\put(9,0){\line(1,0){42}}
\put(0,9){\line(0,1){42}}
\put(60,9){\line(0,1){42}}
\end{picture}}

\newcommand{\ytwoCd}{\begin{picture}(\xmax,\ymax)(-\xx,\yy)
\setlength{\unitlength}{.1mm}
\put(0,60){\circle*{18}}
\put(60,60){\circle*{18}}
\put(0,0){\circle{18}}
\dashline[60]{10}(7,7)(60,60)
\put(60,0){\circle*{18}}
\put(9,60){\line(1,0){42}}
\put(9,0){\line(1,0){42}}
\put(0,9){\line(0,1){42}}
\put(60,9){\line(0,1){42}}
\end{picture}}

\newcommand{\sfourB}{\begin{picture}(\xmax,\ymax)(-\xx,\yy)
\setlength{\unitlength}{.1mm}
\put(0,60){\circle{18}}
\put(60,60){\circle{18}}
\put(0,0){\circle{18}}
\put(60,0){\circle{18}}
\put(9,60){\line(1,0){42}}
\put(9,0){\line(1,0){42}}
\put(0,9){\line(0,1){42}}
\put(60,9){\line(0,1){42}}
\put(7,7){\line(1,1) {46.5}}
\end{picture}}

\newcommand{\SfourB}{\begin{picture}(\xmax,\ymax)(-\xx,\yy)
\setlength{\unitlength}{.1mm}
\put(0,60){\circle*{18}}
\put(60,60){\circle*{18}}
\put(0,0){\circle*{18}}
\put(60,0){\circle*{18}}
\put(9,60){\line(1,0){42}}
\put(9,0){\line(1,0){42}}
\put(0,9){\line(0,1){42}}
\put(60,9){\line(0,1){42}}
\put(7,7){\line(1,1) {46.5}}
\end{picture}}

\newcommand{\soneSthreeBA}{\begin{picture}(\xmax,\ymax)(-\xx,\yy)
\setlength{\unitlength}{.1mm}
\put(0,60){\circle*{18}}
\put(60,60){\circle*{18}}
\put(0,0){\circle{18}}
\put(60,0){\circle*{18}}
\put(9,60){\line(1,0){42}}
\put(9,0){\line(1,0){42}}
\put(0,9){\line(0,1){42}}
\put(60,9){\line(0,1){42}}
\put(7,7){\line(1,1) {46.5}}
\end{picture}}

\newcommand{\soneSthreeBB}{\begin{picture}(\xmax,\ymax)(-\xx,\yy)
\setlength{\unitlength}{.1mm}
\put(0,60){\circle*{18}}
\put(60,60){\circle*{18}}
\put(0,0){\circle*{18}}
\put(60,0){\circle{18}}
\put(9,60){\line(1,0){42}}
\put(9,0){\line(1,0){42}}
\put(0,9){\line(0,1){42}}
\put(60,9){\line(0,1){42}}
\put(7,7){\line(1,1) {46.5}}
\end{picture}}

\newcommand{\stwoStwoBA}{\begin{picture}(\xmax,\ymax)(-\xx,\yy)
\setlength{\unitlength}{.1mm}
\put(0,60){\circle*{18}}
\put(60,60){\circle*{18}}
\put(0,0){\circle{18}}
\put(60,0){\circle{18}}
\put(9,60){\line(1,0){42}}
\put(9,0){\line(1,0){42}}
\put(0,9){\line(0,1){42}}
\put(60,9){\line(0,1){42}}
\put(7,7){\line(1,1) {46.5}}
\end{picture}}

\newcommand{\stwoStwoBB}{\begin{picture}(\xmax,\ymax)(-\xx,\yy)
\setlength{\unitlength}{.1mm}
\put(0,60){\circle*{18}}
\put(60,60){\circle{18}}
\put(0,0){\circle{18}}
\put(60,0){\circle*{18}}
\put(9,60){\line(1,0){42}}
\put(9,0){\line(1,0){42}}
\put(0,9){\line(0,1){42}}
\put(60,9){\line(0,1){42}}
\put(7,7){\line(1,1) {46.5}}
\end{picture}}

\newcommand{\stwoStwoBC}{\begin{picture}(\xmax,\ymax)(-\xx,\yy)
\setlength{\unitlength}{.1mm}
\put(0,60){\circle*{18}}
\put(60,60){\circle{18}}
\put(0,0){\circle{18}}
\put(60,0){\circle*{18}}
\put(9,60){\line(1,0){42}}
\put(9,0){\line(1,0){42}}
\put(0,9){\line(0,1){42}}
\put(60,9){\line(0,1){42}}
\put(7,53){\line(1,-1) {46.5}}
\end{picture}}

\newcommand{\stwoStwoBCtiny}{\begin{picture}(\xmaxtiny,\ymaxtiny)(-\xx,.5\yy)
\setlength{\unitlength}{.05mm}
\put(0,60){\circle*{18}}
\put(60,60){\circle{18}}
\put(0,0){\circle{18}}
\put(60,0){\circle*{18}}
\put(9,60){\line(1,0){42}}
\put(9,0){\line(1,0){42}}
\put(0,9){\line(0,1){42}}
\put(60,9){\line(0,1){42}}
\dottedline{.5}(0,60)(60,0)
\end{picture}}

\newcommand{\stwoStwoBCd}{\begin{picture}(\xmax,\ymax)(-\xx,\yy)
\setlength{\unitlength}{.1mm}
\put(0,60){\circle*{18}}
\put(60,60){\circle{18}}
\put(0,0){\circle{18}}
\put(60,0){\circle*{18}}
\put(9,60){\line(1,0){42}}
\put(9,0){\line(1,0){42}}
\put(0,9){\line(0,1){42}}
\put(60,9){\line(0,1){42}}
\dottedline{5}(0,60)(60,0)
\end{picture}}

\newcommand{\ytwoDd}{\begin{picture}(\xmax,\ymax)(-\xx,\yy)
\setlength{\unitlength}{.1mm}
\put(0,60){\circle*{18}}
\put(60,60){\circle*{18}}
\put(0,0){\circle{18}}
\dashline[60]{10}(7,7)(60,60)
\put(60,0){\circle*{18}}
\put(9,60){\line(1,0){42}}
\put(9,0){\line(1,0){42}}
\put(0,9){\line(0,1){42}}
\put(60,9){\line(0,1){42}}
\put(7,53){\line(1,-1) {46.5}}
\end{picture}}

\newcommand{\sfourC}{\begin{picture}(\xmax,\ymax)(-\xx,\yy)
\setlength{\unitlength}{.1mm}
\put(0,60){\circle{18}}
\put(60,60){\circle{18}}
\put(0,0){\circle{18}}
\put(60,0){\circle{18}}
\put(9,60){\line(1,0){42}}
\put(9,0){\line(1,0){42}}
\put(0,9){\line(0,1){42}}
\put(60,9){\line(0,1){42}}
\put(7,7){\line(1,1) {46.5}}
\put(7,53){\line(1,-1) {46.5}}
\end{picture}}

\newcommand{\soneSthreeC}{\begin{picture}(\xmax,\ymax)(-\xx,\yy)
\setlength{\unitlength}{.1mm}
\put(0,60){\circle*{18}}
\put(60,60){\circle*{18}}
\put(0,0){\circle{18}}
\put(60,0){\circle*{18}}
\put(9,60){\line(1,0){42}}
\put(9,0){\line(1,0){42}}
\put(0,9){\line(0,1){42}}
\put(60,9){\line(0,1){42}}
\put(7,7){\line(1,1) {46.5}}
\put(7,53){\line(1,-1) {46.5}}
\end{picture}}

\newcommand{\stwoStwoC}{\begin{picture}(\xmax,\ymax)(-\xx,\yy)
\setlength{\unitlength}{.1mm}
\put(0,60){\circle*{18}}
\put(60,60){\circle*{18}}
\put(0,0){\circle{18}}
\put(60,0){\circle{18}}
\put(9,60){\line(1,0){42}}
\put(9,0){\line(1,0){42}}
\put(0,9){\line(0,1){42}}
\put(60,9){\line(0,1){42}}
\put(7,7){\line(1,1) {46.5}}
\put(7,53){\line(1,-1) {46.5}}
\end{picture}}

\makeindex             

\begin{document}

\title*{Playing with Marbles: Structural and Thermodynamic Properties of Hard-Sphere Systems}
\titlerunning{Structural and Thermodynamic Properties of Hard-Sphere Systems}
\author{Andr\'es Santos}
\institute{Andr\'es Santos \at Departamento de F\'{\i}sica, Universidad de
Extremadura, Badajoz, E-06071, Spain \at \email{andres@unex.es}}
%
%
\maketitle

\abstract{These lecture notes present an overview of equilibrium statistical mechanics of classical fluids, with special applications to the structural and thermodynamic properties of systems made of particles interacting via the hard-sphere potential or closely related model potentials. The exact statistical-mechanical properties of one-dimensional systems, the  issue of thermodynamic (in)consistency among different routes  in the context of several approximate theories, and the construction of analytical or semi-analytical approximations for the structural properties  are also addressed.
}

\numberwithin{equation}{section}
\numberwithin{figure}{section}
\numberwithin{table}{section}

\section{Introduction}
\label{sec1}

Hard-sphere systems represent a favorite playground in statistical mechanics, both in and out of equilibrium, as they represent the simplest models of  many-body systems of interacting particles \cite{M08}.

Apart from their academic or pedagogical values, hard-sphere models are also important from a more practical point of view. In real fluids, especially at high temperatures and moderate and high densities, the structural and thermodynamic properties are mainly governed by the repulsive forces among molecules and in this context hard-core fluids are very useful as  reference systems \cite{BH76,S13}.

Moreover, the use of the hard-sphere model in the realm of soft condensed matter   has become increasingly  popular \cite{L01}. For instance, the effective
interaction among (sterically stabilized) colloidal particles
can be tuned to match almost perfectly the hard-sphere
model \cite{PZVSPC09}.

\begin{figure}[t]
\sidecaption[t]
\includegraphics[width=.64\columnwidth]{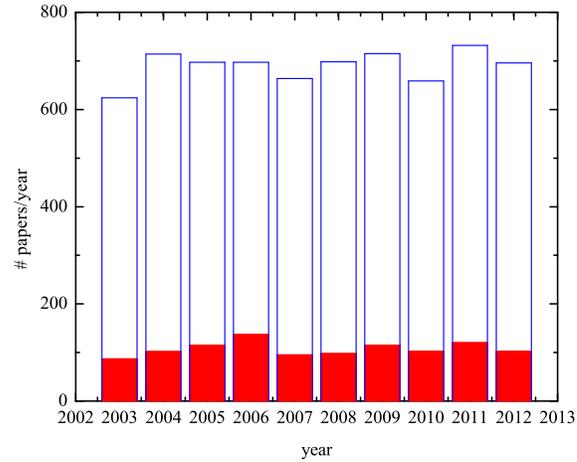}
\caption{Number of papers per year  published in the ten-year period 2003--2012 that include the terms ``hard'' and ``sphere'' as a topic (hollow columns) or in the title (colored columns). \label{fig1.1}}
\end{figure}

As a very imperfect measure of the impact of the hard-sphere model on current  research, Fig.\ \ref{fig1.1} shows the number of papers per year published in the ten-year period 2003--2012 (according to Thomson Reuters' Web of Knowledge) that include the words ``hard'' \emph{and} ``sphere'' as a topic (that is, in the title, in the abstract, or as a keyword). It can be observed that the number is rather stabilized, fluctuating around 700 papers/year. If one constrains the search criterion to papers including ``hard'' and ``sphere'' in the title, about 100 papers/year are found.

Despite the title of this work and the preceding paragraphs, the main aim of these lecture notes is neither restricted to hard-sphere fluids nor focused on the ``state of the art'' of the field. Instead, the notes attempt to present an introduction to the equilibrium statistical mechanics of liquids and non-ideal gases at a graduate-student textbook level, with emphasis on the basics and  fundamentals of the topic. The treatment uses classical (i.e., non-quantum) mechanics and no special prerequisites are required, apart from standard statistical-mechanical ensembles. Most of the content applies to any (short-range) interaction potential, any dimensionality, and (in general) any number of components. On the other hand, some specific applications deal with the  properties of fluids made of particles interacting via the hard-sphere potential or related  potentials. The approach is unavoidably biased toward those aspects the author is more familiarized with. Thus, important topics such as inhomogeneous fluids and density functional theory \cite{E92,TCM08,E10,L10b,L10,R10}, metastable glassy states \cite{PZ10,BB11,K12}, and perturbation theories \cite{BH76,S13} are not  represented in these notes.

Apart from a brief concluding remark, the remainder of these lecture notes is split into the following sections:
\begin{svgraybox}
\begin{itemize}

\item
2. A Brief Survey of Thermodynamic Potentials

\item
3. A Brief Survey of Equilibrium Statistical Ensembles

\item
4. Reduced Distribution Functions

\item
5. Thermodynamics from the Radial Distribution Function

\item
6. One-Dimensional Systems. Exact Solution for Nearest-Neighbor Interactions

\item
7. Density Expansion of the Radial Distribution Function

\item
8. Ornstein--Zernike Relation and Approximate Integral Equation Theories

\item
9. Some Thermodynamic Consistency Relations in Approximate Theories

\item
10. Exact Solution of the Percus--Yevick Equation for Hard Spheres \ldots and Beyond

\end{itemize}

\end{svgraybox}
The core of the notes is made of Sects.\ \ref{sec4}, \ref{sec5}, \ref{sec7}, and \ref{sec8}. They start with the definition of the reduced distribution functions and, in particular, of the radial distribution function $g(r)$ (Sect.\ \ref{sec4}), and continues with the derivation of the main thermodynamic quantities in terms of $g(r)$ (Sect.\ \ref{sec5}). This includes the chemical-potential route, usually forgotten in textbooks. Sections \ref{sec7} and \ref{sec8} are more technical. They have to do with the expansion in powers of density of $g(r)$ and the pressure, the definition of the direct correlation function $c(r)$, and the construction of approximate equations of state and integral-equation theories. Both sections make extensive use of diagrams but several needed theorems and lemmas are justified by simple examples without formal proofs.

In addition to the four core sections mentioned above, there are five more sections that can be seen as optional. Sections \ref{sec2} and \ref{sec3} are included to make the notes as self-contained as possible and to unify the notation, but otherwise can be skipped by the knowledgeable reader. Sections \ref{sec6}, \ref{sec9}, and \ref{sec10} are ``side dishes.'' Whereas one-dimensional systems can be seen as rather artificial, it is undoubtedly important from  pedagogical and illustrative perspectives to derive their exact structural and thermophysical quantities, and this is the purpose of Sect.\ \ref{sec6}. Section \ref{sec9} presents three examples related to the problem of thermodynamic consistency among different routes when an approximate $g(r)$ is employed. Finally, Sect.\ \ref{sec10} derives the exact solution of the Percus--Yevick integral equation for hard spheres as the simplest implementation of a more general class of  approximations.

\section{{A Brief Survey of Thermodynamic Potentials}}
\label{sec2}
Just to fix the notation, this section provides a summary of some of the most important thermodynamic relations.

\subsection{{Isolated Systems. Entropy}}
In a reversible process, the first and second laws of thermodynamics in a fluid mixture can be combined as \cite{Z81,c85}
\beq
\boxed{T\D S
{=}\D E+p\D V -\sum_\nu \mu_\nu \D N_\nu}\;,
\label{2.1}
\eeq
where $S$ is the entropy, $E$ is the internal energy, $V$ is the volume of the fluid, and $N_\nu$ is the number of particles of species $\nu$. All these quantities are \emph{extensive}, i.e., they scale with the size of the system. The coefficients of the differentials in \eqref{2.1} are the conjugate \emph{intensive} quantities: the absolute temperature ($T$), the pressure ($p$), and the chemical potentials ($\mu_\nu$).

Equation \eqref{2.1} shows that the \emph{natural} variables of the entropy are $E$, $V$, and $\{N_\nu\}$, i.e., $S(E,V,\{N_\nu\})$. This implies that $S$ is the right thermodynamic potential in \emph{isolated} systems: at given $E$, $V$, and $\{N_\nu\}$, $S$ is maximal in equilibrium. The respective partial derivatives give the intensive quantities:
\beq
\frac{1}{T}=\left(\frac{\partial S}{\partial E}\right)_{V,\{N_\nu\}}\;,\quad
\frac{p}{T}=\left(\frac{\partial S}{\partial V}\right)_{E,\{N_\nu\}}\;,\quad
\frac{\mu_\nu}{T}=-\left(\frac{\partial S}{\partial N_\nu}\right)_{E,V,\{N_{\gamma\neq \nu}\}}\;.
\label{2.2}
\eeq

The extensive nature of $S$, $E$, $V$, and $\{N_\nu\}$ implies the extensivity condition $S(\lambda E,\lambda V,\{\lambda N_\nu\})=\lambda S(E,V,\{N_\nu\})$. Application of Euler's homogeneous function theorem yields
\beq
S(E,V,\{N_\nu\})=E\left(\frac{\partial S}{\partial E}\right)_{V,\{N_\nu\}} +V\left(\frac{\partial S}{\partial V}\right)_{E,\{N_\nu\}} +\sum_\nu N_\nu\left(\frac{\partial S}{\partial N_\nu}\right)_{E,V,\{N_{\gamma\neq \nu}\}} \;.
\label{2.3}
\eeq
Using \eqref{2.2}, we obtain  the identity
\beq
\boxed{TS=E+pV-\sum_\nu \mu_\nu N_\nu}\;.
\label{2.4}
\eeq
This is the so-called \emph{fundamental equation of thermodynamics}. Differentiating \eqref{2.4} and subtracting \eqref{2.1} one arrives at the Gibbs--Duhem relation
\beq
S\D T-V\D p+\sum_\nu N_\nu \D\mu_\nu=0\;.
\label{2.5}
\eeq

Equation \eqref{2.1}  also shows that $S$, $V$, and $\{N_\nu\}$, are the natural variables of the internal energy $E(S,V,\{N_\nu\})$, so that
\beq
{T}=\left(\frac{\partial E}{\partial S}\right)_{V,\{N_\nu\}}\;,\quad
p=-\left(\frac{\partial E}{\partial V}\right)_{S,\{N_\nu\}}\;,\quad
{\mu_\nu}=\left(\frac{\partial E}{\partial N_\nu}\right)_{S,V,\{N_{\gamma\neq \nu}\}}\;.
\label{2.6}
\eeq

\subsection{{Closed Systems. Helmholtz Free Energy}}
{}From a practical point of view, it is usually more convenient to choose the temperature instead of the internal energy or the entropy as a control variable. In that case,  the adequate thermodynamic potential is no longer either the  entropy or the internal energy, respectively, but the Helmholtz free energy $F$. It is defined from $S$ or $E$ through the Legendre transformation
\beq
F(T,V,\{N_\nu\})=E-TS =-pV+\sum_\nu \mu_\nu N_\nu\;,
\label{2.7}
\eeq
where in the last step use has been made of \eqref{2.4}. {}From \eqref{2.1} we obtain
\beq
\D F=-S \D T-p\D V +\sum_\nu \mu_\nu \D N_\nu\;,
\label{2.8}
\eeq
so that
\beq
{S}=-\left(\frac{\partial F}{\partial T}\right)_{V,\{N_\nu\}}\;,\quad
p=-\left(\frac{\partial F}{\partial V}\right)_{T,\{N_\nu\}}\;,\quad
{\mu_\nu}=\left(\frac{\partial F}{\partial N_\nu}\right)_{T,V,\{N_{\gamma\neq \nu}\}}\;.
\label{2.9}
\eeq
The Helmholtz free energy is the adequate thermodynamic potential in a closed system, that is, a system that cannot exchange mass with the environment but can exchange energy. At fixed $T$, $V$, and $\{N_\nu\}$, $F$ is minimal in equilibrium.

\subsection{{Isothermal-Isobaric Systems. Gibbs Free Energy}}
If, instead of the volume, the independent thermodynamic variable is pressure, we need to perform a Legendre transformation from $F$ to define the Gibbs free energy (or free enthalpy) as
      \beq
G(T,p,\{N_\nu\})=F+pV =\sum_\nu\mu_\nu N_\nu\;.
\label{2.10}
\eeq
The second equality shows that the chemical potential $\mu_\nu$ can be interpreted as the contribution of each particle of species $\nu$ to the total Gibbs free energy. The differential relations now become
\beq
\D G=-S \D T+V\D p +\sum_\nu\mu_\nu \D N_\nu\;,
\label{2.11}
\eeq
\beq
{S}=-\left(\frac{\partial G}{\partial T}\right)_{p,\{N_\nu\}}\;,\quad
V=\left(\frac{\partial G}{\partial p}\right)_{T,\{N_\nu\}}\;,\quad
{\mu_\nu}=\left(\frac{\partial G}{\partial N_\nu}\right)_{T,p,\{N_{\gamma\neq \nu}\}}\;.
\label{2.12}
\eeq
Needless to say, $G$ is minimal in equilibrium if one fixes $T$, $p$, and $\{N_\nu\}$.

\subsection{{Open Systems. Grand Potential}}
In an open system, not only energy but also particles can be exchanged with the environment. In that case, we need to replace $\{N_\nu\}$ by $\{\mu_\nu\}$ as independent variables and define the grand potential $\Omega$ from $F$ via a new Legendre transformation:
     \beq
\Omega(T,V,\{\mu_\nu\})=F-\sum_\nu \mu_\nu N_\nu =-pV.
\label{2.13}
\eeq
Interestingly, the second equality shows that $-\Omega/V$ is not but the pressure, except that it must be seen as a function of temperature and the chemical potentials. Now we have
\beq
\D \Omega=-S \D T-p\D V -\sum_\nu N_\nu \D \mu_\nu\;.
\label{2.14}
\eeq
\beq
{S}=-\left(\frac{\partial \Omega}{\partial T}\right)_{V,\{\mu_\nu\}}\;,\quad
p=-\left(\frac{\partial \Omega}{\partial V}\right)_{T,\{\mu_\nu\}}=-\frac{\Omega}{V}\;,\quad
{N_\nu}=-\left(\frac{\partial \Omega}{\partial \mu_\nu}\right)_{T,p,\{\mu_{\gamma\neq \nu}\}}\;.
\label{2.15}
\eeq

\subsection{{Response Functions}}
We have seen that the thermodynamic variables $E\leftrightarrow T$ (or $S\leftrightarrow T$), $V\leftrightarrow p$, and $N_\nu\leftrightarrow \mu_\nu$ appear as $\text{extensive}\leftrightarrow \text{intensive}$ conjugate pairs. Depending on the thermodynamic potential of interest, one of the members of the pair acts as independent variable and the other one is obtained by differentiation. If an additional  derivative is taken one obtains the so-called \emph{response} functions. For example,  the heat capacities at constant volume and at constant pressure are defined as
\beq
C_V=\left(\frac{\partial E}{\partial T}\right)_{V,\{N_\nu\}}=T\left(\frac{\partial S}{\partial T}\right)_{V,\{N_\nu\}}=-T\left(\frac{\partial^2 F}{\partial T^2}\right)_{V,\{N_\nu\}}\;,
\label{2.16}
\eeq
\beq
C_p=T\left(\frac{\partial S}{\partial T}\right)_{p,\{N_\nu\}}=-T\left(\frac{\partial^2 G}{\partial T^2}\right)_{p,\{N_\nu\}}\;.
\label{2.17}
\eeq
Analogously, it is convenient to define the isothermal compressibility
\begin{eqnarray}
\kappa_T&=&-\frac{1}{V}\left(\frac{\partial V}{\partial p}\right)_{T,\{N_\nu\}}=-\frac{1}{V}\left(\frac{\partial^2 G}{\partial p^2}\right)_{T,\{N_\nu\}}\nn
&=&-\left[{V\left(\frac{\partial p}{\partial V}\right)_{T,\{N_\nu\}}}\right]^{-1}=\left[{V\left(\frac{\partial^2 F}{\partial V^2}\right)_{T,\{N_\nu\}}}\right]^{-1}\;,
\label{2.18}
\end{eqnarray}
and the thermal expansivity
\beq
\alpha_p=\frac{1}{V}\left(\frac{\partial V}{\partial T}\right)_{p,\{N_\nu\}}=\frac{1}{V}\left(\frac{\partial^2 G}{\partial T\partial p}\right)_{\{N_\nu\}}
{=}-\frac{1}{V}\left(\frac{\partial S}{\partial p}\right)_{T,\{N_\nu\}}\;.
\label{2.19}
\eeq
The equivalence between the second and fourth terms in \eqref{2.19} is an example of a Maxwell relation.

\section{A Brief Survey of Equilibrium Statistical Ensembles}
\label{sec3}
In this section a summary of the main equilibrium ensembles is presented, essentially to fix part of the notation that will be needed later on. For simplicity, we will restrict this section to one-component systems, although the extension to mixtures is straightforward.

Let us consider a \emph{classical} system made of $N$ \emph{identical} point particles in $d$ dimensions. In classical mechanics, the dynamical state of the system is characterized by the $N$ vector positions $\{\rr_1,\rr_2,\ldots,\rr_N\}$ and the $N$ vector momenta $\{\pp_1,\pp_2,\ldots,\pp_N\}$. In what follows, we will employ the following short-hand notation
\begin{svgraybox}
 \begin{itemize}
    \item
     $ \rr^N=\{\rr_1,\rr_2,\ldots,\rr_N\}$, $\D \rr^N=\D\rr_1\D\rr_2 \cdots \D\rr_N$,
     \item
     $ \pp^N=\{\pp_1,\pp_2,\ldots,\pp_N\}$, $\D \pp^N=\D\pp_1\D\pp_2 \cdots \D\pp_N$,
     \item
     $\XX^N=\{\rr^N,\pp^N\}$, $\D \XX^N=\D \rr^N \D\pp^N$.
    \end{itemize}
\end{svgraybox}

\begin{figure}[t]
\sidecaption[t]
\includegraphics[width=.5\columnwidth]{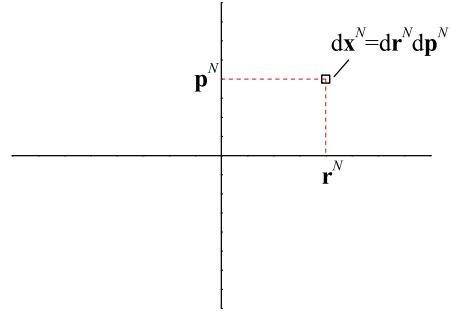}
\caption{Sketch of the phase space of a system of $N$ identical particles. The horizontal axis represents the $d\times N$ position variables ($d$ components for each particle), while the vertical axis represents the $d\times N$ momentum variables. A differential phase-space volume $\D\XX^N$ around a point $\XX^N$ is represented. \label{fig3.1}}
\end{figure}

Thus, the whole \emph{microscopic}  state of the system (\emph{microstate}) is represented by a single point $\XX^N$ in the $(2d\times N)$-dimensional \emph{phase space} (see Fig.~\ref{fig3.1}). The time evolution of the microstate $\XX^N$ is governed by the Hamiltonian of the system $H_N(\XX^N)$ through the classical Hamilton's equations \cite{GSP13}.

Given the practical impossibility of describing the system at a microscopic level, a statistical description is needed. Thus, we define the phase-space probability distribution  function $\varrho_N(\XX^N)$ such that
$\varrho_N(\XX^N)\D\XX^N$ is the probability that the microstate of the system lies inside an infinitesimal
(hyper)volume $\D\XX^N$ around the phase-space point $\XX^N$.

\subsection{{Gibbs Entropy}}
The concept of a phase-space probability distribution  function is valid both out of equilibrium
(where, in general, it changes with time according to the Liouville theorem \cite{B74b,R80}) and in equilibrium (where it is stationary).
In the latter case $\varrho_N(\XX^N)$ can be obtained for isolated, closed, open, \ldots systems by following logical steps and starting from the  \emph{equal a priori probability postulate} for isolated systems. Here we follow an alternative (but equivalent) method based on information-theory arguments \cite{R80,SW71,BN08}.

Let us define the Gibbs entropy \emph{functional}
\beq
\boxed{S[\varrho_N]=-k_B\int\D \mathbf{x}^N\, \varrho_N(\XX^N)\ln \left[C_N\varrho_N(\XX^N)\right]}\;,
\label{3.1}
\eeq
where $k_B$ is the Boltzmann constant and
\beq
C_N\equiv N! h^{dN}\;.
\label{3.2}
\eeq
In \eqref{3.2} the coefficient $h^{dN}$ is introduced to comply with Heisenberg's uncertainty principle and preserve the non-dimensional character of the argument of the logarithm, while the factorial $N!$ accounts for the fact that two apparently different microstates which only differ on the particle labels are physically the same microstate (thus avoiding Gibbs's paradox).

Equation \eqref{3.1} applies to systems with a fixed number of particles $N$. On the other hand, if the system is allowed to exchange particles with the environment, microstates with different $N$ exist, so that one needs to define a a phase-space density $\varrho_N(\XX^N)$ for each $N\geq 0$. In that case, the entropy functional becomes
\beq
S[\{\varrho_N\}]=-k_B\sum_{N=0}^\infty \int\D \mathbf{x}^N\, \varrho_N(\XX^N)\ln \left[C_N\varrho_N(\XX^N)\right]\;.
\label{3.3}
\eeq

Now, the basic postulate consists of asserting that, out of all possible phase-space probability distribution  functions $\varrho_N$ consistent with given \emph{constraints} (which define the \emph{ensemble} of accessible microstates), the \emph{equilibrium} function $\varrho_N^{\text{eq}}$ is the one that \emph{maximizes} the entropy functional $S[\varrho_N]$.
Once $\varrho_N^{\text{eq}}$ is known, connection with thermodynamics is made through the identification of $S^{\text{eq}}=S[\varrho_N^{\text{eq}}]$ as the equilibrium entropy.

\subsection{{Microcanonical Ensemble (Isolated System)}}
The microcanical ensemble describes an isolated system and thus it is characterized by fixed values of $V$, $N$, $E$ (the latter with a tolerance $\varDelta E$, in accordance with the uncertainty principle). Therefore, the basic constraint is the normalization condition
\beq
  \int_{E\leq H_N(\mathbf{x}^N)\leq E+\varDelta E}\D\mathbf{x}^N\,\varrho_N(\mathbf{x}^N)=1\;.
  \label{3.4}
\eeq
Maximization of the entropy functional just says that $\varrho_N(\mathbf{x}^N)=\text{const}$ for all the accessible microstates $E\leq H_N(\mathbf{x}^N)\leq E+\varDelta E$. Thus,
\beq
  \varrho_N(\mathbf{x}^N)=\begin{cases}
    \frac{1}{\omega_{\varDelta E}(E,N,V)}\;,& E\leq H_N(\mathbf{x}^N)\leq E+\varDelta E\;,\\
    0\;,& \text{otherwise}\;.
  \end{cases}
\label{3.5}
\eeq
The normalization function
\beq
  \omega_{\varDelta E}(E,N,V)=\int_{E\leq H_N(\mathbf{x}^N)\leq E+\varDelta E}\D\mathbf{x}^N\;
\label{3.6}
\eeq
is the phase-space volume comprised between the hyper-surfaces $H_N(\mathbf{x}^N)= E$ and $H_N(\mathbf{x}^N)= E+\varDelta E$. By insertion of \eqref{3.5} into \eqref{3.1} one immediately sees that $\omega_{\varDelta E}(E,N,V)$ is directly related to the equilibrium entropy as
\beq
S(E,N,V)=k_B\ln\frac{\omega_{\varDelta E}(E,N,V)}{N! h^{dN}}\;.
\label{3.7}
\eeq
In this expression the specific value of $\varDelta E$ becomes irrelevant in the thermodynamic limit (as long as $\varDelta E \ll E$).

\subsection{{Canonical Ensemble (Closed System)}}
Now the system can have \emph{any} value of the total energy $E$. However, we are free to prescribe a given value of the \emph{average} energy $\llangle E\rrangle$. Therefore, the constraints in the canonical ensemble are
  \beq
  \int\D\XX^N\,\varrho_N(\XX^N)=1\;,\quad \int\D\XX^N\,H_N(\XX^N)\varrho_N(\XX^N)=\llangle E\rrangle\;.
\label{3.8}
  \eeq
The maximization of the entropy functional subject to the constraints \eqref{3.8} can be carried out through the Lagrange multiplier method with the result
  \beq
  \varrho_N(\mathbf{x}^N)=\frac{\E^{-\beta H_N(\mathbf{x}^N)}}{N! h^{dN}\mathcal{Z}_N(\beta,V)}\;,
\label{3.9}
\eeq
where $\beta$ is the Lagrange multiplier associated with $\llangle E\rrangle$ and the \emph{partition function} $\mathcal{Z}_N$ is determined from the normalization condition as
  \beq
  \mathcal{Z}_N(\beta,V)=\frac{1}{N! h^{dN}}\int \D\mathbf{x}^N\, \E^{-\beta H_N(\mathbf{x}^N)}\;.
  \label{3.10}
  \eeq
Substitution of \eqref{3.9} into \eqref{3.1} and use of \eqref{3.8} yields
  \beq
  S=k_B\left(\ln \mathcal{Z}_N+\beta{\llangle E\rrangle}\right)\;.
  \label{3.11}
  \eeq
Comparison with \eqref{2.7} (where now the internal energy is represented by $\llangle E\rrangle$) allows one to identify
  \beq
\beta= \frac{1}{k_BT}\;,\quad F(T,N,V)=-k_B T\ln  \mathcal{Z}_N(\beta,V)\; .
\label{3.12}
  \eeq
Therefore, in the canonical ensemble the connection with thermodynamics is conveniently established via the Helmholtz free energy rather than via the entropy.

As an average of a phase-space dynamical variable, the internal energy can be directly obtained from  $\ln  \mathcal{Z}_N$ as
  \beq
  \llangle E\rrangle= -\frac{\partial \ln \mathcal{Z}_N}{\partial \beta}\;.
  \label{3.13}
  \eeq
Moreover, we can obtain the energy fluctuations:
  \beq
  \llangle E^2\rrangle-\llangle E\rrangle^2=\frac{\partial^2 \ln \mathcal{Z}_N}{\partial \beta^2}=k_BT^2 C_V\;.
  \label{3.14}
  \eeq
In the last step, use has been made of \eqref{2.16}.

\subsection{{Grand Canonical Ensemble (Open System)}}
In an open system neither the energy nor the number of particles is determined but we can choose to fix their average values. As a consequence, the constraints are
 \beq
  \sum_{N=0}^\infty \int\D\XX^N\,\varrho_N(\XX^N)=1\;,\quad \sum_{N=0}^\infty \int\D\XX^N\,H_N(\XX^N)\varrho_N(\XX^N)=\llangle E\rrangle\;,
  \label{3.15}
  \eeq
  \beq
  \sum_{N=0}^\infty N \int\D\XX^N\,\varrho_N(\XX^N)=\llangle N\rrangle\;.
  \label{3.16}
  \eeq
The solution to the maximization problem is
       \beq
  \varrho_N(\mathbf{x}^N)=\frac{\E^{-\alpha N}\E^{-\beta H_N(\mathbf{x}^N)}}{N! h^{dN}\Xi(\beta,\alpha,V)}\;,
  \label{3.17}
     \eeq
where $\alpha$ and $\beta$ are Lagrange multipliers and the \emph{grand partition function} is
  \beq
  \Xi(\beta,\alpha,V)=\sum_{N=0}^\infty \frac{\E^{-\alpha N}}{N! h^{dN}}\int \D\mathbf{x}^N\, \E^{-\beta H_N(\mathbf{x}^N)}=\sum_{N=0}^\infty \E^{-\alpha N}\mathcal{Z}_N(\beta,V)\;.
  \label{3.18}
  \eeq
In this case the equilibrium entropy becomes
  \beq
  S=k_B\left(\ln\Xi+\beta{\llangle E\rrangle}{T}+\alpha\llangle N\rrangle\right)\;.
  \eeq
{}From comparison with the first equality of \eqref{2.13} we can identify
  \beq
 \beta= \frac{1}{k_BT}\;,\quad \alpha= -\beta\mu\;,\quad\Omega(T,\mu,V)=-k_BT\ln \Xi(\beta,\alpha,V)\;.
 \label{3.19}
  \eeq
The  average  and fluctuation relations are
  \beq
  \llangle E\rrangle= -\frac{\partial \ln \Xi}{\partial \beta}\;,\quad
  \llangle E^2\rrangle-\llangle E\rrangle^2=\frac{\partial^2 \ln \Xi}{\partial \beta^2}=k_BT^2 C_V\;,
  \label{3.20}
  \eeq
  \beq
  \llangle N\rrangle= -\frac{\partial \ln \Xi}{\partial \alpha}\;,
  \label{3.21}
  \eeq
   \beq
  \llangle N^2\rrangle-\llangle N\rrangle^2=\frac{\partial^2 \ln \Xi}{\partial \alpha^2}=k_BT \frac{\llangle N\rrangle^2}{V}\kappa_T\;.
  \label{3.22}
  \eeq
The second equality of \eqref{3.22} requires the use of thermodynamic relations and mathematical properties of partial derivatives.

\subsection{{Isothermal-Isobaric Ensemble}}
In this ensemble the volume is a fluctuating quantity and only its average value is fixed. Thus, similarly to the grand canonical ensemble, the constraints are
  \beq
  \int_{0}^\infty \D V \int\D\XX^N\,\varrho_N(\XX^N)=1\;,
  \quad \int_{0}^\infty \D V \int\D\XX^N\,H_N(\XX^N)\varrho_N(\XX^N)=\llangle E\rrangle\;,
  \label{3.23}
  \eeq
   \beq
 \int_{0}^\infty \D V\,V \int\D\XX^N\,\varrho_N(\XX^N)=\llangle V\rrangle\;.
 \label{3.24}
  \eeq
Not surprisingly, the solution  is
      \beq
  \varrho_N(\mathbf{x}^N)=\frac{\E^{-\gamma V}\E^{-\beta H_N(\mathbf{x}^N)}}{V_0 N! h^{dN}\Delta_N(\beta,\gamma)}\;,
  \label{3.25}
    \eeq
where   $V_0$ is an arbitrary volume scale factor (needed to keep the correct dimensions), $\gamma$ and $\beta$ are again Lagrange multipliers, and the isothermal-isobaric partition function is
    \beq
  \Delta_N(\beta,\gamma)=\frac{1}{V_0N! h^{dN}}\int_0^\infty \D V\,{\E^{-\gamma V}}\int \D\mathbf{x}^N\, \E^{-\beta H_N(\mathbf{x}^N)}=\frac{1}{V_0}\int_0^\infty \D V\,{\E^{-\gamma V}} \mathcal{Z}_N(\beta,V)\;.
  \label{3.26}
  \eeq
As expected, the entropy becomes
  \beq
  S=k_B\left(\ln\Delta_N+\beta{\llangle E\rrangle}+\gamma\llangle V\rrangle\right)\;.
  \label{3.27}
  \eeq
{}From comparison with \eqref{2.10} we conclude that
  \beq
  \beta= \frac{1}{k_BT}\;,\quad \gamma= \beta p\;,\quad G(T,p,N)=-k_BT\ln \Delta_N(\beta,\gamma)\;.
  \label{3.27b}
  \eeq
The main average  and fluctuation relations are
  \beq
  \llangle E\rrangle= -\frac{\partial \ln \Delta_N}{\partial \beta}\;,\quad
   \llangle E^2\rrangle-\llangle E\rrangle^2=\frac{\partial^2 \ln \Delta_N}{\partial \beta^2}=k_BT^2 C_V\;,
   \label{3.28}
  \eeq
   \beq
  \llangle V\rrangle= -\frac{\partial \ln \Delta_N}{\partial \gamma}\;,
  \label{3.29}
  \eeq
  \beq
  \llangle V^2\rrangle-\llangle V\rrangle^2=\frac{\partial^2 \ln \Delta_N}{\partial \gamma^2}=k_BT{\llangle V\rrangle}\kappa_T\;.
  \label{3.30}
  \eeq
Equations \eqref{3.22} and \eqref{3.30} are equivalent. Both show that the density fluctuations are proportional to the isothermal compressibility and decrease as the size of the system increases. In \eqref{3.22} the volume is constant, so that the density fluctuations are due to fluctuations in the number of particles, while the opposite happens in \eqref{3.30}.

\subsection{{Ideal Gas}}
The exact evaluation of the normalization functions \eqref{3.6}, \eqref{3.10}, \eqref{3.18}, and \eqref{3.26} is in general a formidable task due to the involved dependence of the Hamiltonian on the coordinates of the particles. However, in the case of non-interacting particles (ideal gas), the Hamiltonian  depends only on the momenta:
\beq
H_N(\XX^N)\to H_N^\id(\mathbf{p}^N)=\sum_{i=1}^N \frac{p_i^2}{2m}\;,
\label{3.31}
\eeq
where $m$ is the mass of a particle. In this case the $N$-body Hamiltonian is just the sum over all the particles of the one-body Hamiltonian $p_i^2/2m$ and the exact statistical-mechanical results can be easily  obtained. The expressions for the normalization function, the thermodynamic potential, and the first derivatives of the latter for each one of the four ensembles considered above are the following ones:

\begin{itemize}

\item
\emph{Microcanonical ensemble}
\beq
\ln \omega^\id_{\Delta E}(E,N,V)=N\left\{\frac{d}{2}+\ln\left[{V}\left(\frac{4\pi m E}{dN}\right)^{d/2}\right]\right\}\;,
\label{3.32}
\eeq
\beq
S^\id(E,N,V)=Nk_B\left\{\frac{d+2}{2}+\ln\left[\frac{V}{N}\left(\frac{4\pi m E}{dN h^2}\right)^{d/2}\right]\right\}\;,
\label{3.33}
\eeq
\beq
k_BT^\id=\frac{2}{d}\frac{E}{N}\;,\quad p^\id=\frac{2}{d}\frac{E}{V}\;,\quad
\mu^\id=-\frac{2}{d}\frac{E}{N}\ln\left[\frac{V}{N}\left(\frac{4\pi m E}{dN h^2}\right)^{d/2}\right]\;,
\label{3.34}
\eeq
\item
\emph{Canonical ensemble}
\beq
\mathcal{Z}_N^\id(\beta,V)=\frac{\left[\zeta(\beta,V)\right]^N}{N!}\;,\quad \zeta(\beta,V)=\frac{V}{\left[\Lambda(\beta)\right]^d}\;,\quad
\Lambda(\beta)\equiv\frac{h}{\sqrt{2\pi m k_BT}}\;,
\label{3.35}
\eeq
\beq
F^\id(T,N,V)=Nk_BT \left[\ln\frac{\Lambda^d(\beta)}{V/N}-1\right]\;,
\label{3.36}
\eeq
\beq
\llangle E\rrangle^\id=\frac{d}{2}Nk_BT\;,\quad p^\id=\frac{N}{V}k_BT\;,\quad \mu^\id=k_BT\ln\left[\frac{\Lambda^d(\beta)}{V/N}\right]\;,
\label{3.37}
\eeq

\item
\emph{Grand canonical ensemble}
\beq
\Xi^\id(\beta,\alpha,V)=\sum_{N=0}^\infty \E^{-\alpha N}\frac{\left[\zeta(\beta,V)\right]^N}{N!}=\E^{z\zeta}\;,\quad
z\equiv \E^{-\alpha}=\E^{\beta\mu}\;,
\label{3.38}
\eeq
\beq
\Omega^\id(\beta,\alpha,V)=-p^\id V=-k_BT \E^{-\alpha} \zeta(\beta,V)\;,
\label{3.39}
\eeq
\beq
\llangle E\rrangle^\id=\frac{d}{2}k_BT \E^{-\alpha}\zeta(\beta,V)\;,
\quad
\llangle N\rrangle^\id= \E^{-\alpha}\zeta(\beta,V)\;,
\label{3.40}
\eeq

\item
\emph{Isothermal-isobaric ensemble}
\beq
\Delta_N^\id(\beta,\gamma)=\frac{1}{V_0N![\Lambda(\beta)]^{dN}}\int_0^\infty \D V\, {V^N}\E^{-\gamma V}=\frac{\gamma^{-(N+1)}}{V_0[\Lambda(\beta)]^{dN}}\;.
\label{3.41}
\eeq
\beq
G^\id(N,p,T)=\mu^\id N=Nk_BT \ln\left[\beta p \Lambda^d(\beta)\right]\;,
\label{3.42}
\eeq
\beq
\llangle E\rrangle^\id=\frac{d}{2}N k_BT\;,
\quad
\llangle V\rrangle^\id= \frac{N k_BT}{p}\;.
\label{3.43}
\eeq
\end{itemize}

In \eqref{3.35} $\zeta$ is the one-particle partition function and $\Lambda$ is the thermal de Broglie wavelength. In \eqref{3.38} $z$ is the \emph{fugacity}. Note that \eqref{3.33}, \eqref{3.36}, \eqref{3.39}, and \eqref{3.42} are equivalent. Likewise, \eqref{3.34}, \eqref{3.37}, \eqref{3.40}, and \eqref{3.43} are also equivalent. This a manifestation of the ensemble equivalence in the thermodynamic limit, the only difference lying in the choice of independent and dependent variables.

\subsection{{Interacting Systems}}
Of course, particles do interact in real systems, so the Hamiltonian has the form
\beq
H_N(\XX^N)= H_N^\id(\mathbf{p}^N)+\Phi_N(\rr^N)\;,
\label{3.44}
\eeq
where $\Phi_N$ denotes the total potential energy. As a consequence, the partition function factorizes into  its ideal and non-ideal parts:
 \beq
\boxed{\mathcal{Z}_N(\beta,V)=\mathcal{Z}_N^\id(\beta,V)Q_N(\beta,V)\;,\quad
Q_N(\beta,V)=V^{-N}\int\D\rr^N\, \E^{-\beta\Phi_N(\rr^N)}}\;.
\label{3.45}
\eeq
The non-ideal part $Q_N$ is the \emph{configuration integral}. In the canonical ensemble, $Q_N$ is responsible for the \emph{excess} contributions $\llangle E\rrangle^\ex=\llangle E\rrangle-\llangle E\rrangle^\id$, $p^\ex=p-p^\id$, $\mu^\ex=\mu-\mu^\id$:
\beq
\llangle E\rrangle^\ex=-\frac{\partial\ln Q_N}{\partial\beta}\;,\quad
p^\ex=k_BT\frac{\partial\ln Q_N}{\partial V}\;,\quad
\mu^\ex=-k_BT\frac{\partial\ln Q_N}{\partial N}\;.
\label{3.46}
\eeq

The grand partition function does not factorize but can be written as
\beq
\boxed{\Xi(\beta,\alpha,V)=1+\sum_{N=1}^\infty \frac{V^NQ_N(\beta,V)}{N!}[\zz(\beta,\alpha)]^N}\;,
\label{3.47}
\eeq
where
\beq
\zz(\beta,\alpha)\equiv \frac{z(\alpha)}{\left[\Lambda(\beta)\right]^d}
\label{3.47b}
\eeq
is a sort of modified fugacity and we have taken into account that $Q_0=1$. Thus, the configuration integrals are related to the coefficients in the expansion of the grand partition function in powers of the quantity $\zz$.

\section{{Reduced Distribution Functions}}
\label{sec4}
The $N$-body probability distribution function $\varrho_N(\XX^N)$ contains all the statistical-mechanical information about the system. On the other hand, partial information embedded in \emph{marginal} few-body distributions are usually enough for the most relevant quantities. Moreover, it is much simpler to introduce useful approximations at the level of the marginal distributions than at the $N$-body level.

\begin{figure}[t]
\sidecaption[t]
\includegraphics[width=.5\columnwidth]{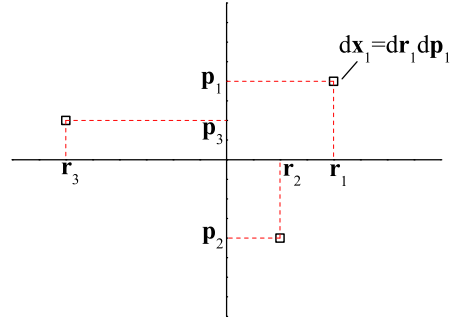}
\caption{Sketch of the one-body phase space. The horizontal axis represents the $d$ position coordinates, while the vertical axis represents the $d$ momentum components. Three points ($\XX_1$, $\XX_2$, and $\XX_3$) are represented. \label{fig4.1}}
\end{figure}

Let us introduce the $s$-body \emph{reduced distribution function} $f_s(\XX^s)$ such that
$f_s(\XX^s)\D\XX^s$ is the (average) number of groups of $s$ particles such that one particle lies inside a volume $\D\XX_1$  around the (1-body) phase-space point $\XX_1$, other particle lies inside a volume $\D\XX_2$ around the (1-body) phase-space point $\XX_2$, \ldots and so on (see Fig.\ \ref{fig4.1} for $s=3$).
More explicitly,
   \beqa
f_s(\mathbf{x}^s)&=&
\sum_{i_1\neq i_2\neq\cdots \neq i_s}\int\D {\XX'}^N\, \delta(\XX_{i_1}'-\XX_1) \cdots \delta(\XX_{i_s}'-\XX_s)\varrho_N({\XX'}^N)\nn
&=&
\frac{N!}{(N-s)!}\int \D\mathbf{x}_{s+1}\int \D\mathbf{x}_{s+2}\cdots \int \D\mathbf{x}_{N}\,
\varrho_N(\mathbf{x}^N)\;.
\label{4.1}
\eeqa
In most situations it is enough to take $s=2$ and integrate out the momenta. Thus, we define the  \emph{configurational} two-body distribution function as
\beq
n_2(\mathbf{r}_1, \mathbf{r}_2)=\int \D\mathbf{p}_{1}\int \D\mathbf{p}_{2}\,f_2(\mathbf{x}_1, \mathbf{x}_2)\;.
\label{4.2}
\eeq
Obviously, its normalization condition is
\beq
\int \D\mathbf{r}_{1}\int \D\mathbf{r}_{2}\,n_2(\mathbf{r}_1, \mathbf{r}_2)=N(N-1)\;.
\label{4.3}
\eeq
The importance of $n_2$  arises especially when one is interested in evaluating the average of a dynamical variable of the form
\beq
A(\mathbf{r}^N)=\frac{1}{2}\sum_{i\neq j} A_2(\mathbf{r}_{i}, \mathbf{r}_{j})\;.
\label{4.4}
\eeq
In that case, it is easy to obtain
\beq
\llangle A\rrangle\equiv\int \D\mathbf{x}^N \, A(\mathbf{r}^N) \varrho_N(\mathbf{x}^N)=\frac{1}{2}\int \D\mathbf{r}_{1}\int \D\mathbf{r}_{2}\,A_2(\mathbf{r}_1, \mathbf{r}_2)n_2(\mathbf{r}_1,
 \mathbf{r}_2)\;.
 \label{4.5}
\eeq

The quantities \eqref{4.1} and \eqref{4.2} can be defined both out of and in equilibrium. In the latter case, however, we can benefit from the (formal) knowledge of $\varrho_N$. In particular, in the canonical ensemble [see \eqref{3.9} and \eqref{3.45}] one has
  \beq
 n_2(\mathbf{r}_1,  \mathbf{r}_2)=\frac{N(N-1)}{V^N Q_N}\int \D\mathbf{r}_{3}\cdots \int \D\mathbf{r}_{N}\,\E^{-\beta \Phi_N(\mathbf{r}^N)}\;.
 \label{4.6}
  \eeq
In the absence of interactions ($\Phi_N=0$),
\beq
  n_2^\id=\frac{N(N-1)}{V^2}\approx n^2\;,\quad n\equiv \frac{N}{V}\;.
  \label{4.7}
  \eeq

In the grand canonical ensemble the equations equivalent to \eqref{4.3}, \eqref{4.6}, and \eqref{4.7} are
\beq
\int \D\mathbf{r}_{1}\int \D\mathbf{r}_{2}\,n_2(\mathbf{r}_1, \mathbf{r}_2)=\llangle N(N-1)\rrangle\;,
\label{4.8}
\eeq
  \beq
  n_2(\mathbf{r}_1,  \mathbf{r}_2)=\frac{1}{\Xi}\sum_{N=2}^\infty\frac{[\zz(\beta,\alpha)]^N}{N!}N(N-1)\int \D\mathbf{r}_{3}\cdots \int \D\mathbf{r}_{N} \,\E^{-\beta \Phi_N(\mathbf{r}^N)}\;,
  \label{4.9}
  \eeq
  \beq
  n_2^\id=\frac{\llangle N(N-1)\rrangle}{V^2}\approx n^2,\quad n\equiv \frac{\llangle N\rrangle}{V}\;.
  \label{4.10}
  \eeq

\subsection{{Radial Distribution Function}}
Taking into account \eqref{4.7} and \eqref{4.10}, we define the \emph{pair} correlation function $g(\mathbf{r}_1, \mathbf{r}_2)$ by
\beq
n_2(\mathbf{r}_1, \mathbf{r}_2)=n^2 {g(\mathbf{r}_1, \mathbf{r}_2)}\;.
\label{4.11}
\eeq
Thus, according to \eqref{4.6},
\beq
\boxed{g(\mathbf{r}_1, \mathbf{r}_2)=\frac{V^{-(N-2)}}{Q_N}\int\D\rr_3\cdots\int\D\rr_N\, \E^{-\beta\Phi_N(\rr^N)}}
\label{4.12}
\eeq
in the canonical ensemble.

Now, taking into account the translational invariance property of the system, one has $g(\mathbf{r}_1, \mathbf{r}_2)=g(\mathbf{r}_1-\mathbf{r}_2)$.
Moreover, a fluid is rotationally invariant, so that (assuming central forces), $g(\mathbf{r}_1-\mathbf{r}_2)=g(r_{12})$, where $r_{12}\equiv |\mathbf{r}_1-\mathbf{r}_2|$ is the distance between the points $\rr_1$ and $\rr_2$.
In such a case, the function $g(r)$ is called {\emph{radial distribution function}} and will play a very important role henceforth.

An interesting normalization relation holds in the grand canonical ensemble. Inserting \eqref{4.11} into \eqref{4.8} we get
\beq
\boxed{V^{-1}\int \D\mathbf{r}\, g(r)= \frac{\llangle N(N-1)\rrangle}{\llangle N\rrangle^2}=\frac{\llangle N^2\rrangle}{\llangle N\rrangle^2}
-\frac{1}{\llangle N\rrangle}}\;.
\label{4.12b}
\eeq
In the thermodynamic limit ($\llangle N\rrangle\to\infty$ and $V\to\infty$ with $n=\text{const}$), we know that ${\llangle N^2\rrangle}/{\llangle N\rrangle^2}\to 1$ [see \eqref{3.22}] (except near the critical point, where $\kappa_T$ diverges). This implies that $V^{-1}\int \D\mathbf{r}\, g(r)\approx 1$, meaning that $g(r)\approx 1$ for \emph{macroscopic} distances $r$, which are those dominating the value of the integral. In other words, $\int \D\mathbf{r}\, \left[g(r)-1\right]\ll V$.

\begin{figure}[t]
\includegraphics[width=.39\columnwidth]{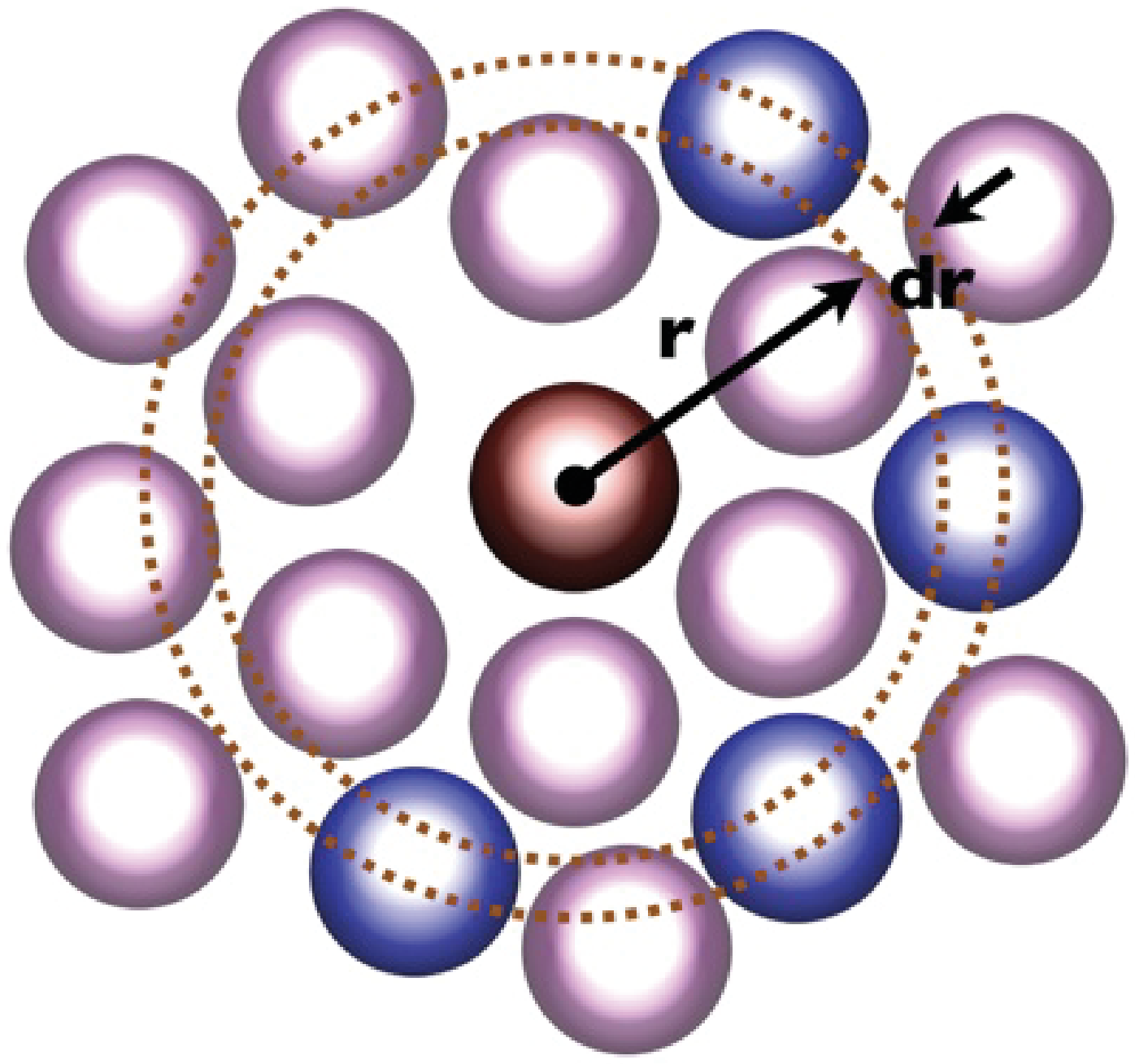}
\includegraphics[width=.6\columnwidth]{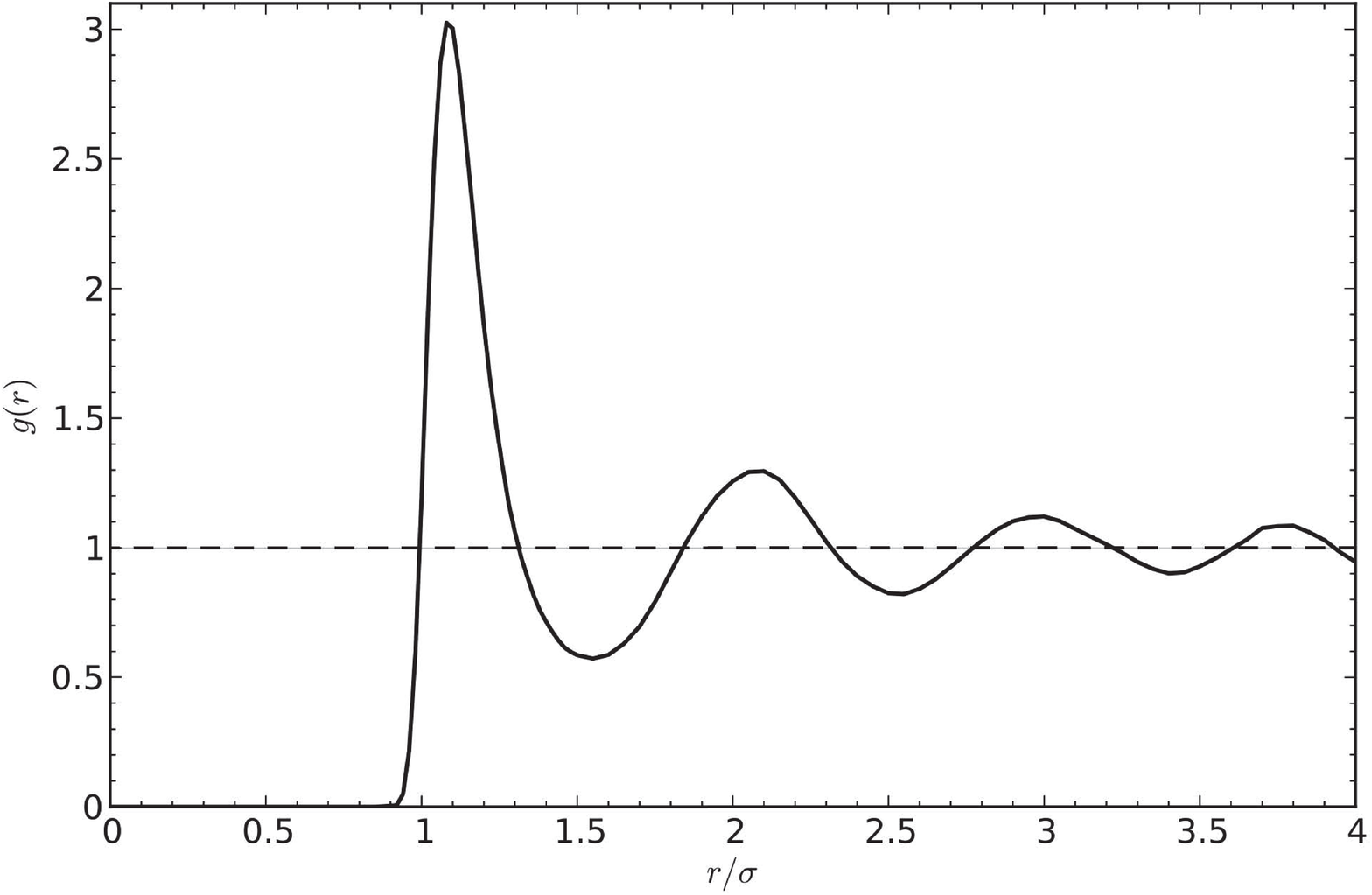}
\caption{Left panel: Schematic view of how $g(r)$ is determined. The red particle is the reference one and  the blue particles are those whose centers are at a distance between $r$ and $r+\D r$. The average number of blue particles, divided by $n 4\pi r^2 \D r$ (in three dimensions) gives $g(r)$. Right panel: {Radial distribution function for a Lennard-Jones fluid at a reduced temperature $T^*=0.71$ and a reduced density $n^*=0.844$, as obtained from Monte Carlo simulations.} Source:
\url{http://en.wikipedia.org/wiki/Radial\_distribution\_function}.
\label{fig4.2}}
\end{figure}

Apart from the formal definition provided by \eqref{4.11} and \eqref{4.12}, it is important to have a more intuitive physical interpretation of $g(r)$. Two simple equivalent interpretations are:
\begin{svgraybox}
\begin{itemize}
\item
$g(r)$ is the probability of finding a particle at a distance $r$ away from a given reference particle, \emph{relative} to the probability for an ideal gas.
\item
If a given reference particle is taken to be at the origin,  then the \emph{local} average density at a distance $r$  from that particle is $ng(r)$.
\end{itemize}
  \end{svgraybox}
Figure \ref{fig4.2} illustrates the meaning of $g(r)$ and depicts the typical shape of the function for a (three-dimensional) fluid of particles interacting via the Lennard-Jones (LJ) potential
\beq
  \phi(r)=4\varepsilon\left[\left(\frac{\sigma}{r}\right)^{12}-\left(\frac{\sigma}{r}\right)^{6}\right]
  \label{4.13}
  \eeq
at the reduced temperature $T^*\equiv k_BT/\varepsilon=0.71$ and the reduced density $n^*=n\sigma^3=0.844$.
The Lennard-Jones potential is characterized by a scale distance $\sigma$ and well depth $\epsilon$, and is repulsive for $r<2^{1/6}\sigma$ and attractive for $r>2^{1/6}\sigma$. As we see from Fig.\ \ref{fig4.2}, $g(r)$ is practically zero in the region $0\leq r\lesssim \sigma$ (due to the strongly repulsive force exerted by the reference particle at those distances), presents a very high peak at $r\approx \sigma$, oscillates thereafter, and eventually tends to unity for long distances as compared with $\sigma$. Thus, a liquid may present a strong structure captured by $g(r)$.

\begin{figure}[t]
\includegraphics[width=.7\columnwidth]{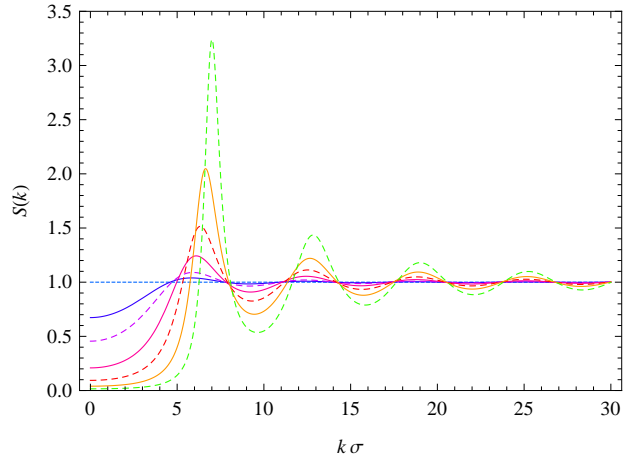}
\caption{Structure factor of a three-dimensional hard-sphere fluid (as obtained from the Percus--Yevick approximation) at several values of the packing fraction $\eta\equiv (\pi/6)n\sigma^3=0.05$, $0.1$, $0.2$, $0.3$, $0.4$, and $0.5$, in increasing order of complexity.\label{fig4.3}}
\end{figure}

Some functions related to the radial distribution function $g(r)$ can be defined. The first one is simply the so-called \emph{total correlation function}
\beq
\boxed{h(r)=g(r)-1}\;.
\label{4.14}
\eeq
Its Fourier transform
\beq
\widetilde{h}(k)\equiv \int\D \rr\, \E^{-\I\mathbf{k}\cdot\rr} h(r)\
\label{4.15a}
\eeq
is directly connected to the (static) \emph{structure factor}:
\beq
      \boxed{S(k)=1+n\widetilde{h}(k)}\;.
      \label{4.15}
\eeq
The typical shape of $S(k)$  at several densities is illustrated in Fig.\ \ref{fig4.3} for the hard-sphere (HS) potential \cite{M08}
\beq
\phi(r)=\begin{cases}
\infty\;,&r<\sigma\;,\\
0\;,&r>\sigma\;,
\end{cases}
\label{4.15HS}
\eeq
where $\sigma$ is the diameter of the spheres.

The structure factor is a very important quantity because it is experimentally accessible by elastic scattering of radiation (x-rays or neutrons) by the fluid \cite{B74b,HM06}. Thus, while $g(r)$ can be measured directly in simulations (either Monte Carlo or molecular dynamics) \cite{AT87,FS02}, it can be obtained indirectly in experiments from a numerical inverse Fourier transform of $S(k)-1$.

\section{{Thermodynamics from the Radial Distribution Function}}
\label{sec5}
As shown by \eqref{3.7}, \eqref{3.12}, \eqref{3.19}, and \eqref{3.27b}, the knowledge of any of the ensemble normalization functions allows one to obtain the full thermodynamic information about the system. But now imagine that instead of the normalization function (for instance, the partition function in the canonical ensemble), we are given (from experimental measures, computer simulations, or a certain theory) the radial distribution function $g(r)$. Can we have access to thermodynamics directly from $g(r)$? As we will see in this section, the answer is affirmative in the case of pairwise interactions.

\subsection{Compressibility Route}
The most straightforward route to thermodynamics from $g(r)$ is provided by choosing the grand canonical ensemble and simply combining \eqref{3.22} and \eqref{4.12b} to obtain
\beq
\boxed{\chi\equiv n k_BT \kappa_T=k_BT \left(\frac{\partial n}{\partial p}\right)_T=1+ n \int \D\mathbf{r}\, h(r)=S(0)}\;,
\label{5.1}
\eeq
where $\chi$ is the isothermal susceptibility and we recall that the total correlation function is defined by \eqref{4.14} and in the last step use has been made of \eqref{4.15}. Therefore, the zero wavenumber limit of the structure factor (see Fig.\ \ref{fig4.2}) is directly related to the isothermal compressibility.

Equation \eqref{5.1} is usually known as the \emph{compressibility equation of state} or the \emph{compressibility route} to thermodynamics.
\subsection{{Energy Route}}
Equation \eqref{5.1} applies regardless of the specific form of the potential energy function $\Phi_N(\mathbf{r}^N)$. {}From now on, however, we assume that the interaction is \emph{pairwise additive}, i.e., $\Phi_N$ can be expressed as a sum over all pairs of a certain function (interaction potential) $\phi$ that depends on the distance between the two particles of the pair. In mathematical terms,
\beq
\boxed{\Phi_N(\mathbf{r}^N)=\sum_{i=1}^{N-1}\sum_{j=i+1}^N\phi(r_{ij})=\frac{1}{2}\sum_{i\neq j}\phi(r_{ij})}\;.
\label{5.2}
\eeq
We have previously encountered two particular examples [see \eqref{4.13} and \eqref{4.15HS}] of interaction potentials.

The pairwise additivity condition \eqref{5.2} implies that $\Phi_N$ is a dynamical variable of the form \eqref{4.4}. As a consequence, we can apply the property \eqref{4.5} to the average potential energy:
\beq
\llangle E\rrangle^\ex=\llangle \Phi_N(\mathbf{r}^N)\rrangle=\frac{1}{2}\int\D\rr_1\int\D\rr_2 \,n_2(\rr_1,\rr_2)\phi(r_{12})\;.
\label{5.3}
\eeq
Adding the ideal-gas term [see \eqref{3.37}] and taking into account \eqref{4.11}, we finally obtain
\beq
\boxed{\llangle E\rrangle=N\left[\frac{d}{2}k_BT+\frac{n}{2}\int \D\mathbf{r}\, \phi(r) g(r)\right]}\;,
\label{5.4}
\eeq
where we have used the general property $\int \D \rr_1\int \D \rr_2\, \mathcal{F}(r_{12})=V\int \D \rr \, \mathcal{F}(r)$, $\mathcal{F}(r)$ being an arbitrary function.

Equation \eqref{5.4} defines the \emph{energy route} to thermodynamics. It can be equivalently written in terms of the so-called \emph{cavity function}
\beq
\boxed{y(r)\equiv g(r)\E^{\beta\phi(r)}}\;.
\label{5.5}
\eeq
The result is
\beq
\boxed{\llangle E\rrangle=N\left[\frac{d}{2}k_BT+\frac{n}{2}\int \D\mathbf{r}\, \phi(r){\E^{-\beta\phi(r)}}y(r)\right]}\;.
\label{5.6}
\eeq

\begin{figure}[t]
\sidecaption[t]
\includegraphics[width=.62\columnwidth]{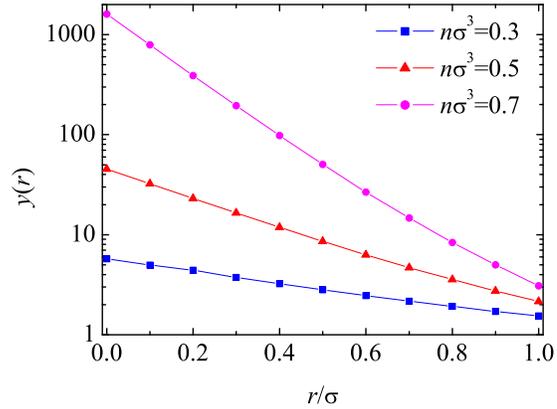}
\caption{Cavity function in the overlapping region $r<\sigma$ for a hard-sphere fluid at three different densities, as obtained from Monte Carlo simulations \protect\cite{LM84}.\label{fig5.1}}
\end{figure}

The cavity function $y(r)$ is much more regular than the radial distribution function $g(r)$. It is continuous even if the interaction potential is discontinuous or diverges. In the case of hard spheres, for instance, while $g(r)=0$ if $r<\sigma$, $y(r)$ is well defined in that region, as illustrated by Fig.\ \ref{fig5.1}.

\subsection{Virial Route}
Now we consider the pressure, which is the quantity more directly related to the equation of state. In the canonical ensemble,  the excess pressure is proportional to $\partial \ln Q_N/\partial V$ [see \eqref{3.46}] and thus it is not the average of a dynamical variable of type \eqref{4.5}. To make things worse, the volume $V$ appears in the configurational integral [see \eqref{3.45}] both explicitly and \emph{implicitly} through the integration limits. Let us make this more evident by writing
\beq
Q_N(V)=V^{-N}\int_{V^N}\D\rr^N\, \E^{-\beta\Phi_N(\rr^N)}\;.
\label{5.7}
\eeq

To get rid of this difficulty, we imagine now that the system is a sphere of volume $V$ and the origin of coordinates is chosen at the center of the sphere. If the whole system is blown up by a factor $\lambda$ \cite{B74b}, the volume changes from $V$ to $\lambda^d V$ and the configurational integral changes from $Q_N(V)$ to $Q_N(\lambda^d V)$ with
\beq
Q_N(\lambda^d V)=(\lambda^d V)^{-N}\int_{(\lambda^d V)^N}\D\rr^N\, \E^{-\beta\Phi_N(\rr^N)}=V^{-N}\int_{V^N}\D{\rr'}^N\, \E^{-\beta\Phi_N(\lambda^N{\rr'}^N)}\;,
\label{5.8}
\eeq
where in the last step the change  $\rr_i\to\rr_i'=\rr_i/\lambda$ has been performed. We see that $Q_N(\lambda^d V)$ depends on $\lambda$ explicitly through the argument of the interaction potential. Next, taking into account the identity $\partial \ln Q_N(\lambda^d V)/\partial V=(\lambda/Vd) \partial \ln Q_N(\lambda^d V)/\partial \lambda$, we can write
\beq
\frac{\partial \ln Q_N(V)}{\partial V}=\frac{1}{Vd}\left. \frac{\partial \ln Q_N(\lambda^dV)}{\partial \lambda}\right|_{\lambda=1}\;,
\label{5.9}
\eeq
so that
\beqa
\left. \frac{\partial \ln Q_N(\lambda^dV)}{\partial \lambda}\right|_{\lambda=1}&=&
-\beta\llangle \left. \frac{\partial \Phi_N(\lambda^N \rr^N)}{\partial \lambda}\right|_{\lambda=1}\rrangle\nn
&=&-\frac{\beta}{2}\int\D\rr_1\int\D\rr_2\, n_2(\rr_1,\rr_2)\left.
\frac{\partial \phi(\lambda r_{12})}{\partial \lambda}\right|_{\lambda=1}\nn
&=&-\frac{\beta}{2}n^2V\int\D\rr\,g(r)\left. \frac{\partial \phi(\lambda r)}{\partial \lambda}\right|_{\lambda=1}\;.
\label{5.10}
\eeqa
In the second equality use has been made of \eqref{4.5}. Finally, a mathematical property similar to \eqref{5.9} is
\beq
\left. \frac{\partial \phi(\lambda r)}{\partial \lambda}\right|_{\lambda=1}=r\frac{\D \phi( r)}{\D r}\;.
\label{5.11}
\eeq
Inserting \eqref{5.11} into \eqref{5.10}, and using \eqref{5.9}, we obtain the sought result:
\beq
\boxed{Z\equiv\frac{p}{n k_BT}=1-\frac{n\beta}{2d}\int \D\mathbf{r}\, r \frac{\D\phi(r)}{\D r} g(r)}\;.
\label{5.12}
\eeq
This is known as the pressure route or \emph{virial route} to the equation of state, where $Z$ is the \emph{compressibility factor}. Expressed in terms of the cavity function \eqref{5.5}, the virial route becomes
\beq
\boxed{Z\equiv\frac{p}{n k_BT}=1+\frac{n}{2d}\int \D\mathbf{r}\, y(r) r \frac{\partial \E^{-\beta\phi(r)}}{\partial r}}\;.
\label{5.13}
\eeq

\subsection{Chemical-Potential Route}
A look at \eqref{3.46} shows that we have already succeeded in expressing the first two derivatives of $\ln Q_N$ in terms of integrals involving the radial distribution function. The third derivative involves the chemical potential and is much more delicate. First, noting that $N$ is actually a discrete variable, we can rewrite
\beq
\beta\mu^\ex=-\frac{\partial\ln Q_N}{\partial N}\to \ln \frac{Q_{N}(\beta,V)}{Q_{N+1}(\beta,V)}\;.
\label{5.14a}
\eeq
Thus, the (excess) chemical potential is related to the response of the system to the addition of one  more particle without changing either temperature or volume.

The $N$-body potential energy is expressed by \eqref{5.2}. Now we add an extra particle (labeled as $i=0$), so that the $(N+1)$-body potential energy becomes
\beq
\Phi_{N+1}(\rr^{N+1})=\sum_{i=1}^{N-1}\sum_{j=i+1}^N \phi(r_{ij})+\Cred{\sum_{j=1}^N\phi(r_{0j})}\;.
\label{5.14}
\eeq
The trick now consists of introducing the extra particle (the ``solute'') little by little through a \emph{charging process} \cite{B74b,R80,O33,H56,RFL59,MR75,S12b,SR13}. We do so by introducing a \emph{coupling parameter} $\xi$ such that its value $0\leq\xi\leq 1$ controls the strength of the interaction of particle $i=0$ to the rest of particles (the ``solvent''):
\beq
\phi^\xxx(r_{0j})=\begin{cases}
  0\;,&\xi=0\;,\\
  \phi(r_{0j})\;,&\xi=1\;.
\end{cases}
\label{5.15}
\eeq
The associated total potential energy and configuration integral are
\beq
\Phi_{N+1}^\xxx(\rr^{N+1})=\Phi_N(\rr^N)+\Cred{\sum_{j=1}^{N}\phi^\xxx(r_{0j})}\;,
\label{5.16}
\eeq
\beq
Q_{N+1}^\xxx(\beta,V)=V^{-(N+1)}\int \D\rr^{N+1}\,\E^{-\beta \Phi_{N+1}^\xxx(\rr^{N+1})}\;.
\label{5.17}
\eeq
   Thus, assuming that $Q_{N+1}^\xxx$ is a smooth function of $\xi$, \eqref{5.14a} becomes
    \beq
\beta\mu^\ex=-\int_0^1\D\xi\,\frac{\partial \ln Q_{N+1}^\xxx(\beta,V)}{\partial \xi}\;.
\label{5.18}
\eeq
Since the dependence of $Q_{N+1}^\xxx$ on $\xi$ takes place through the extra summation in \eqref{5.16} and all the solvent particles are assumed to be identical,
\beq
\frac{\partial \ln Q_{N+1}^\xxx}{\partial \xi}
=-\frac{n\beta V^{-N}}{Q_{N+1}^\xxx}
\int \D\rr^{N+1}\, \E^{-\beta \Phi_{N+1}^\xxx(\rr^{N+1})} \frac{\partial
\phi^\xxx(r_{01})}{\partial \xi}\;.
\label{5.19}
\eeq
Now we realize that, similarly to \eqref{4.12}, the solute-solvent radial distribution function is defined as
\beq
g^\xxx(r_{01})=\frac{V^{-(N-1)}}{Q_{N+1}^\xxx}\int\D\rr_2\cdots\int\D\rr_N\, \E^{-\beta\Phi_{N+1}^\xxx(\rr^{N+1})}\;.
\label{5.20}
\eeq
This allows us to rewrite \eqref{5.19} in the form
\beq
\frac{\partial \ln Q_{N+1}^\xxx}{\partial \xi}
=-\frac{n\beta}{V}\int \D\rr_0\int \D\rr_1\,g^\xxx(r_{01})\frac{\partial
\phi^\xxx(r_{01})}{\partial \xi}\;.
\label{5.21}
\eeq
  Finally,
\beq
\boxed{\mu=k_BT\ln\left(n\Lambda^d\right)+n\int_0^1 \D\xi \int \D\rr\, g^\xxx(r)\frac{\partial \phi^\xxx(r)}{\partial\xi}}\;,
\label{5.22}
\eeq
or, equivalently,
\beq
\boxed{\beta\mu=\ln\left(n\Lambda^d\right)-n\int_0^1 \D\xi \int \D\rr\, y^\xxx(r)\frac{\partial \E^{-\beta \phi^\xxx(r)}}{\partial\xi}}\;.
\label{5.23}
\eeq
In contrast to the other three conventional routes [see \eqref{5.1}, \eqref{5.4}, and \eqref{5.12}], the \emph{chemical-potential route} \eqref{5.22} requires the knowledge of the solute-solvent correlation functions for all the values $0\leq \xi\leq 1$ of the coupling parameter $\xi$.

\subsection{{Extension to Mixtures}}
 In a multicomponent system the main quantities are
\begin{svgraybox}
 \begin{itemize}
    \item
    Number of particles of species $\alpha$: $N_\alpha$.

    \item
    Total number of particles: $N=\sum_\alpha N_\alpha$.

    \item
    Mole fraction of species $\alpha$: $x_\alpha=N_\alpha/N$, $\sum_\alpha x_\alpha=1$.

    \item
    Interaction potential between a particle of species $\alpha$ and a particle of species $\gamma$: $\phi_{\alpha\gamma}(r)$.

    \item
    Radial distribution function for the pair $\alpha\gamma$: $g_{\alpha\gamma}(r)$
\end{itemize}
\end{svgraybox}
All the previous thermodynamic routes can be generalized to mixtures.

\subsubsection{Compressibility Route}
The generalization of \eqref{5.1} to mixtures is not trivial \cite{BT70}. The result is
\beq
\boxed{\chi^{-1}\equiv\left(\frac{\partial \beta p}{\partial n}\right)_T=
{\sum_{\alpha,\gamma}\sqrt{x_\alpha x_\gamma}\left(\mathsf{I}+\widehat{\mathsf{h}}\right)^{-1}_{\alpha\gamma}}}\;,
\label{5.24}
\eeq
{where the element $\widehat{h}_{\alpha\gamma}$ of the matrix $\widehat{\mathsf{h}}$ is proportional to
the zero wavenumber limit of the Fourier transform of the total correlation function $h_{\alpha\gamma}({r})=g_{\alpha\gamma}({r})-1$, namely
\beq
\widehat{h}_{\alpha\gamma}=n\sqrt{x_\alpha x_\gamma}\int \D\mathbf{r}\,h_{\alpha\gamma}\left({r}\right)\;.
\label{5.25}
\eeq

    \subsubsection{Energy Route}
In this case, \eqref{5.6} is simply generalized as
  \beq
\boxed{\llangle E\rrangle=N\left[\frac{d}{2}k_BT+\frac{n}{2}
{\sum_{\alpha,\gamma}x_\alpha x_\gamma\int \D\mathbf{r}\, \phi_{\alpha\gamma}(r) \E^{-\beta \phi_{\alpha\gamma}(r)}y_{\alpha\gamma}(r)}
\right]}\;.
\label{5.26}
\eeq
\subsubsection{Virial Route}
Likewise, the generalization of \eqref{5.13} to mixtures reads
\beq
\boxed{Z\equiv\frac{p}{n k_BT}=1+\frac{n}{2d}
{\sum_{\alpha,\gamma}x_\alpha x_\gamma\int \D\mathbf{r}\, y_{\alpha\gamma}(r) r \frac{\partial \E^{-\beta\phi_{\alpha\gamma}(r)}}{\partial r}}}\;.
\label{5.27}
\eeq

\subsubsection{Chemical-Potential Route}
In this case, there exists a chemical potential associated with each species and the generalization of \eqref{5.23} is \cite{SR13}
\beq
\boxed{
{\beta\mu_\nu=\ln\left(nx_\nu\Lambda_\nu^d\right)-n\sum_\alpha x_\alpha\int_0^1 \D\xi \int \D\rr\, y_{\nu\alpha}^\xxx(r)\frac{\partial \E^{-\beta \phi_{\nu\alpha}^\xxx(r)}}{\partial\xi}}}\;.
\label{5.28}
\eeq
Here, the solute particle $i=0$ is  coupled to a particle of species $\alpha$ via an interaction potential $\phi^\xxx_{\nu\alpha}(r)$ such that
\beq
\phi_{\nu\alpha}^\xxx(r)=\begin{cases}
  0\;,&\xi=0\;,\\
  \phi_{\nu\alpha}(r)\;,&\xi=1\;,
\end{cases}
\label{5.29}
\eeq
so that it becomes a particle of species $\nu$ at the end of the charging process.
The associated radial distribution and cavity functions are
$g_{\nu\alpha}^\xxx(r)$ and $y_{\nu\alpha}^\xxx(r)=g_{\nu\alpha}^\xxx(r)\E^{\beta \phi_{\nu\alpha}^\xxx(r)}$, respectively.

The Helmholtz and Gibbs free energies can be obtained from $\mu_\nu$ as [see \eqref{2.10}]
\beq
\sum_\nu \mu_\nu N_\nu={G}=-V^2\frac{\partial\left( F/V\right)}{\partial V}\;.
\label{5.30}
\eeq

\subsection{Hard Spheres}

\begin{figure}[t]
\sidecaption[t]
\includegraphics[width=.5\columnwidth]{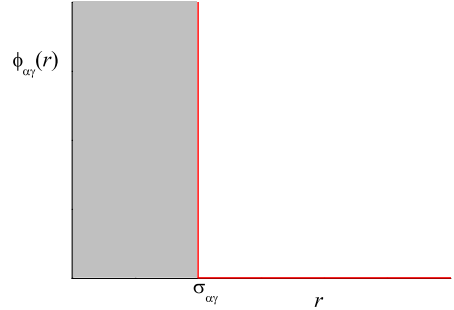}
\caption{{Hard-sphere interaction potential. The potential is equal to infinity in the shaded region and zero otherwise. \label{fig5.2}}}
\end{figure}

Let us now particularize the above expressions for multicomponent hard-sphere fluids \cite{RS82}. The interaction potential function is given by the form \eqref{4.15HS} for any pair of species, namely (see Fig.\ \ref{fig5.2})
\beq
\phi_{\alpha\gamma}(r)=\begin{cases}
\infty\;,&r<\sigma_{\alpha\gamma}\;,\\
0\;,&r>\sigma_{\alpha\gamma}\;.
\end{cases}
\label{5.31}
\eeq
Here, $\sigma_{\alpha\gamma}$ is the closest possible distance between the center of a sphere of species $\alpha$ and the center of a sphere of species $\gamma$. If we call $\sigma_{\alpha}=\sigma_{\alpha\alpha}$ to the closest distance between two spheres of the same species $\alpha$, it is legitimate to refer to $\sigma_\alpha$ as the \emph{diameter} of a sphere of species $\alpha$. However, that does not necessarily mean that two spheres of different type repel each other with a distance equal to the sum of their radii. Depending on that, one can classify hard-sphere mixtures into additive or nonadditive:
\begin{svgraybox}
  \begin{itemize}
    \item
    Additive mixtures: $\sigma_{\alpha\gamma}=\frac{1}{2}(\sigma_\alpha+\sigma_\gamma)$ for all pairs
    $\alpha\gamma$.
     \item
    Nonadditive mixtures: $\sigma_{\alpha\gamma}\neq\frac{1}{2}(\sigma_\alpha+\sigma_\gamma)$ for at least one pair     $\alpha\gamma$.
  \end{itemize}
\end{svgraybox}
As a consequence of \eqref{5.31},
\beq
\E^{-\beta\phi_{\alpha\gamma}(r)}=\Theta(r-\sigma_{\alpha\gamma})\;,
\quad
    \frac{\partial \E^{-\beta\phi_{\alpha\gamma}}(r)}{\partial r}=\delta\left(r-\sigma_{\alpha\gamma}\right)\;,
    \label{5.32}
\eeq
where $\Theta(x)$ and $\delta(x)$ are the Heaviside step function and the Dirac delta function, respectively.

The compressibility route \eqref{5.24} does not include the interaction potential explicitly and so it is not simplified in the hard-sphere case. As for the energy route, the integral \eqref{5.26} vanishes because $\phi_{\alpha\gamma}(r)\E^{-\beta \phi_{\alpha\gamma}(r)}\to 0$ both for $r<\sigma_{\alpha\gamma}$ and $r>\sigma_{\alpha\gamma}$, while $y_{\alpha\gamma}(r)$ is finite even in the region $r<\sigma_{\alpha\gamma}$ (see Fig.\ \ref{fig5.1}). Therefore,
\beq
\llangle E\rrangle=N\frac{d}{2}k_BT\;.
\label{5.33}
\eeq
But this is  the ideal-gas internal energy! This is an expected result since the hard-sphere potential is only different from zero when two particles overlap but those configurations are forbidden by the Boltzmann factor $\E^{-\beta \phi_{\alpha\gamma}(r)}$.

The generic virial route \eqref{5.27} is highly simplified for hard spheres. First, one changes to spherical coordinates and takes into account that the total $d$-dimensional solid angle (area of a $d$-dimensional sphere of unit radius) is
\beq
\int\D\widehat{\rr}=d 2^d v_d\;,
\label{5.34}
\eeq
where
\beq
v_d=\frac{(\pi /4)^{d/2}}{\Gamma (1+d/2)}
\label{5.35}
\eeq
is the volume of a $d$-dimensional sphere of unit diameter. Next, using the property \eqref{5.32}, we obtain
\beq
\boxed{
\frac{p}{n k_BT}=1+
{2^{d-1}nv_d\sum_{\alpha,\gamma}x_\alpha x_\gamma\sigma_{\alpha\gamma}^d y_{\alpha\gamma}(\sigma_{\alpha\gamma})}}\;.
\label{5.36}
\eeq

The same method works for the chemical-potential route \eqref{5.28} with the choice
\beq
\E^{-\beta\phi_{\nu\alpha}^\xxx(\rr)}=\Theta(r-\sigma_{\nu\alpha}^\xxx)\;,
\label{5.37}
\eeq
where $\sigma_{\nu\alpha}^{(0)}=0$ and $\sigma_{\nu\alpha}^{(1)}=\sigma_{\nu\alpha}$.
Changing the integration variable in \eqref{5.28} from $\xi$ to $\sigma_{\nu\alpha}^\xxx$, one gets
\beq
\boxed{
\beta\mu_\nu=\ln\left(nx_\nu\Lambda_\nu^d\right)
{+d 2^dnv_d\sum_\alpha x_\alpha\int_{0}^{\sigma_{\nu\alpha}} \D\sigma_{0\alpha} \,\sigma_{0\alpha}^{d-1}y_{0\alpha}(\sigma_{0\alpha})}}\;,
\label{5.38}
\eeq
where the notation has been simplified as $\sigma_{\nu\alpha}^\xxx\to \sigma_{0\alpha}$ and $y_{\nu\alpha}^\xxx\to y_{0\alpha}$.

If $\sigma_{\alpha\gamma}\geq \frac{1}{2}\left(\sigma_\alpha+\sigma_\gamma\right)$ (positive or zero \emph{nonadditivity}), then it can be proved  \cite{SR13} that
\beq
d 2^dnv_d\sum_\alpha x_\alpha\int_{0}^{\frac{1}{2}\sigma_{\alpha}} \D\sigma_{0\alpha} \,\sigma_{0\alpha}^{d-1}y_{0\alpha}(\sigma_{0\alpha})=-\ln(1-\eta)\;,
\label{5.39}
\eeq
where
\beq
\boxed{\eta\equiv n v_d\sum_\alpha x_\alpha \sigma_\alpha ^d}
\label{5.40}
\eeq
is the total \emph{packing} fraction. In that case, \eqref{5.38} can be rewritten as
\beq
\boxed{
\beta\mu_\nu=\ln\frac{nx_\nu\Lambda_\nu^d}{1-\eta}
{+d 2^dnv_d\sum_\alpha x_\alpha\int_{\frac{1}{2}\sigma}^{\sigma_{\nu\alpha}} \D\sigma_{0\alpha} \,\sigma_{0\alpha}^{d-1}y_{0\alpha}(\sigma_{0\alpha})}}\;.
\label{5.41}
\eeq

\subsection{The Thermodynamic Inconsistency Problem}

\begin{figure}[t]
\sidecaption[t]
\includegraphics[width=.64\columnwidth]{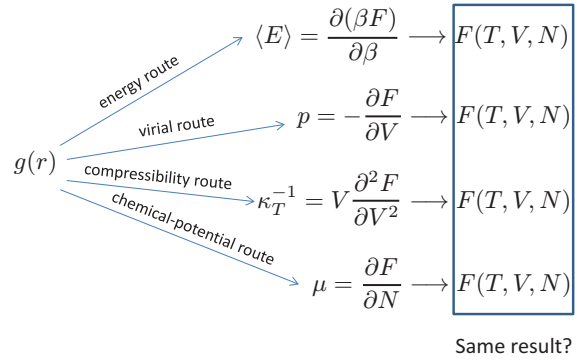}
\caption{Schematic view of the thermodynamic inconsistency problem. \label{fig5.3}}
\end{figure}

Going back to the case of an arbitrary interaction potential, we have seen that the knowledge of the radial distribution function $g(r)$ (where, for simplicity, we are using the one-component language) allows one to obtain four important thermodynamic quantities: the internal energy, the pressure, the isothermal compressibility, and the chemical potential. By integration, one could in principle derive the free energy of the system (except for functions playing the role of integration constants) from any of those routes, as sketched in Fig.\ \ref{fig5.3}. The important question is, would one obtain consistent results?

Since all the thermodynamic routes are derived from formally exact statistical-mechanical formulas, it is obvious that  the use of the \emph{exact} radial distribution function $g(r)$ must lead to the same exact free energy $F(T,V,N)$, regardless of the route followed.
On the other hand, if an \emph{approximate} $g(r)$ is used, one must be prepared to obtain (in general) a different approximate $F(T,V,N)$ from each separate route. This is known as the \emph{thermodynamic (in)consistency problem}. Which route is more accurate, i.e., which route is more effective in concealing the deficiencies of an approximate $g(r)$, depends on the approximation, the potential, and the thermodynamic state.

\section{One-Dimensional Systems. Exact Solution for Nearest-Neighbor Interactions}
\label{sec6}

As is apparent from \eqref{4.12}, the evaluation of $g(r)$ is a formidable task, comparable to that of the evaluation of the configuration integral itself. However, in the case of one-dimensional systems ($d=1$) of particles which only interact with their nearest neighbors, the problem can be exactly solved \cite{SZK53,LZ71,HC04,S07,BNS09}.

Let us consider a one-dimensional system of $N$ particles in a box of length $L$ (so the number density is $n=N/L$) subject to an interaction potential $\phi(r)$  such that
\begin{svgraybox}
    \begin{enumerate}
      \item $\lim_{r\to 0}\phi(r)=\infty$. This implies that the \emph{order} of the particles in the line does not change.
            \item
      $\lim_{r\to \infty}\phi(r)=0$. The interaction has a \emph{finite} range.
            \item
      Each particle  interacts \emph{only} with its two nearest neighbors.
    \end{enumerate}
    \end{svgraybox}
    The total potential energy is then
\beq
\Phi_N(\rr^N)=\sum_{i=1}^{N-1}\phi(x_{i+1}-x_i)\;.
\label{6.1a}
\eeq

\subsection{{Nearest-Neighbor and Pair Correlation Functions}}

\begin{figure}[t]
\includegraphics[width=.9\columnwidth]{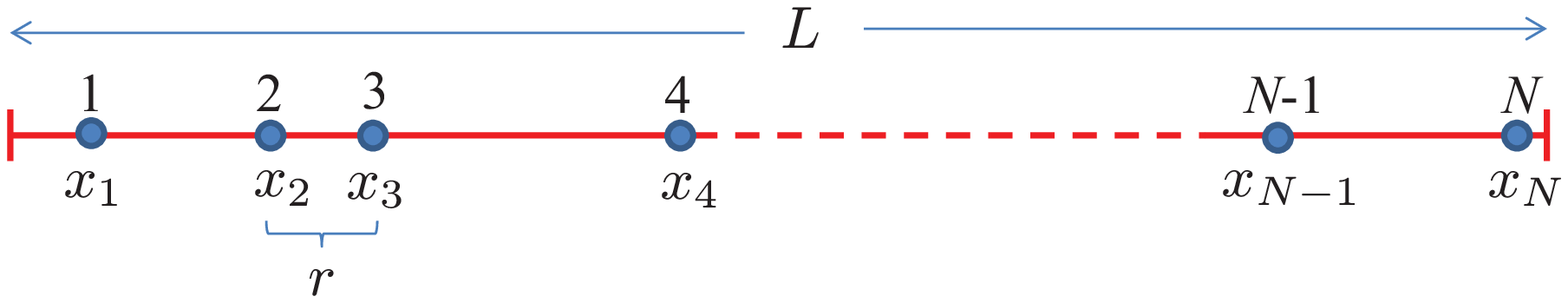}\\
\includegraphics[width=.9\columnwidth]{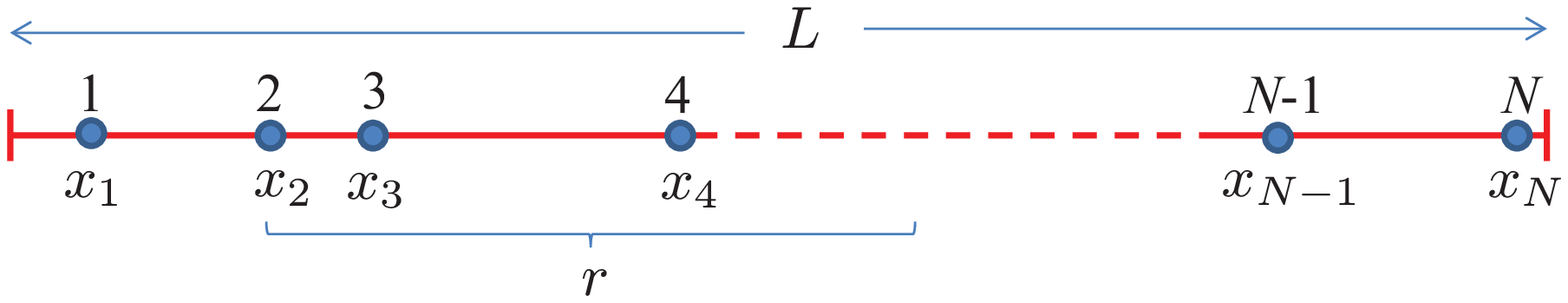}\\
\includegraphics[width=.9\columnwidth]{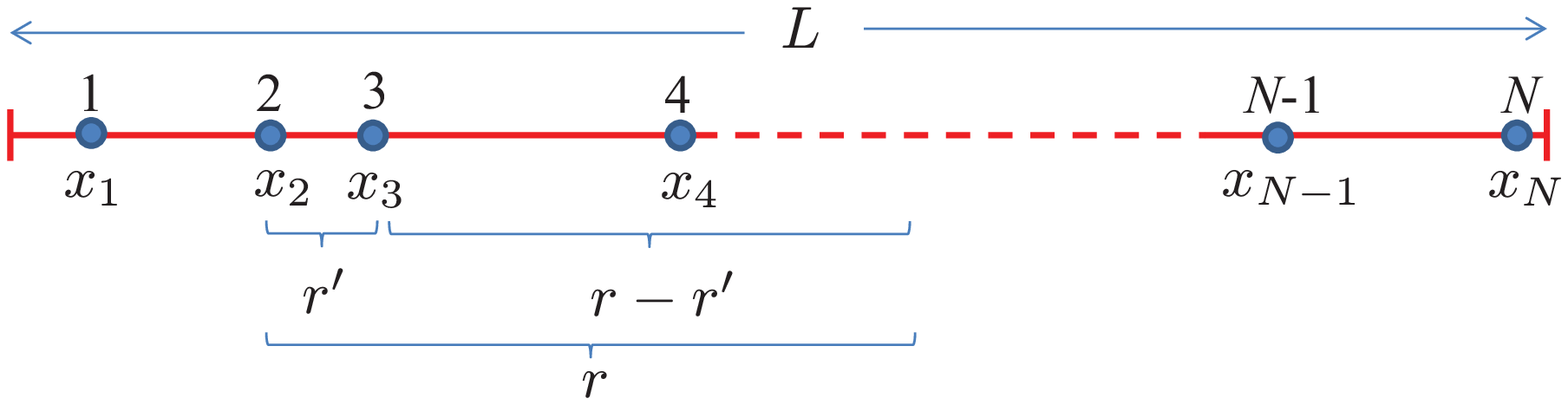}
\caption{Top panel: Two nearest-neighbor particles separated a distance $r$. Middle panel: Two $\ell$th-order neighbors separated a distance $r$. Bottom panel: Illustration of the convolution property.\label{fig6.1}}
\end{figure}

Given a particle  at a certain position, let $p^{(1)}(r)\D r$ be the \emph{conditional} probability of finding  its (right) \emph{nearest neighbor} at a distance between $r$ and $r+\D r$ (see Fig.\ \ref{fig6.1}, top panel). More in general, we can define $p^{(\ell)}(r)\D r$ as the conditional probability of finding  its (right) $\ell$th
neighbor  ($1\leq \ell\leq N-1$) at a distance between $r$ and $r+\D r$ (see Fig.\ \ref{fig6.1}, middle panel). Since the $\ell$th neighbor must be somewhere, the normalization condition is
    \beq
    \int_0^\infty \D r\, p^{(\ell)}(r)=1 \;.
    \label{6.1}
    \eeq
In making the upper limit equal to infinity, we are implicitly assuming   the thermodynamic limit ($L\to\infty$, $N\to \infty$, $n=\text{const}$). Moreover, periodic boundary conditions are supposed to be applied when needed.

As illustrated by the bottom panel of Fig.\ \ref{fig6.1}, the following recurrence relation holds
    \beq
    p^{(\ell)}(r)=\int_0^r\D r'\, p^{(1)}(r')p^{(\ell-1)}(r-r')\;.
    \label{6.2}
    \eeq
The convolution structure of the integral invites one to introduce the Laplace transform
\beq
  P^{(\ell)}(s)=\int_0^\infty \D r \; \E^{-rs} p^{(\ell)}(r)\;,
  \label{6.3}
  \eeq
so that \eqref{6.2} becomes
\beq
   P^{(\ell)}(s)= P^{(1)}(s) P^{(\ell-1)}(s)\Rightarrow  P^{(\ell)}(s)=\left[P^{(1)}(s)\right]^\ell\;.
\label{6.4}
   \eeq
The normalization condition \eqref{6.1} is equivalent to
\beq
 P^{(\ell)}(0)=1\;.
 \label{6.4b}
 \eeq

Now, given a reference particle  at a certain position, let $n g(r)\D r$ be the \emph{number of particles}  at a distance between $r$ and $r+\D r$, regardless of whether those particles are the nearest neighbor, the next-nearest neighbor, \ldots of the reference particle.    Thus,
    \beq
    n g(r)=\sum_{\ell=1}^{N-1} p^{(\ell)}(r)\stackrel{N\to\infty}{\longrightarrow}\sum_{\ell=1}^{\infty} p^{(\ell)}(r)\;.
    \label{6.5}
    \eeq
Introducing the  Laplace transform
  \beq
   G(s)=\int_0^\infty \D r \; \E^{-rs} g(r)\;,
   \label{6.6}
  \eeq
and using \eqref{6.4}, we have
   \beq
   G(s)=\frac{1}{n}\sum_{\ell=1}^\infty \left[P^{(1)}(s)\right]^\ell=\frac{1}{n}\frac{P^{(1)}(s)}{1-P^{(1)}(s)}\;.
   \label{6.7}
   \eeq
Thus, the determination of the radial distribution function $g(r)$ reduces to the determination of the nearest-neighbor distribution function $p^{(1)}(r)$. To that end, we take advantage of the ensemble equivalence in the thermodynamic limit and use the isothermal-isobaric ensemble.

\subsection{Nearest-Neighbor Distribution. Isothermal-Isobaric Ensemble}

      \begin{figure}[t]
\includegraphics[width=0.9\columnwidth]{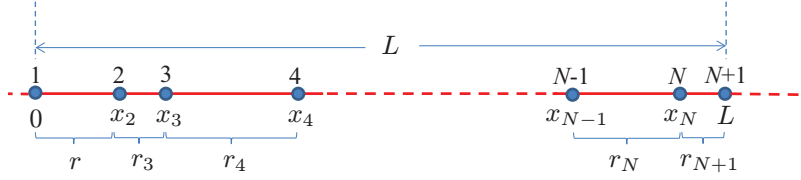}
\caption{Illustration of the evaluation of $ p^{(1)}(r)$ in the isothermal-isobaric ensemble. \label{fig6.2}}
\end{figure}

The isothermal-isobaric ensemble is described by \eqref{3.25}. The important point is that the $N$-body probability distribution function in configuration space is proportional to $\E^{-\beta p V-\beta \Phi_N(\rr^N)}$
Therefore, in this ensemble the one-dimensional nearest-neighbor probability distribution function is
  \beq
  p^{(1)}(r)\propto \int_r^\infty \D L \;\E^{-\beta p L}\int_{x_2}^L\D x_3 \int_{x_3}^L\D x_4 \cdots \int_{x_{N-1}}^L\D x_N \,
    \E^{-\beta\Phi_N(\rr^N)}\;,
    \label{6.8}
  \eeq
where we have identified the volume $V$ with the length $L$ and have taken the particles $i=1$ (at $x_1=0$) and $i=2$ (at $x_2=r$) as the canonical nearest-neighbor pair (see Fig.\ \ref{fig6.2}). Next, using \eqref{6.1a} and applying periodic boundary conditions,
  \beqa
  p^{(1)}(r)&\propto& \E^{-\beta \phi(r)}\int_r^\infty \D L \;\E^{-\beta p L}\int_{0}^{L-r}\D r_3 \, \E^{-\beta \phi(r_3)}
  \int_{0}^{L-r-r_3}\D r_4 \,\E^{-\beta \phi(r_4)}\nn
  &&\times  \cdots \int_{0}^{L-r-r_3-\cdots -r_{N-1}}\D r_N \, \E^{-\beta \phi(r_N)}\E^{-\beta \phi(r_{N+1})}\;,
  \label{6.9}
  \eeqa
where a change of variables $x_i\to r_i=x_i-x_{i-1}$ ($i=3,\ldots, N$) has been carried out and $r_{N+1}=L-r-r_3-r_4-\cdots -r_{N}$. Finally, the change of variable $L\to L'=L-r$ shows that a factor $\E^{-\beta p r}$ comes out of the integrals, the latter being independent of $r$.    In summary,
    \beq
   p^{(1)}(r)=K \E^{-\beta \phi(r)} \E^{-\beta p r}\;,
   \label{6.10}
   \eeq
   where the proportionality constant $K$ will be determined by normalization. The Laplace transform of \eqref{6.10} is
    \beq
   P^{(1)}(s)=K \Omega(s+\beta p)\;
   \label{6.11}
   \eeq
 where
 \beq
  \boxed{\Omega(s)\equiv \int_0^\infty \D r\, \E^{-r s} \E^{-\beta\phi(r)}}
  \label{6.12}
  \eeq
  is the Laplace transform of the Boltzmann factor $\E^{-\beta\phi(r)}$. The normalization condition \eqref{6.4b} yields
   \beq
     K=\frac{1}{\Omega(\beta p)}\;.
     \label{6.13}
    \eeq

\subsection{Exact Radial Distribution Function and Equation of State}
Insertion of \eqref{6.13} into \eqref{6.7} gives the exact radial distribution function (in Laplace space):
    \beq
    \boxed{G(s)=\frac{1}{n}\frac{\Omega(s+\beta p)}{\Omega(\beta p)-\Omega(s+\beta p)}}\;.
    \label{6.14}
    \eeq
To fully close the problem, it remains to relate the pressure $p$, the density $n$, and the temperature $T$ (equation of state).    To do that, we apply the consistency condition
    \beq
    \lim_{r\to\infty}g(r)=1\Rightarrow \lim_{s\to 0} s G(s)=1\;.
    \label{6.14b}
    \eeq
Expanding   $\Omega(s+\beta p)$ in powers of $s$ and imposing \eqref{6.14b}, we obtain
    \beq
    \boxed{n(p,T)=-\frac{\Omega(\beta p)}{\Omega'(\beta p)}}\;,\quad \Omega'(s)\equiv \frac{\partial \Omega(s)}{\partial s}\;.
    \label{6.15}
    \eeq

As a consistency test, let  us prove that the equation of state \eqref{6.15} is equivalent to the compressibility route \eqref{5.1}. First, according to \eqref{6.15}, the isothermal susceptibility is
\beq
  \chi=\left(\frac{\partial n}{\partial\beta p}\right)_\beta=-1+\frac{\Omega(\beta p)\Omega''(\beta p)}{\left[\Omega'(\beta p)\right]^2}\;.
  \label{6.16}
  \eeq
Alternatively, the Laplace transform of  $h(r)$ is $H(s)=G(s)-s^{-1}$,  and thus the Fourier transform can be obtained as
\beq
\widetilde{h}(k)=\left[H(s)+H(-s)\right]_{s=\I k}=\left[G(s)+G(-s)\right]_{s=\I k}\;.
\label{6.17a}
\eeq
In particular, the zero wavenumber limit is
\beq
  \int \D \rr\, h(r)=2\lim_{s\to 0}\left[G(s)-\frac{1}{s}\right]=\frac{\Omega'(\beta p)}{\Omega(\beta p)}-\frac{\Omega''(\beta p)}{2\Omega'(\beta p)}\;,
  \label{6.17}
  \eeq
so that
\beqa
1+n\int \D \rr \, h(r)&=&1-2\frac{\Omega(\beta p)}{\Omega'(\beta p)}\left[\frac{\Omega'(\beta p)}{\Omega(\beta p)}-\frac{\Omega''(\beta p)}{2\Omega'(\beta p)}\right]\nn
    &=&-1+\frac{\Omega(\beta p)\Omega''(\beta p)}{\left[\Omega'(\beta p)\right]^2}\;.
    \label{6.18}
  \eeqa
Comparison between   \eqref{6.16} and \eqref{6.18} shows that \eqref{5.1} is indeed satisfied.

\subsection{Extension to Mixtures}
In the case of one-dimensional mixtures the arguments outlined above can be extended without special difficulties \cite{HC04,S07,BNS09}. Now, instead of $p^{(\ell)}(r)\D r$ one defines  $p_{\alpha\gamma}^{(\ell)}(r)\D r$ as the conditional probability that the $\ell$th neighbor to the right of a reference particle of species $\alpha$  is located at a distance between $r$ and $r+\D r$ \emph{and} belongs to species $\gamma$. The counterparts of \eqref{6.1}, \eqref{6.2}, and \eqref{6.5} are
\beq
   \sum_\gamma \int_0^\infty \D r\, p_{\alpha\gamma}^{(\ell)}(r)=1 \;,
    \label{6.19}
    \eeq
    \beq
    p_{\alpha\gamma}^{(\ell)}(r)=\sum_\nu\int_0^r\D r'\, p_{\alpha\nu}^{(1)}(r')p_{\nu\gamma}^{(\ell-1)}(r-r')\;,
    \label{6.20}
    \eeq
    \beq
    n x_\gamma g_{\alpha\gamma}(r)=\sum_{\ell=1}^{\infty} p_{\alpha\gamma}^{(\ell)}(r)\;.
    \label{6.21}
    \eeq
Next, by defining the Laplace transforms $P_{\alpha\gamma}^{(\ell)}(s)$ and $G_{\alpha\gamma}(s)$ of  $p_{\alpha\gamma}^{(\ell)}(r)$ and $g_{\alpha\gamma}(r)$, respectively, one easily arrives at
    \beq
    G_{\alpha\gamma}(s)=\frac{1}{n x_\gamma}\left(\mathsf{P}^{(1)}(s)\cdot\left[\mathsf{I}-\mathsf{P}^{(1)}(s)\right]^{-1}\right)_{\alpha\gamma}\;,
    \label{6.22}
    \eeq
where $\mathsf{P}^{(1)}(s)$ is the matrix of elements $P_{\alpha\gamma}^{(1)}(s)$.

The nearest-neighbor probability distribution is again derived in the isothermal-isobaric ensemble with the result
    \beq
    p^{(1)}_{\alpha\gamma}(r)=x_\gamma K_{\alpha\gamma}\E^{-\beta\phi_{\alpha\gamma}(r)}\E^{-\beta p r}\;,
    \label{6.23}
    \eeq
    so that
    \beq
   P^{(1)}_{\alpha\gamma}(s)=x_\gamma K_{\alpha\gamma}\Omega_{\alpha\gamma}(s+\beta p)\;,
   \label{6.24}
    \eeq
 where  $\Omega_{\alpha\gamma}(s)$ is the Laplace transform of $\E^{-\beta\phi_{\alpha\gamma}(r)}$. The normalization condition \eqref{6.19} imposes the following relationship for the constants
 $K_{\alpha\gamma}=K_{\gamma\alpha}$:
    \beq
     \sum_{\gamma}x_\gamma K_{\alpha\gamma}\Omega_{\alpha\gamma}(\beta p)=1\;.
     \label{6.25}
    \eeq
To complete the determination of  $K_{\alpha\gamma}$, we can make use of the physical condition stating that $\lim_{r\to\infty} p_{\alpha\gamma}^{(1)}(r)/p_{\alpha\nu}^{(1)}(r)$ must be independent of the identity $\alpha$ of the species the reference particle belongs to, so that $K_{\alpha\gamma}/K_{\alpha\nu}$ is independent of $\alpha$. It is easy to see that such a condition implies
    \beq
    K_{\alpha\gamma}^2=K_{\alpha\alpha}K_{\gamma\gamma}\;.
    \label{6.26}
    \eeq

Finally, the equation of state $n(p,T)$ is determined, as in the one-component case, from the condition $\lim_{r\to\infty}g_{\alpha\gamma}(r)=1\Rightarrow \lim_{s\to 0} s G_{\alpha\gamma}(s)=1$.

\subsubsection{Binary Case}
As a more explicit situation, here we particularize to a binary mixture. In that case, \eqref{6.22} yields
    \beq
    G_{11}(s)=\frac{Q_{11}(s)\left[1-Q_{22}(s)\right]+Q_{12}^2(s)}{n x_1 D(s)}\;,
    \label{6.27}
    \eeq
     \beq
    G_{22}(s)=\frac{Q_{22}(s)\left[1-Q_{11}(s)\right]+Q_{12}^2(s)}{n x_2 D(s)}\;,
    \label{6.28}
    \eeq
     \beq
    G_{12}(s)=\frac{Q_{12}(s)}{n \sqrt{x_1x_2} D(s)}\;,
    \label{6.29}
    \eeq
  where
  \beq
  Q_{\alpha\gamma}(s)\equiv \sqrt{\frac{x_\alpha}{x_\gamma}}P_{\alpha\gamma}^{(1)}(s)=\sqrt{x_\alpha x_\gamma}K_{\alpha\gamma}\Omega_{\alpha\gamma}(s+\beta p)\;,
  \label{6.30}
  \eeq
  \beq
  D(s)\equiv \left[1-Q_{11}(s)\right]\left[1-Q_{22}(s)\right]-Q_{12}^2(s)\;.
  \label{6.31}
  \eeq

The parameters $K_{\alpha\gamma}$ are obtained from \eqref{6.25} and \eqref{6.26}. First, $K_{11}$ and $K_{22}$ can be expressed in terms of $K_{12}$ as
    \beq
    K_{11}=\frac{1-x_2 K_{12}\Omega_{12}(\beta p)}{x_1\Omega_{11}(\beta p)}\;, \quad K_{22}=\frac{1-x_1 K_{12}\Omega_{12}(\beta p)}{x_2\Omega_{22}(\beta p)}\;.
    \label{6.32}
    \eeq
 The remaining parameter $K_{12}$ satisfies a quadratic equation whose solution is
 \beq
 K_{12}=\frac{1-\sqrt{1-4x_1x_2 R(\beta p)}}{2x_1 x_2 R(\beta p)\Omega_{12}(\beta p)}\;,\quad R(\beta p)\equiv 1-\frac{\Omega_{11}(\beta p)\Omega_{22}(\beta p)}{\Omega_{12}^2(\beta p)}\;.
 \label{6.33}
 \eeq
Finally, the equation of state becomes
    \beq
    n(p,T)=-\frac{1}{x_1^2 K_{11}\Omega_{11}'(\beta p)+x_2^2 K_{11}\Omega_{11}'(\beta p)+2x_1x_2 K_{12}\Omega_{12}'(\beta p)}\;.
    \label{6.34}
    \eeq

\subsection{Examples}
\subsubsection{Sticky Hard Rods}

\begin{figure}[t]
\includegraphics[width=.5\columnwidth]{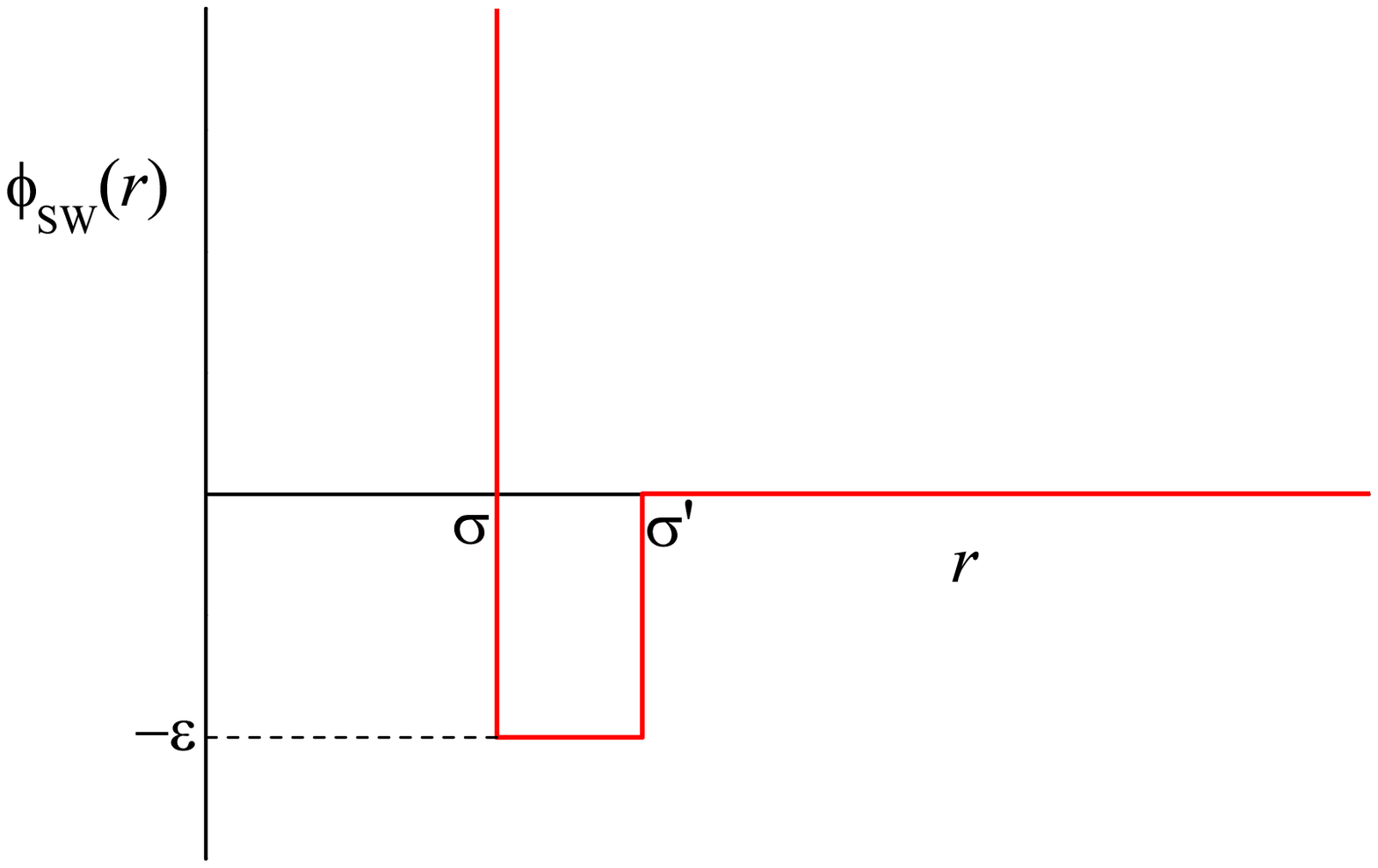}
\includegraphics[width=.5\columnwidth]{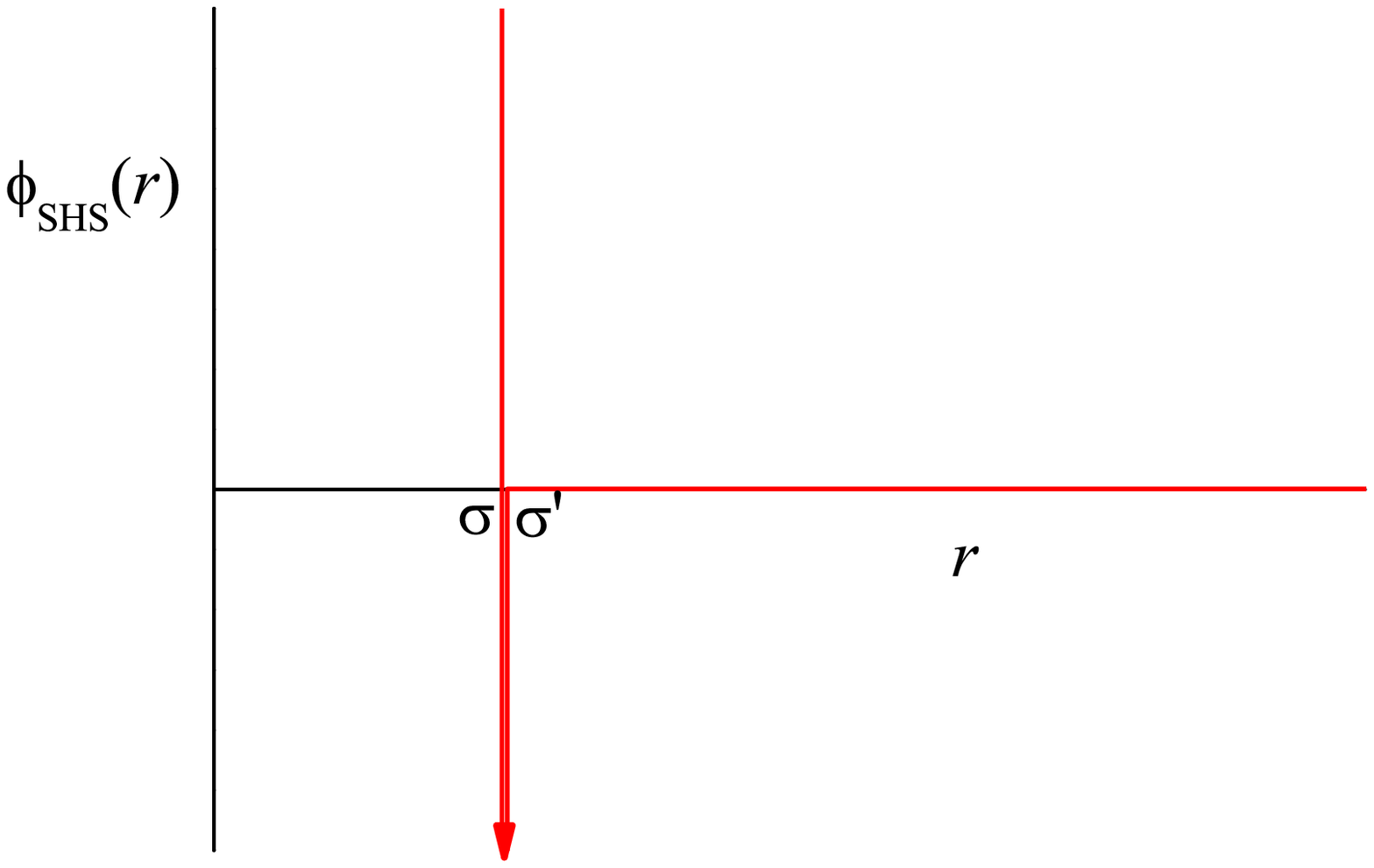}
\caption{Left panel: Square-well potential. Right panel: Sticky-hard-sphere potential.\label{fig6.3}}
\end{figure}

As an application,  we consider here the sticky-hard-rod fluid, which is the one-dimensional version of the so-called sticky-hard-sphere (SHS) fluid. Let us first introduce the square-well (SW) potential (see Fig.\ \ref{fig6.3}, left panel)
\beq
\phi_\sw(r)=\begin{cases}
\infty\;,&r<\sigma\;,\\
{-\varepsilon}\;,&\sigma<r<\sigma'\;,\\
0\;,&r>\sigma'\;.
\end{cases}
\label{6.35}
\eeq
The associated Boltzmann factor is
\beq
\E^{-\beta\phi_\sw(r)}=\begin{cases}
0\;,&r<\sigma\;,\\
{\E^{\beta \varepsilon}}\;,&\sigma<r<\sigma'\;,\\
1\;,&r>\sigma'\;,
\end{cases}
\label{6.36}
\eeq
whose Laplace transform is
  \beq
  \Omega(s)=\frac{1}{s}\left[\E^{\beta\varepsilon}\left(\E^{-\sigma s}-\E^{-\sigma's}\right)+\E^{-\sigma's}\right]\;.
  \label{6.37}
  \eeq
In order to apply the exact results for one-dimensional systems, we must prevent the square-well interaction from extending beyond nearest neighbors. This implies the constraint $\sigma'\leq 2\sigma$.

Now we take the sticky-hard-sphere limit \cite{B68} (see Fig.\ \ref{fig6.3}, right panel)
\beq
\sigma'\to\sigma\;,\quad\varepsilon\to\infty\;,\quad
\tau^{-1}\equiv (\sigma'-\sigma)\E^{\beta\varepsilon}=\text{finite}\;,
\label{6.38}
\eeq
where the temperature-dependent parameter $\tau^{-1}$ measures the ``stickiness'' of the interaction.
In this limit, \eqref{6.36} and \eqref{6.37} become
\beq
\E^{-\beta\phi(r)}=\Theta(r-\sigma)+\tau^{-1}\delta(r-\sigma)\;,
\label{6.39}
\eeq
\beq
  \Omega(s)=\left(\tau^{-1}+\frac{1}{s}\right)\E^{-\sigma s}\;.
  \label{6.40}
  \eeq
The equation of state \eqref{6.15} expresses the density as a function of temperature and pressure. Solving the resulting quadratic equation for the pressure one simply gets
    \beq
    {Z}\equiv\frac{\beta p}{n}=\frac{\sqrt{1+4\tau^{-1} n/(1-n\sigma)-1}}{2\tau^{-1} n}\;.
    \label{6.41}
    \eeq
In the hard-rod  special case ($\tau^{-1}\to 0$), the equation of state becomes $Z=(1-n\sigma)^{-1}$.

  \begin{figure}[t]
\includegraphics[width=.5\columnwidth]{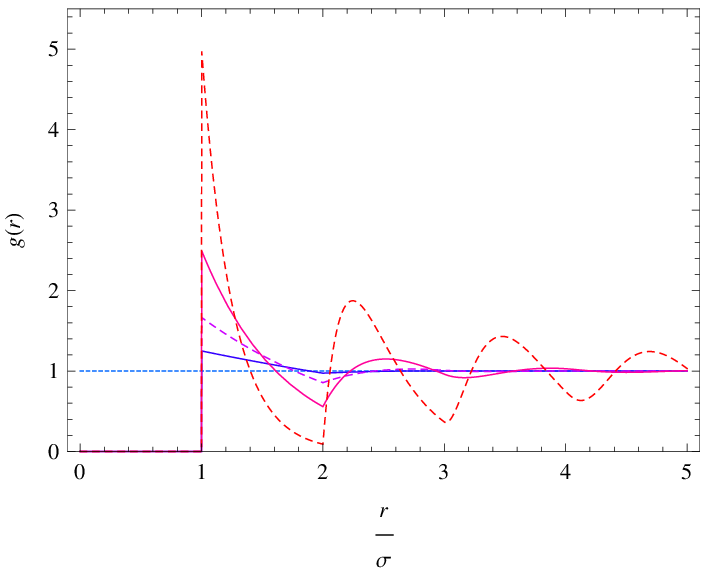}
\includegraphics[width=.5\columnwidth]{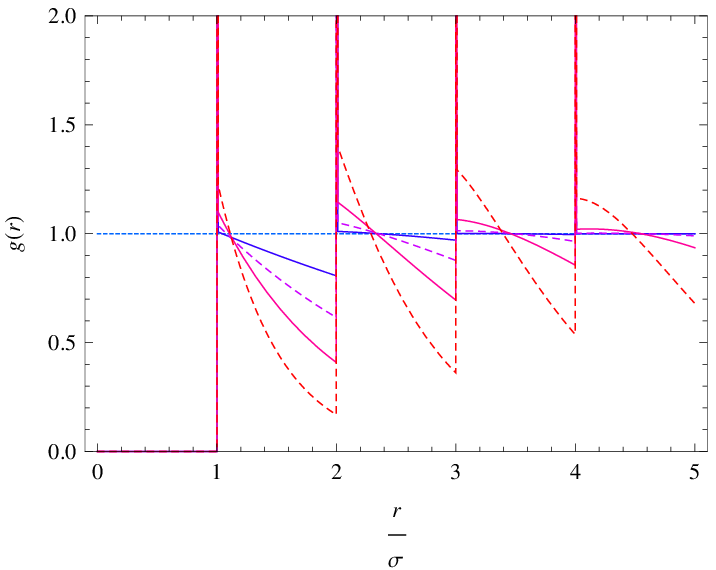}
\caption{Radial distribution function for hard rods ($\tau^{-1}=0$, left panel) and sticky hard rods ($\tau^{-1}/\sigma=0.5$, right panel) at several values of the packing fraction $\eta\equiv n\sigma=0.2$, $0.4$, $0.6$, and $0.8$, in increasing order of complexity.\label{fig6.4}}
\end{figure}

As for the radial distribution function, application of \eqref{6.14} gives
  \beq
    G(s)=\frac{1}{n}\frac{\left(\tau^{-1}+\frac{1}{s+\beta p}\right)\E^{-\sigma s}}{\tau^{-1}+\frac{1}{\beta p}-\left(\tau^{-1}+\frac{1}{s+\beta p}\right)\E^{-\sigma s}}
    =\frac{1}{n}\sum_{\ell=1}^\infty \frac{\left(\tau^{-1}+\frac{1}{s+\beta p}\right)^\ell}{\left(\tau^{-1}+\frac{1}{\beta p}\right)^\ell}\E^{-\ell\sigma s}\;.
    \label{6.42}
    \eeq
    The last equality allows one to perform the inverse Laplace transform term by term with the result
    \beq
    g(r)=\sum_{\ell=1}^\infty \Psi_\ell(r-\ell\sigma)\Theta(r-\ell\sigma)\;,
    \label{6.43}
    \eeq
where
    \beq
    \Psi_\ell(r)=\frac{1}{n(\tau^{-1}+1/\beta p)^\ell}\left[\tau^{-\ell}\delta(r)+\sum_{k=1}^\ell \binom{\ell}{k}\frac{\tau^{-(\ell-k)}}{(k-1)!}r^{k-1}\E^{-\beta pr}
    \right]\;.
    \label{6.44}
    \eeq
Note that, although an infinite number of terms formally appear in \eqref{6.43}, only the first $j$ terms are needed if one is interested in $g(r)$ in the range $1\leq r/\sigma < j+1$.
Figure \ref{fig6.4} shows $g(r)$ for hard rods ($\tau^{-1}=0$) and a representative case of a sticky-hard-rod fluid ($\tau^{-1}/\sigma=0.5$) at several densities \cite{note_13_08}.

Using \eqref{6.39}, it is straightforward to see that the radial distribution function and the cavity functions are related by
  \beq
  g(r)=\tau^{-1} y(\sigma)\delta(r-\sigma)+y(r)\Theta(r-\sigma)\;.
  \label{6.45}
    \eeq
This, together with \eqref{6.43} and \eqref{6.44}, implies the contact value
\beq
y(\sigma)=\frac{1}{n(\tau^{-1}+1/\beta p)}\;.
\eeq
 This value is useful to obtain the mean potential energy per particle,
     \beq
     \frac{\llangle E\rrangle^\ex}{N\varepsilon}=-n\tau^{-1} y(\sigma)=-\frac{1}{1+\tau/\beta p}\;,
     \label{6.46}
     \eeq
where the energy route \eqref{5.6} has been particularized to our system.

\subsubsection{Mixtures of Nonadditive Hard Rods}

\begin{figure}[t]
\sidecaption[t]
\includegraphics[width=.5\columnwidth]{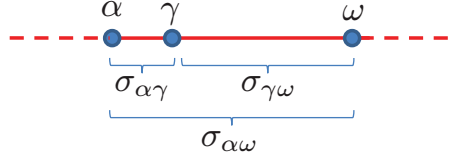}
\caption{Threshold situation ($\sigma_{\alpha\omega}=\sigma_{\alpha\gamma}+\sigma_{\gamma\omega}$) for nearest-neighbor interaction. \label{fig6.5}}
\end{figure}

As a representative example of a one-dimensional mixture, we consider here a nonadditive hard-rod binary mixture [see \eqref{5.31} and Fig.\ \ref{fig5.2}]. The nearest-neighbor interaction condition requires $\sigma_{\alpha\omega}\leq \sigma_{\alpha\gamma}+\sigma_{\gamma\omega}$, $\forall (\alpha,\gamma,\omega)$, as illustrated by Fig.\ \ref{fig6.5}. In the binary case, this condition implies $2\sigma_{12}\geq \max(\sigma_1,\sigma_2)$.

 \begin{figure}[t]
   \includegraphics[width=.7\columnwidth]{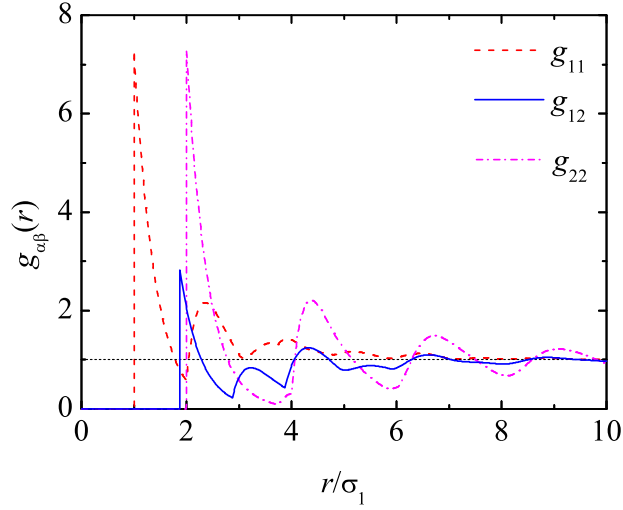}
   \caption{Radial distribution functions for a one-dimensional hard-rod binary mixture with $x_1=x_2=\frac{1}{2}$,  $\sigma_{2}=2\sigma_{1}$, $\sigma_{12}=\frac{15}{8}\sigma_{1}$, and $n\sigma_{1}=\frac{1}{2}$. \label{fig6.6}}
   \end{figure}

The Laplace transform of $\E^{-\beta\phi_{\alpha\gamma}(r)}$ is
 \beq
  \Omega_{\alpha\gamma}(s)=\frac{\E^{-\sigma_{\alpha\gamma}s}}{s}\;.
  \label{6.47}
  \eeq
 The recipe
described by \eqref{6.27}--\eqref{6.34} can be easily implemented.  In order to  obtain the pair
correlation functions $g_{\alpha\gamma}(r)$ in real space, we first note that, according to \eqref{6.31},
\beq
\frac{1}{D(s)}=\sum_{m=0}^\infty
\left[Q_{11}(s)+Q_{22}(s)+Q_{12}^2(s)-Q_{11}(s)Q_{22}(s)\right]^m\;.
\label{6.48}
\eeq
When this is inserted into \eqref{6.27}--\eqref{6.29},
one can express $G_{\alpha\gamma}(s)$ as linear combinations of terms of the
form
\beq
Q_{11}^{k_{11}}(s)Q_{22}^{k_{22}}(s)Q_{12}^{k_{12}}(s)=\frac{\E^{-a(s+\beta p)}}{(s+\beta p)^k}\left(x_1
K_{11}\right)^{k_{11}+k_{12}/2} \left(x_2
K_{22}\right)^{k_{22}+k_{12}/2}\;,
\label{6.49}
\eeq
where $a\equiv k_{11}
\sigma_{1}+k_{22}\sigma_{2}+k_{12}\sigma_{12}$ and $k\equiv
k_{11}+k_{22}+k_{12}$. The inverse Laplace transforms
$g_{\alpha\gamma}(r)=\mathcal{L}^{-1}\left[G_{\alpha\gamma}(s)\right]$ are readily
evaluated by using the property
\beq
\mathcal{L}^{-1}\left[\frac{\E^{-a(s+\beta p)}}{(s+\beta p)^k}\right]=\E^{-\beta p r}\frac{(r-a)^{k-1}}{(k-1)!}\Theta(r-a)\;.
\label{6.50}
\eeq
Analogously to the case of \eqref{6.43}, only the terms with $\{k_{11},k_{22},k_{12}\}$ such that $a<r_{\max}$ are needed if one is interested in distances $r<r_{\max}$.
Figure \ref{fig6.6} shows $g_{\alpha\gamma}(r)$ for a particular binary mixture \cite{S07}.

\section{Density Expansion of the Radial Distribution Function}
\label{sec7}
Except for one-dimensional systems with nearest-neighbor interactions, the exact evaluation of the radial distribution function $g(r)$ or the equation of state $p(n,T)$ by theoretical tools for arbitrary interaction potential $\phi(r)$, density $n$, and temperature $T$ is simply not possible. However, the problem can be controlled if one gives up the ``arbitrary density'' requirement and is satisfied with the low-density regime. In such a case, a series expansion in powers of density is the adequate tool:
\beq
g(r)=g_0(r)+g_1(r) n+g_2(r) n^2+\cdots\;,
\label{7.1}
\eeq
\beq
Z\equiv\frac{p}{nk_BT}=1+B_2(T) n+B_3(T) n^2+\cdots\;.
\label{7.1.2}
\eeq
Therefore, our aim in this section is to derive expressions for the {\emph{virial} coefficients} $g_k(r)$ and $B_k(T)$ as functions of $T$ for any (short-range) interaction potential $\phi(r)$. First, a note of caution: although for an ideal gas one has $g^\id(r)=1$ (and $Z^\id=B_1=1$), in a real  gas $g_0(r)\neq 1$. This is because even, if the density is extremely small,  interactions create correlations among particles. For instance, in a hard-sphere fluid, $g(r)=0$ for $r<\sigma$, no matter how large or small the density is.

What is the  basic idea behind the virial expansions? This is very clearly stated by E. G. D. Cohen in a recent work \cite{C13}:
\begin{quotation}
\begin{svgraybox}

The virial or density expansions reduce the intractable $N (\sim 10^{23})$--particle problem
of a macroscopic gas in a volume $V$ to a sum of an increasing number of tractable isolated few ($1$, $2$, $3$, \ldots) particle problems, where each group of particles \emph{moves} alone in the volume $V$ of the system.

Density expansions will then appear, since the number of single particles, pairs of particles, triplets of particles, \ldots, in the system are proportional to $n$, $n^2$, $n^3$, \ldots, respectively, where $n = N/V$ is the number density of the particles.
\end{svgraybox}
\end{quotation}

In order to attain the goals \eqref{7.1} and \eqref{7.1.2}, it is convenient to work with the grand canonical ensemble. This is because in that ensemble we already have a natural series power expansion for free: the grand canonical partition function is expressed as a series in powers of fugacity [see \eqref{3.47}]. Let us consider a generic quantity $X$ that can be obtained from $\Xi$ by taking its logarithm, by differentiation, etc. Then, from the expansion in \eqref{3.47} one could in principle obtain
\beq
X=\sum_{\ell=0}^\infty {\bar{X}_\ell}\zz^\ell\;,
\label{7.1.3}
\eeq
where the coefficients $\bar{X}_\ell$ are related to the configuration integrals $Q_N$ and depend on the choice of $X$. In particular, in the case of the average density $n=\llangle N\rrangle/V$, we can write
\beq
\boxed{n=\sum_{\ell=1}^\infty \ell \bb_\ell\zz^\ell}\;.
\label{7.1.4}
\eeq
Now, eliminating the (modified) fugacity $\zz$ between \eqref{7.1.3} and \eqref{7.1.4} one can express $X$ in powers of $n$:
\beq
X=\sum_{k=0}^\infty {X}_k n^k\;.
\label{7.1.5}
\eeq
The first few relations are
\beq
X_0=\bar{X}_0\;,\quad X_1=\frac{\bar{X}_1}{\bb_1}\;,\quad X_2=\frac{\bar{X}_2}{\bb_1^2}-\frac{2\bb_2}{\bb_1^3}\bar{X}_1\;\quad
X_3=\frac{\bar{X}_3}{\bb_1^3}-\frac{4\bb_2}{\bb_1^4}\bar{X}_2-\left(\frac{3\bb_3}{\bb_1^4}-\frac{8\bb_2^2}{\bb_1^5}\right)\bar{X}_1\;,
\label{7.1.6}
\eeq
\beq
X_4=\frac{\bar{X}_4}{\bb_1^4}-\frac{6\bb_2}{\bb_1^5}\bar{X}_3-2\left(\frac{3\bb_3}{\bb_1^5}-\frac{10\bb_2^2}{\bb_1^6}\right)\bar{X}_2-
2\left(\frac{2\bb_4}{\bb_1^5}+\frac{20\bb_2^3}{\bb_1^7}-\frac{15\bb_2\bb_3}{\bb_1^6}\right)\bar{X}_1\;.
\label{7.1.7}
\eeq

\subsection{Mayer Function and Diagrams}
As we have seen many times before, the key quantity related to the interaction potential is the Boltzmann factor $\E^{-\beta \phi(r)}$. Since it is equal to unity in the ideal-gas case, a convenient way of measuring deviations from the ideal gas is by means of the \emph{Mayer function}
   \beq
   f(r)\equiv \E^{-\beta \phi(r)}-1\;.
   \label{7.2}
   \eeq

 \begin{figure}[t]
\includegraphics[width=.4\columnwidth]{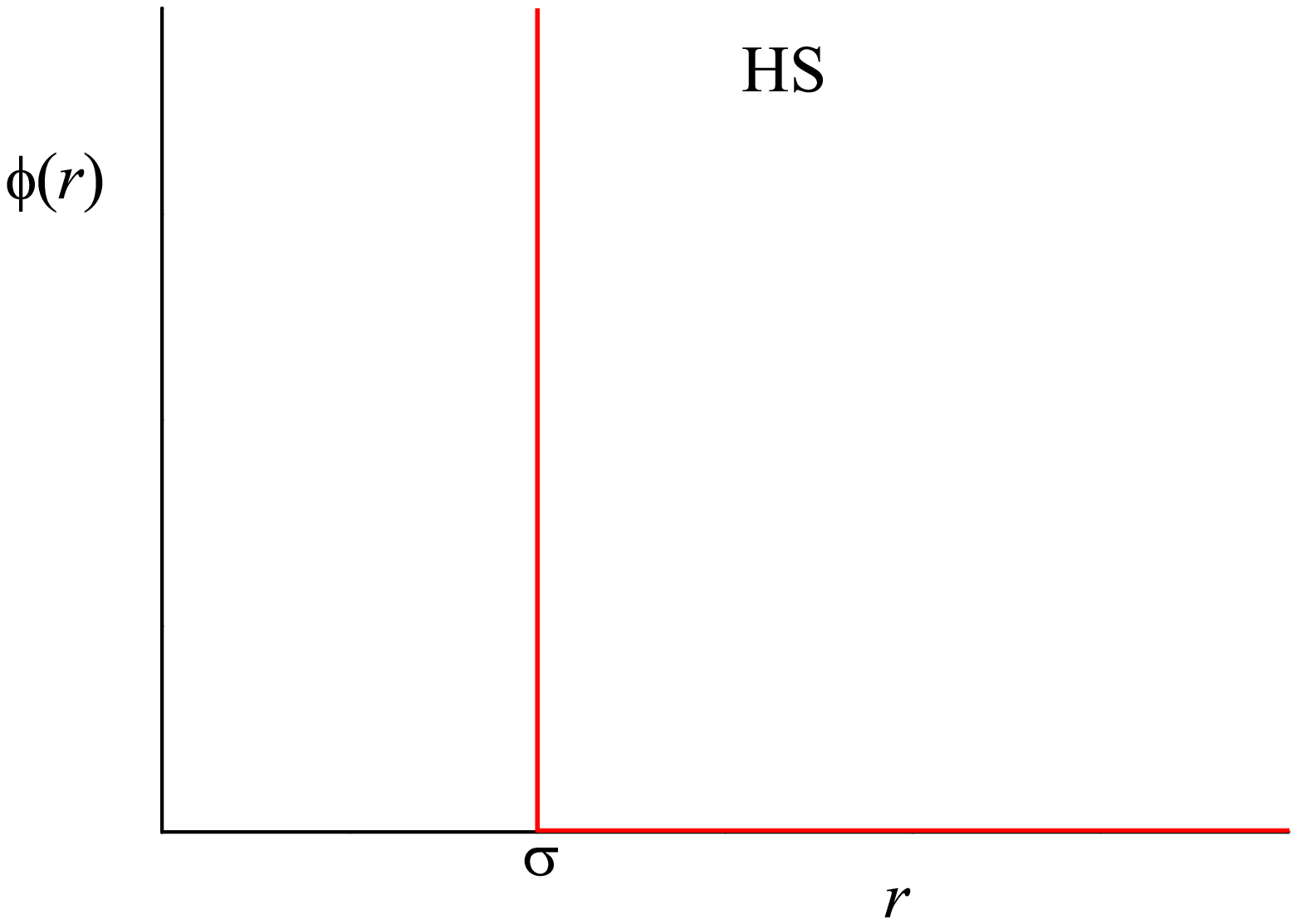}\hfill
\includegraphics[width=.45\columnwidth]{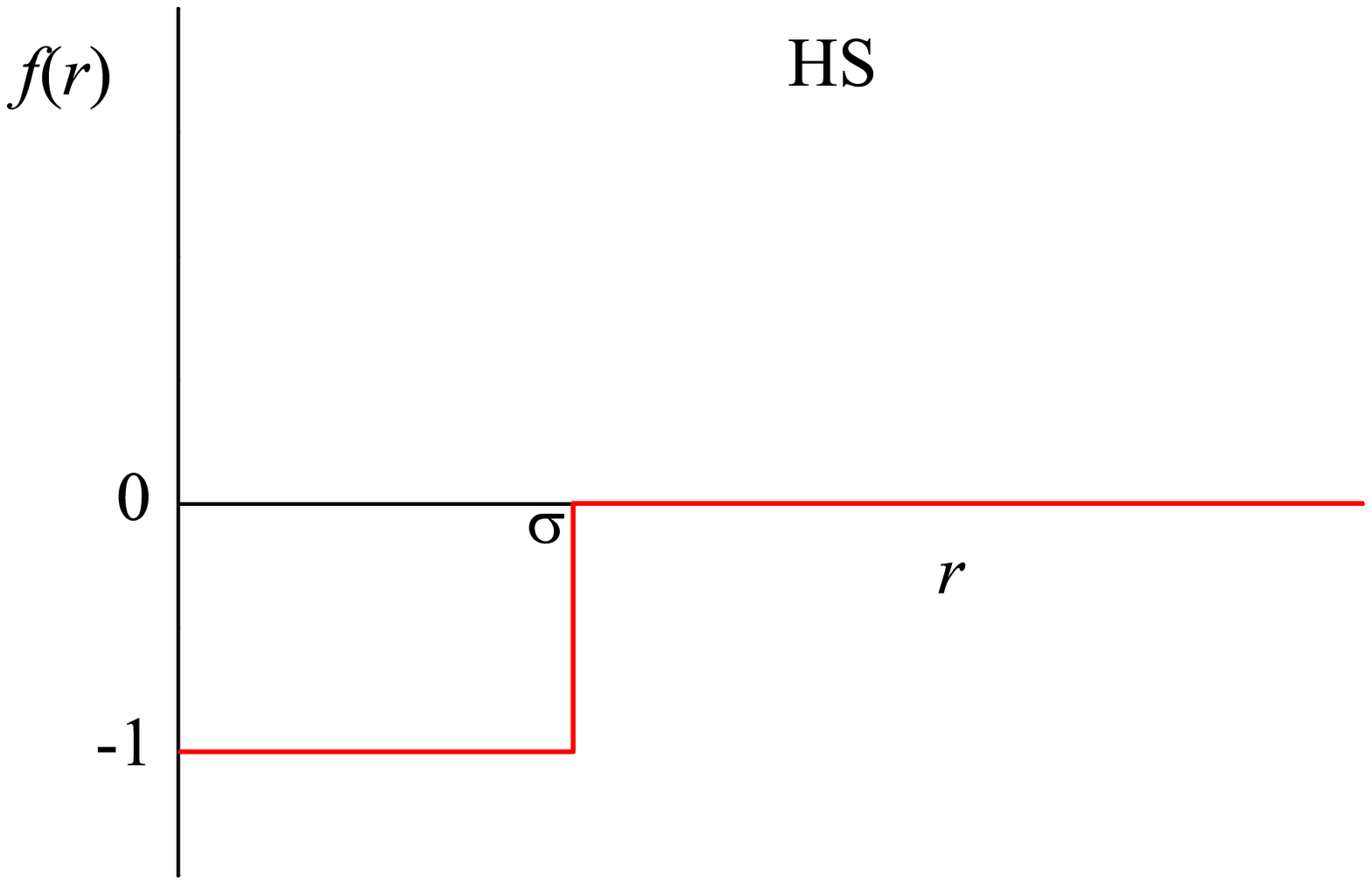}\\

\vspace{.3cm}

\includegraphics[width=.45\columnwidth]{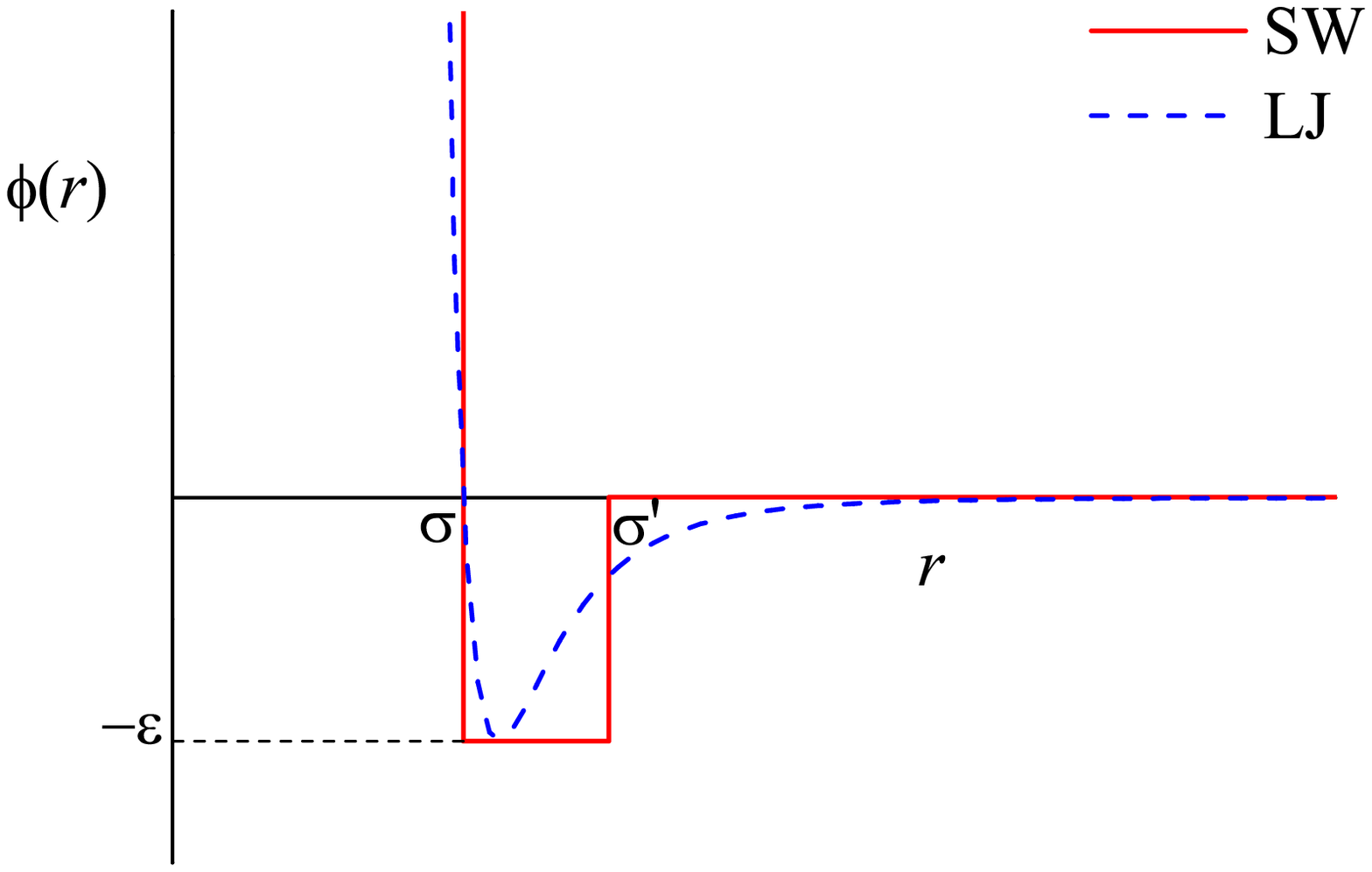}\hfill
\includegraphics[width=.45\columnwidth]{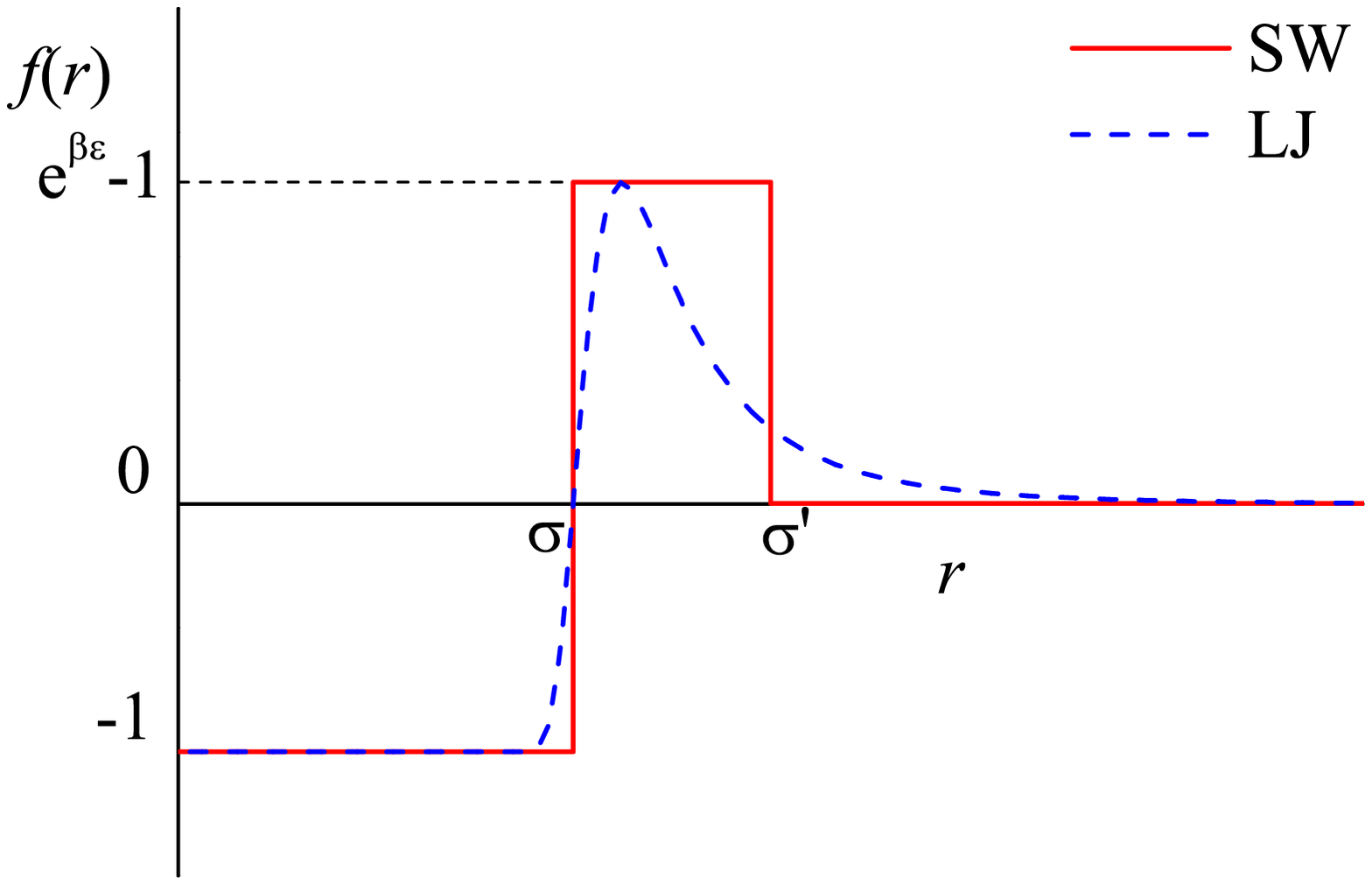}
\caption{Left panels: Hard-sphere, square-well, and Lennard-Jones potentials. Right panels: Respective Mayer functions. \label{fig7.1}}
\end{figure}

The shape of the Mayer function for the hard-sphere potential \eqref{4.15HS}, the Lennard-Jones potential \eqref{4.13}, and the square-well potential \eqref{6.35} is shown in Fig.\ \ref{7.1}.

Let us now rewrite \eqref{3.47} as
\beq
  \Xi=1+\sum_{N=1}^\infty \frac{\zz^N}{N!}\int \D\mathbf{r}^N\, W_N(1,2,\ldots, N)\;,
  \label{7.3}
  \eeq
  where
  \beq
    W_N(1,2,\ldots,N)\equiv W_N(\rr^N)= \E^{-\beta \Phi_N(\rr^N)}=\prod_{1\leq i<j\leq N}(1+f_{ij})\;,\quad f_{ij}\equiv f(r_{ij})\;,
    \label{7.4}
  \eeq
and use has been made of \eqref{3.45} and of the pairwise additivity property \eqref{5.2}.
When expanding the product, $2^{N(N-1)/2}$ terms appear in $W_N$. To manage those terms, it is very convenient to represent them with diagrams. Each diagram contributing to $W_N$ is made of $N$ open circles (representing the $N$ particles), some of them joined by a bond (representing a factor $f_{ij}$).
The diagrams contributing to $W_1$--$W_4$ are
     \beq
  W_1(1)=1=\sone\;,
  \label{7.5}
  \eeq
   \beq
  W_2(1,2)=1+f_{12}=\dtwo+\stwo\;,
  \label{7.6}
  \eeq
   \beqa
  W_3(1,2,3)&=&(1+f_{12})(1+f_{13})(1+f_{23})\nn
  &=&\dthreeA+3\dthreeB+3\rthree+\sthree\;,
  \label{7.7}
  \eeqa
   \beqa
  W_4(1,2,3,4)&=&(1+f_{12})(1+f_{13})(1+f_{14})(1+f_{23})(1+f_{24})(1+f_{34})\nn
  &=&\dfourA+6\dfourB+12\dfourC+3\dfourD+4\dfourE+12\rfourA\nn
  &&+4\rfourB+12\rfourC+3\sfourA+6\sfourB+\sfourC\;.
  \label{7.8}
  \eeqa
The numerical coefficients before some diagrams refer to the number of diagrams topologically equivalent, i.e., those that differ only in the particle labels associated with each circle. Some of the diagrams are \emph{disconnected}  (i.e., there exists at least one  particle isolated from the remaining ones), while the other ones are \emph{connected} diagrams or \emph{clusters} (i.e., it is possible to go from any particle to any other particle by following a path made of bonds). Therefore, in general,
 \begin{svgraybox}
   $$
{W_N(1,2,\ldots,N)=\sum}\text{all (connected and disconnected) diagrams of $N$ particles}.
$$
 \end{svgraybox}

As we will see, in our goal of obtaining the coefficients in the expansions \eqref{7.1} and \eqref{7.1.2}, we will follow a \emph{distillation} process upon which we will get rid of the least relevant diagrams at each stage, keeping only those containing more information. The first step consists of taking the logarithm of the grand partition function:
  \beq
\ln \Xi= \sum_{\ell=1}^\infty \frac{\zz^\ell}{\ell!} \int \D\mathbf{r}^\ell\, U_\ell(1,2,\ldots, \ell)\;,
\label{7.9}
\eeq
where the functions $U_\ell(1,2,\ldots,\ell)$ are called \emph{cluster} (or Ursell) functions. They are obviously related to the functions $W_N(1,2,\ldots,N)$. In fact, by comparing \eqref{7.3} and \eqref{7.9}, one realizes that the relationship between $\{W_N\}$ and $\{U_\ell\}$ is exactly the same as that between \emph{moments} and \emph{cumulants} of a certain probability distribution \cite{R80}. In that analogy, $\Xi$ plays the role of the \emph{characteristic function} (or Fourier transform of the probability distribution) and $-\I \zz$ plays the role of the Fourier variable. The first few relations are
    \beq
W_1(1)=U_1(1)\;,
\label{7.10}
\eeq
\beq
W_2(1,2)=U_1(1)U_1(2)+U_2(1,2)\;,
\label{7.11}
\eeq
\beq
W_3(1,2,3)=U_1(1)U_1(2)U_1(3)+3U_1(1)U_2(2,3)+U_3(1,2,3)\;,
\label{7.12}
\eeq
\beqa
\hspace{-.5cm}W_4(1,2,3,4)&=&U_1(1)U_1(2)U_1(3)U_1(4)+6U_1(1)U_1(2)U_2(3,4)
\nn
&&+3U_2(1,2)U_2(3,4)+4U_1(1)U_3(2,3,4)+U_4(1,2,3,4)\;.
\label{7.13}
\eeqa
Again, each numerical factor represents the number of terms equivalent (except for particle labeling) to the indicated canonical term. Using \eqref{7.5}--\eqref{7.8}, one finds
    \beq
  U_1(1)=1=\sone\;,
  \label{7.14}
  \eeq
   \beq
  U_2(1,2)=f_{12}=\stwo\;,
  \label{7.15}
  \eeq
     \beq
  U_3(1,2,3)=3\rthree+\sthree\;,
  \label{7.16}
  \eeq
  \beqa
  U_4(1,2,3,4)&=&12\rfourA+4\rfourB+12\rfourC+3\sfourA\nn
  &&+6\sfourB+\sfourC\;.
  \label{7.17}
  \eeqa
We observe that all the disconnected diagrams have gone away. In  general,
\begin{svgraybox}
$$
 U_\ell(1,2,\ldots, \ell)=\sum\text{all \emph{connected}  diagrams (i.e.,``clusters'') of $\ell$ particles}.
$$
\end{svgraybox}

For later use, it is important to classify the clusters into \emph{reducible} and \emph{irreducible}. The first class is made of those clusters having at least one \emph{articulation point}, i.e., a point that, if removed together with its bonds, the resulting diagram becomes disconnected. Examples of reducible clusters are
\beq
\rthreeART\;, \quad
       \rfourAART\;, \quad
       \rfourBART\;, \quad
      \rfourCART\;,
      \label{7.18}
 \eeq
 where the articulation points are surrounded by circles.
Irreducible clusters (also called \emph{stars}) are those clusters with no articulation point. For instance,
        \beq
   \sthree\;, \quad \sfourA\;, \quad \sfourB\;, \quad \sfourC\;.
   \label{7.19}
   \eeq

\subsection{External Force. Functional Analysis}
As can be seen from \eqref{3.20}--\eqref{3.22}, the thermodynamic quantities can be obtained in the grand canonical ensemble from derivatives of $\ln\Xi$. On the other hand, the pair correlation function $n_2(\rr_1,\rr_2)$ is given by \eqref{4.9} and is not obvious at all how it can be related to a derivative of $\ln\Xi$. This is possible, however, by means of a trick consisting of assuming  that an \emph{external} potential $u(\mathbf{r})$ is added to the system.
In that case,
  \beq
  \Phi_N(\mathbf{r}^N)\to \Phi_N(\mathbf{r}^N|u)=\Phi_N(\mathbf{r}^N)+\sum_{i=1}^N u(\mathbf{r}_i)\;,
  \label{7.20}
  \eeq
  \beq
U_\ell(\mathbf{r}^\ell|\theta)=U_\ell(\mathbf{r}^\ell)\prod_{i=1}^\ell \theta(\mathbf{r}_i)\;,\quad \theta(\rr)\equiv \E^{-\beta u(\rr)}\;,
\label{7.21}
\eeq
   \beq
\ln \Xi(\alpha,\beta,V|\theta)= \sum_{\ell=1}^\infty \frac{\zz^\ell}{\ell!} \int \D\mathbf{r}^\ell \, U_\ell(1,2,\ldots, \ell|\theta)\;.
\label{7.22}
\eeq
Thus, the quantities $U_\ell$ and $\ln\Xi$ become \emph{functionals} of the free function $\theta(r)$.

To proceed, we will need a few  simple functional derivatives:
   \beq
\frac{\delta}{\delta \theta(\mathbf{r})}\theta(\mathbf{r}_1)=\delta(\mathbf{r}_1-\mathbf{r})\;,
\label{7.23}
\eeq
\beq
\frac{\delta}{\delta \theta(\mathbf{r})}\prod_{k=1}^N \theta(\mathbf{r}_k)=\left[\prod_{k=1}^N \theta(\mathbf{r}_k)\right]\sum_{i=1}^N\frac{\delta(\mathbf{r}_i-\mathbf{r})}{\theta(\mathbf{r}_i)}\;,
\label{7.24}
\eeq
\beq
\frac{\delta^2}{\delta \theta(\mathbf{r})\delta \theta(\mathbf{r}')}\prod_{k=1}^N \theta(\mathbf{r}_k)=\left[\prod_{k=1}^N \theta(\mathbf{r}_k)\right]\sum_{i\neq j}\frac{\delta(\mathbf{r}_i-\mathbf{r})\delta(\mathbf{r}_j-\mathbf{r}')}{\theta(\mathbf{r}_i)\theta(\mathbf{r}_j)}\;.
\label{7.25}
\eeq
    It is then straightforward to obtain the $s$-body reduced distribution function $n_s$ in the absence of external force as the $s$th-order functional derivative of $\Xi(\theta)$ at $\theta=1$, divided by $\Xi$. In particular,
   \beq
n_1(\mathbf{r}_1)=\frac{1}{\Xi}\left.\frac{\delta \Xi(\theta)}{\delta \theta(\mathbf{r}_1)}\right|_{\theta=1}=\left.\frac{\delta \ln \Xi(\theta)}{\delta \theta(\mathbf{r}_1)}\right|_{\theta=1}\;,
\label{7.26}
\eeq
\beqa
n_2(\mathbf{r}_1,\mathbf{r}_2)&=&
\frac{1}{\Xi}\left.\frac{\delta^2 \Xi(\theta)}{\delta \theta(\mathbf{r}_1) \delta \theta(\mathbf{r}_2)}\right|_{\theta=1}\nn
&=&\left.\frac{\delta^2 \ln \Xi(\theta)}{\delta \theta(\mathbf{r}_1) \delta \theta(\mathbf{r}_2)}\right|_{\theta=1}
+\left.\frac{\delta \ln \Xi(\theta)}{\delta \theta(\mathbf{r}_1)}\frac{\delta \ln \Xi(\theta)}{\delta \theta(\mathbf{r}_2)}\right|_{\theta=1}\nn
&=&n_1(\mathbf{r}_1)n_1(\mathbf{r}_2)+\left.\frac{\delta^2 \ln \Xi(\theta)}{\delta \theta(\mathbf{r}_1) \delta \theta(\mathbf{r}_2)}\right|_{\theta=1}\;.
\label{7.27}
\eeqa
In \eqref{7.26} and \eqref{7.27}, $n_1(\rr)=n=\llangle N\rrangle /V$ is actually independent of the position $\rr$ of the particle, but it is convenient to keep the notation $n_1(\rr)$ for the moment.

\subsection{Root and Field Points}
Taking into account \eqref{7.21}, application of \eqref{7.24} and \eqref{7.25} yields
  \beq
\left.\frac{\delta}{\delta \theta(\alert{\mathbf{r}})}\int \D\mathbf{r}^\ell U_\ell(\mathbf{r}^\ell|\theta)\right|_{\theta=1}
=\ell \int \D\mathbf{r}_2\cdots \D\mathbf{r}_\ell\, U_\ell(\alert{\mathbf{r}};\mathbf{r}_2,\ldots,\mathbf{r}_\ell)\;,
\label{7.28}
\eeq
\beq
\left.\frac{\delta^2}{\delta \theta(\alert{\mathbf{r}})\delta \theta(\alert{\mathbf{r}'})}\int \D\mathbf{r}^\ell U_\ell(\mathbf{r}^\ell|\theta)\right|_{\theta=1}
=\ell(\ell-1) \int \D\mathbf{r}_3\cdots \D\mathbf{r}_\ell\, U_\ell(\alert{\mathbf{r}},\alert{\mathbf{r}'};\mathbf{r}_3,\ldots,\mathbf{r}_\ell)\;.
\label{7.29}
\eeq
In the above two equations we have distinguished between position variables that are integrated out and those which are not. We will call \emph{field} points to the former and \emph{root} points to the latter. Thus,
\begin{svgraybox}
  $$ U_\ell(\alert{\mathbf{r}};\mathbf{r}_2,\ldots,\mathbf{r}_\ell):\text{ Ursell function with \alert{1 root point} and $\ell-1$ field points}\;,
  $$
  $$
U_\ell(\alert{\mathbf{r}},\alert{\mathbf{r}'};\mathbf{r}_3,\ldots,\mathbf{r}_\ell):\text{ Ursell function with \alert{2 root points} and $\ell-2$ field points}\;.
$$
\end{svgraybox}

    Therefore, using \eqref{7.22}, \eqref{7.26}, and \eqref{7.27}, we have
    \beq
n_1(\alert{\mathbf{r}_1})=\zz+\sum_{\ell=2}^\infty\frac{\zz^\ell}{(\ell-1)!} \int \D\mathbf{r}_{2}\cdots d\mathbf{r}_\ell\, U_\ell(\alert{1};2,\ldots,\ell)\;,
\label{7.30}
\eeq
\beq
n_2(\alert{\mathbf{r}_1},\alert{\mathbf{r}_2})=n_1(\alert{\mathbf{r}_1})n_1(\alert{\mathbf{r}_2})+\zz^2
U_2(\alert{1},\alert{2})+\sum_{\ell=3}^\infty\frac{\zz^\ell}{(\ell-2)!} \int \D\mathbf{r}_{3}\cdots d\mathbf{r}_\ell\, U_\ell(\alert{1},\alert{2};3,\ldots,\ell)\;.
\label{7.31}
\eeq

{}From \eqref{7.14}--\eqref{7.17} we see that the first few \alert{one-root} cluster diagrams are
       \beq
  \bb_1\equiv U_1(\alert{1})=\sone\;,
  \label{7.32}
  \eeq
   \beq
  2\bb_2\equiv \int \D\mathbf{r}_2\,U_2(\alert{1};2)=\soneSone\;,
  \label{7.33}
  \eeq
     \beq
6\bb_3\equiv  \int \D\mathbf{r}_2\int \D\mathbf{r}_3\, U_3(\alert{1};2,3)=\roneRtwoA+2 \roneRtwoB+\soneStwo\;,
 \label{7.34}
  \eeq
   \beqa
  24\bb_4\equiv\int \D\mathbf{r}_2\int \D\mathbf{r}_3\int \D\mathbf{r}_4\,U_4(\alert{1};2,3,4)&=&6\roneRthreeAA+6\roneRthreeAB+\roneRthreeBA +3\roneRthreeBB\nn
  &&+3\roneRthreeCA+3\roneRthreeCB+6\roneRthreeCC  \nn
  && +3\soneSthreeA
  +3\soneSthreeBA\;.
  \label{7.35}
  \eeqa
Now a filled circle means that the integration over that field point is carried out. As a consequence, some of the diagrams  in \eqref{7.16} and \eqref{7.17} that were topologically equivalent need to be disentangled in \eqref{7.34} and \eqref{7.35} since the new diagrams are invariant under the permutation of two field points but not under the permutation $\text{root}\leftrightarrow\text{field}$. We observe from \eqref{7.30} that the expansion of density in powers of fugacity has the structure \eqref{7.1.4} with
\begin{svgraybox}
$$
\bb_\ell=\frac{1}{\ell!}\sum \text{all clusters with $1$ root and $\ell-1$ field points.}
$$
\end{svgraybox}

Analogously, the first few \alert{two-root} cluster diagrams are
     \beq
  U_2(\alert{1},\alert{2})=\alert{\stwo}\;,
    \label{7.36}
  \eeq
     \beq
 \int \D\mathbf{r}_3\, U_3(\alert{1},\alert{2};3)=\rtwoRoneA+2 \alert{\rtwoRoneB}+\alert{\stwoSone}\;,
   \label{7.37}
   \eeq
   \beqa
  \int \D\mathbf{r}_3\int \D\mathbf{r}_4\,U_4(\alert{1},\alert{2};3,4)&=&2\rtwoRtwoAA+4\alert{\rtwoRtwoAB}+2\alert{\rtwoRtwoAC}+4\rtwoRtwoAD
  \nn
  &&+2\alert{\rtwoRtwoBA} +2\rtwoRtwoBB+2\alert{\rtwoRtwoCA}+4\alert{\rtwoRtwoCB}\nn
  &&+4\rtwoRtwoCC+2\alert{\rtwoRtwoCD} +2\alert{\stwoStwoAA}+\stwoStwoAB\nn
  &&+4\alert{\stwoStwoBA}+\alert{\stwoStwoBB}+\stwoStwoBC+\alert{\stwoStwoC}\;.
    \label{7.38}
  \eeqa
In \eqref{7.36}--\eqref{7.38} we have colored those diagrams in which a direct bond between the root particles \alert{1} and \alert{2} exists. We will call them \alert{\emph{closed}} clusters. The other clusters in which the two root particles are not directly linked will be called \emph{open} clusters.

Closed clusters  factorize into $\alert{\stwo}$ times an \emph{open} cluster.
     For instance,
    \beq
\alert{\rtwoRoneB}=\alert{\stwo}\times \dtwoDoneB\;,
\label{7.39}
\eeq
\beq
\alert{\stwoSone}=\alert{\stwo}\times\rtwoRoneA\;,
\label{7.40}
\eeq
\beq
\alert{\rtwoRtwoAB}=\alert{\stwo}\times\dtwoDtwoCB\;,
\label{7.41}
\eeq
\beq
\alert{\rtwoRtwoAC}=\alert{\stwo}\times\dtwoDtwoD\;.
\label{7.42}
\eeq
In some cases, the root particles \alert{1} and \alert{2} become isolated \emph{after} factorization.

\subsection{Expansion of $n_2(\mathbf{r}_1,\mathbf{r}_2)$ in Powers of  Fugacity}
According to \eqref{7.31}, the coefficients of the expansion of $n_2(1,2)$ come from two sources: the product $n_1(1)n_1(2)$ and the two-root clusters. The first class is represented by two-root diagrams where particles $1$ and $2$ are fully isolated. The second class includes open and closed clusters, the latter ones factorizing as in \eqref{7.39}--\eqref{7.42}. Taking into account all of this, one realizes that the first few coefficients can be factorized as
\beq
\zz^2: 1+\alert{\stwo}=\E^{-\beta\phi_{12}}\;,
\label{7.43}
\eeq
\beqa
\label{7.44}
\zz^3:&&(1+\alert{\stwo})\underbrace{\left(2\dtwoDoneB+\rtwoRoneA\right)}\;,\\
&& \hspace{3.4cm}\alpha_3 \nonumber
\eeqa
\beqa
\label{7.45}
\zz^4:(1+\alert{\stwo})&&\frac{1}{2}\left(2\dtwoDtwoD+2\dtwoDtwoCA+4\dtwoDtwoCB+2\dtwoDtwoE+2\rtwoRtwoAA\right.\nn
&&\underbrace{\left.+4\rtwoRtwoAD+2\rtwoRtwoBB+4\rtwoRtwoCC+\stwoStwoAB+\stwoStwoBC\right)}\;.\nn
&&\hspace{3.8cm}\alpha_4
\eeqa
It can be proved that this factorization scheme extends to all the orders. Thus,
in general,
\beq
\boxed{n_2(\mathbf{r}_1,\mathbf{r}_2)=\E^{-\beta\phi(\mathbf{r}_1,\mathbf{r}_2)}\sum_{\ell=2}^\infty \alpha_{\ell}(\mathbf{r}_1,\mathbf{r}_2) \zz^\ell}\;,
\label{7.46}
\eeq
where
\begin{svgraybox}
  $$
 \alpha_{\ell}(\mathbf{r}_1,\mathbf{r}_2)=\frac{1}{(\ell-2)!}\sum\text{all \emph{open} clusters with 2 root points and $\ell-2$ field points.}
$$
\end{svgraybox}
A note of caution about the nomenclature employed is in order. We say that the diagrams in $\alpha_\ell$ are \emph{open} because the two root particles are not directly linked. But they are also \emph{clusters} because either the group of $\ell$ particles are connected or they would be connected if we imagine a bond between the two roots.
Having this in mind, we can classify the (open) clusters into (open) reducible clusters and (open) irreducible clusters (or stars), as done in \eqref{7.18} and \eqref{7.19}. Of course, all open clusters with particles 1 and 2 isolated are reducible. The open reducible clusters factorize into products of open irreducible clusters. For instance,
  \beq
  \dtwoDoneB=\soneSone\;, \quad  \dtwoDtwoD=(\soneSone)^2\;,
  \label{7.47}
  \eeq
    \beq
  \dtwoDtwoCA=(\soneSone)^2\;,\quad \dtwoDtwoCB=(\soneSone)^2\;,
  \label{7.48}
  \eeq
    \beq
  \dtwoDtwoE=\soneStwo\;,
  \label{7.49}
  \eeq
  \beq
  \rtwoRtwoAD=\soneSone\times \rtwoRoneA\;, \quad \rtwoRtwoBB=\soneSone\times \rtwoRoneA\;.
  \label{7.50}
  \eeq

   Examples of two-root open \emph{irreducible} clusters (``stars'') are
\beq
\rtwoRoneA\;,\quad \rtwoRtwoAA\;, \quad\rtwoRtwoCC\;, \quad\stwoStwoAB\;, \quad\stwoStwoBC\;.
\label{7.51}
\eeq

\subsection{Expansion in Powers of Density}
Equation \eqref{7.46} has the structure of \eqref{7.1.3} with $\bar{X}_0=\bar{X}_1=0$ and $\bar{X}_\ell=\E^{-\beta\phi}\alpha_\ell$.
Elimination of fugacity in favor of density, as in \eqref{7.1.5}, allows us to write
    \beq
    \boxed{n_2(\rr_1,\rr_2)=\E^{-\beta\phi(\mathbf{r}_1,\mathbf{r}_2)}\sum_{k=2}^\infty \gamma_k(\rr_1,\rr_2) n^k}\;,
    \label{7.52}
    \eeq
where ${X}_0={X}_1=0$ and ${X}_k=\E^{-\beta\phi}\gamma_k$. Using \eqref{7.1.6} and \eqref{7.1.7}, we obtain
  \beq
  \gamma_{2}=1\;,
  \label{7.53}
  \eeq
  \beq
  \gamma_{3}=\alpha_{3}-4\bb_2=\rtwoRoneA\;,
  \label{7.54}
  \eeq
  \beqa
 \gamma_{4}&=&\alpha_{4}-6\alpha_{3}\bb_2+20\bb_2^2-6\bb_3\nn
  &=&\frac{1}{2}\left(2\rtwoRtwoAA+4\rtwoRtwoCC+\stwoStwoAB+\stwoStwoBC\right)\;.
  \label{7.55}
  \eeqa
Here we have taken into account that $\bb_1=\alpha_2=1$. The explicit diagrams displayed in \eqref{7.54} and \eqref{7.55} are the ones surviving after considering \eqref{7.33}, \eqref{7.34}, \eqref{7.44}, \eqref{7.45}, and the factorization properties \eqref{7.47}--\eqref{7.50}.
    In general,
    \begin{svgraybox}
      $$
      \gamma_{k}(\mathbf{r}_1,\mathbf{r}_2)=\frac{1}{(k-2)!}\sum\text{all {open} \emph{stars} with 2 root points and $k-2$ field points.}
      $$
    \end{svgraybox}

\begin{table}[t]
\caption{Summary of diagrams contributing to different quantities}
\label{table7.1}       
\begin{tabular}{p{1.3cm}p{3.cm}p{1.5cm}p{4.cm}p{1.5cm}}
\hline\noalign{\smallskip}
Quantity & Expansion in powers of & Coefficient & Diagrams &Equation \\
\noalign{\smallskip}\svhline\noalign{\smallskip}
$\Xi$ & fugacity ($\zz$)  & $W_N$ &All (disconnected+clusters)&\protect\eqref{7.3}\\
$\ln \Xi$ & fugacity ($\zz$) & $U_\ell$ & Clusters (reducible+stars)&\protect\eqref{7.9}\\
$n_2$ &fugacity ($\zz$)  & $\alpha_\ell$ & Open clusters (reducible+stars)&\protect\eqref{7.46}\\
$n_2$ & density ($n$)  & $\gamma_k$ & Open stars&\protect\eqref{7.52}\\
\noalign{\smallskip}\hline\noalign{\smallskip}
\end{tabular}
\end{table}

A summary of the ``distillation'' process leading to \eqref{7.52} is presented in Table \ref{table7.1}.
Taking into account the definitions \eqref{4.11} and \eqref{5.5} of the radial distribution function and the cavity function, respectively, \eqref{7.52} can be rewritten as
\beq
\boxed{g(r)=\E^{-\beta\phi(r)}\left[1+\sum_{k=1}^\infty \gamma_{k+2}(r) n^k\right]}\;,\quad \boxed{y(r)=1+\sum_{k=1}^\infty \gamma_{k+2}(r) n^k}\;.
 \label{7.56}
\eeq
Thus, the functions $g_k(r)$ in \eqref{7.1} are given by $g_k(r)=\E^{-\beta\phi(r)}\gamma_{k+2}(r)$.
In particular, in the limit $n\to 0$, $g(r)\to g_0(r)=\E^{-\beta\phi(r)}$, which differs from the ideal-gas function $g^\id(r)=1$, as anticipated. However, $\lim_{n\to 0}y(r)=1$.

The formal extension of the result $g_0(r)=\E^{-\beta\phi(r)}$ to any order in density \emph{defines} the so-called \emph{potential of mean force} $\psi(r)$ from
\beq
\boxed{g(r)=\E^{-\beta\psi(r)}}\Rightarrow \boxed{\psi(r)=-k_BT\ln g(r)}\;.
\label{7.56b}
\eeq
Obviously, $\psi(r)\neq \phi(r)$, except in the limit $n\to 0$.  In general,
\beq
\beta\psi(r)=\beta \phi(r)-\ln y(r)\;.
\label{7.56c}
\eeq

 \begin{figure}[t]
 \sidecaption[t]
\includegraphics[width=.64\columnwidth]{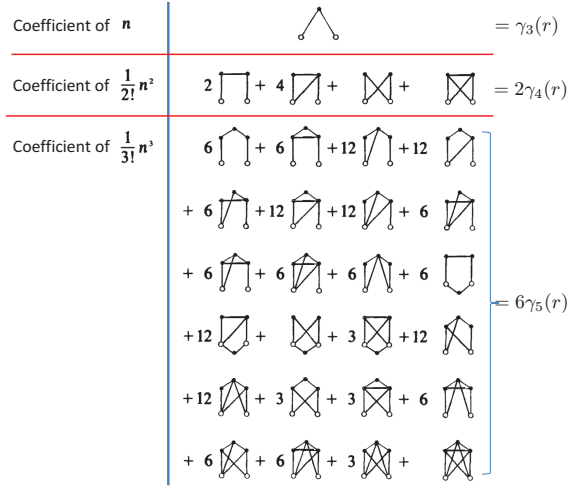}
\caption{Diagrams contributing to $\gamma_3(r)$, $\gamma_4(r)$, and $\gamma_5(r)$. Adapted from Table 8.3.1 of  \protect\cite{B74b}.\label{fig7.2}}
\end{figure}

The diagrams representing the functions $\gamma_3(r)$ and $\gamma_4(r)$ are given by \eqref{7.54} and \eqref{7.55}, respectively. As the order $k$ increases, the number of diagrams and their complexity increase dramatically. This is illustrated by Fig.\ \ref{fig7.2}.

The simplest diagram (of course, apart from $\gamma_2=1$) is the one corresponding to $\gamma_3$. More explicitly,
\beq
\gamma_3(r_{12})=\int\D \rr_3\, f(r_{13})f(r_{23})\;.
\label{7.56bb}
\eeq
In the special case of hard spheres, where $f(r)=-\Theta(\sigma-r)$ (see Fig.\ \ref{fig7.1}), $\gamma_3(r)$ is the overlap volume of two spheres of radius $\sigma$ whose centers are separated a distance $r$. In $d$ dimensions, the result is \cite{BC87}
\beq
\gamma_3(r)=\frac{2^{d-1}\left(\pi/4\right)^{(d-1)/2}}{\Gamma\left(\frac{d+1}{2}\right)}\sigma^d \Theta(2\sigma-r)\text{B}_{1-r^2/4\sigma^2}\left(\frac{d+1}{2},\frac{1}{2}\right)\;,
\label{7.57}
\eeq
where
\beq
\text{B}_x(a,b)=\int_0^x \D t\, t^{a-1}(1-t)^{b-1}
\label{7.58}
\eeq
is the incomplete beta function \cite{AS72}. In particular, for three-dimensional systems,
\beq
\gamma_3(r)=\frac{\pi}{12}(2-r)^2(4+r)\Theta(2-r)\;,
\label{7.59}
\eeq
where $r$ is assumed to be measured in units of $\sigma$.
For this system, each one of the diagrams contributing to $\gamma_4(r)$ has also been evaluated \cite{NvH52,RKM66,SM07}. The results are
\beqa
\rtwoRtwoAA&=&\frac{\pi^2}{36}\frac{3}{35r}(r-1)^4(r^3+4r^2-53r-162)\Theta(1-r)\nn
&&-\frac{\pi^2}{36}\frac{1}{35r}(r-3)^4(r^3+12r^2+27r-6)\Theta(3-r)\;,
\label{7.60}
\eeqa
\beqa
\rtwoRtwoCC&=&-\frac{\pi^2}{36}\frac{2}{35r}(r-1)^4(r^3+4r^2-53r-162)\Theta(1-r)+\frac{\pi^2}{36}\frac{1}{35r}\nn
&&\times(r-2)^2(r^5+4r^4-51r^3-10r^2+479
r-81)\Theta(2-r)\;,
\label{7.61}
\eeqa
\beq
\stwoStwoAB=\left[\gamma_3(r)\right]^2\;,
\label{7.61b}
\eeq
\beq
\stwoStwoBC=\chi_A(r)\Theta(1-r)+\chi_B(r)\Theta(\sqrt{3}-r)-\left[\gamma_3(r)\right]^2\;,
\label{7.62}
\eeq
where
\beqa
\chi_A(r)&=&\frac{\pi^2}{630}(r-1)^4\left(r^2+4r-53-162
r^{-1}\right)\nn&& -2\pi\left(\frac{3 r^6}{560}-\frac{r^4}{15}
+\frac{r^2}{2}-\frac{2r}{15}+\frac{9}{35r}\right)
\cos^{-1}\frac{-r^2+r+3}{\sqrt{3(4-r^2)}}\;,
\label{7.63}
\eeqa
\beqa
\chi_B(r)&=&\pi\left[-r^2\left(\frac{3r^2}{280}-\frac{41}{420}\right)\sqrt{3-r^2}
-\left(\frac{23}{15}r-\frac{36}{35r}\right)\cos^{-1}\frac{r}{\sqrt{3(4-r^2)}}
  \right.\nn
&&    +\left(\frac{3
r^6}{560}-\frac{r^4}{15}+\frac{r^2}{2}+\frac{2r}{15}-\frac{9}{35r}\right)
\cos^{-1}\frac{r^2+r-3}{\sqrt{3(4-r^2)}}\nn &&
\left.+\left(\frac{3
r^6}{560}-\frac{r^4}{15}
+\frac{r^2}{2}-\frac{2r}{15}+\frac{9}{35r}\right)
\cos^{-1}\frac{-r^2+r+3}{\sqrt{3(4-r^2)}}\right]\;,
\label{7.64}
\eeqa

\subsection{Equation of State. Virial Coefficients}
The knowledge of the coefficients $\gamma_k(r)$ allows us to obtain the virial coefficients $B_k(T)$ defined in \eqref{7.1.2}. As long as all the exact diagrams in $\gamma_k(r)$ are incorporated, it does not matter which route is employed to get the virial coefficients. The most straightforward route is the virial one [see \eqref{5.13}].
    Therefore,
    \beq
    B_k(T)=\frac{1}{2d}\int \D\mathbf{r}\, \gamma_{k}(r) r \frac{\partial f(r)}{\partial r}\;,
    \label{7.65}
    \eeq
where we have taken into account that $\partial \E^{-\beta\phi(r)}/\partial r=\partial f(r)/\partial r$.
     In particular, the second virial coefficient is
      \beq
     B_2(T)=\frac{1}{2d}\int \D\mathbf{r}\,  r \frac{\partial f(r)}{\partial r}=2^{d-1}v_d\int_0^\infty \D r\, r^d \frac{\partial f(r)}{\partial r}=-d 2^{d-1} v_d \int_0^\infty \D r\, r^{d-1}  f(r)\;,
     \label{7.66}
     \eeq
     where we have passed to spherical coordinates [see \eqref{5.34} and \eqref{5.35}] and have integrated by parts. Going back to a volume integral,
     \beq
     \boxed{B_2(T)=-\frac{1}{2}\int \D\mathbf{r}\, f(r)}\;.
     \label{7.67}
     \eeq

     In general, it can be proved that \cite{R80}
     \begin{svgraybox}
     $$
        B_k(T)=-\frac{k-1}{k!}\sum\text{all {open} \emph{stars} with 1 root  and $k-1$ field points.}
        $$
     \end{svgraybox}
     The first few cases are
     \beq
   B_2(T)=-\frac{1}{2}\soneSone\;,\quad
   B_3(T)=-\frac{1}{3}\soneStwo\;,
   \label{7.69}
   \eeq
   \beq
   B_4(T)=-\frac{1}{8}\left(3\soneSthreeA+6\soneSthreeBA+\soneSthreeC\right)\;.
   \label{7.70}
   \eeq

\subsubsection{Second Virial Coefficient}
For $d$-dimensional hard spheres, the second virial coefficient is simply
\beq
B_2=2^{d-1}v_d \sigma^d\;,
\label{7.71}
\eeq
so that the equation of state truncated after $B_2$ is
\beq
Z\equiv\frac{p}{nk_BT}=1+2^{d-1}\eta+\cdots\;,
\label{7.72}
\eeq
where
\beq
\boxed{\eta\equiv n v_d\sigma^d}
\label{7.73}
\eeq
is the packing fraction [see \eqref{5.40} for its definition in the multicomponent case].

The hard-sphere Mayer function is independent of temperature (see Fig.\ \ref{fig7.1}) and so are the hard-sphere virial coefficients. On the other hand, in general $B_2(T)$ is a function of temperature. As a simple example, the result for the square-well potential [see \eqref{6.35} and Fig.\ \ref{fig7.1}] is
\beq
B_2(T)=2^{d-1}v_d \sigma^d\left\{1-\left(\E^{\beta\varepsilon}-1\right)\left[(\sigma'/\sigma)^d-1\right]\right\}\;.
\label{7.74}
\eeq

The evaluation is less straightforward in the case of continuous potentials like the Lennard-Jones one [see \eqref{4.13}]. Let us consider the more general case of the Lennard-Jones ($2s$-$s$) potential (with $s>d$):
  \beq
  \phi(r)=4\varepsilon\left[\left(\frac{\sigma}{r}\right)^{2s}-\left(\frac{\sigma}{r}\right)^{s}\right]\;.
  \label{7.75}
  \eeq
{}Starting from the last equality in \eqref{7.66} and introducing the change of variable $r\to t\equiv \sqrt{8\beta\epsilon}(\sigma/r)^s$, one has
\beq
B_2(T)=-2^{d-1}v_d\sigma^d \frac{d}{s}\left(8\beta\epsilon\right)^{d/2s}\int_0^\infty \D t\, t^{-d/s-1}\left(\E^{-t^2/2+\sqrt{2\beta\epsilon} t}-1\right)\;.
\label{7.77}
\eeq
The integral can be compared with the following integral representation of the parabolic cylinder function \cite{AS72}:
\beq
\text{D}_a(z)=\frac{\E^{-z^2/4}}{\Gamma(-a)}\int_0^\infty \D t\, t^{-a-1}\left(\E^{-t^2/2-zt}-1\right)\;,\quad 0<\text{Re}(a)<1\;.
\label{7.78}
\eeq
Thus, \eqref{7.77} becomes
  \beq
 \frac{B_2(T)}{B_2^\hs}=\Gamma\left(1-\frac{d}{s}\right)\left(\frac{8}{T^*}\right)^{d/2s}\E^{1/2T^*}\text{D}_{d/s}\left(-\sqrt{\frac{2}{T^*}}\right)\;,
 \label{7.80}
  \eeq
where $ T^*\equiv{k_BT}/{\varepsilon}$ and $B_2^\hs$ is given by \eqref{7.71}.
To the best of the author's knowledge, the compact expression \eqref{7.80} has not been published before.

\begin{figure}[t]
\includegraphics[width=.5\columnwidth]{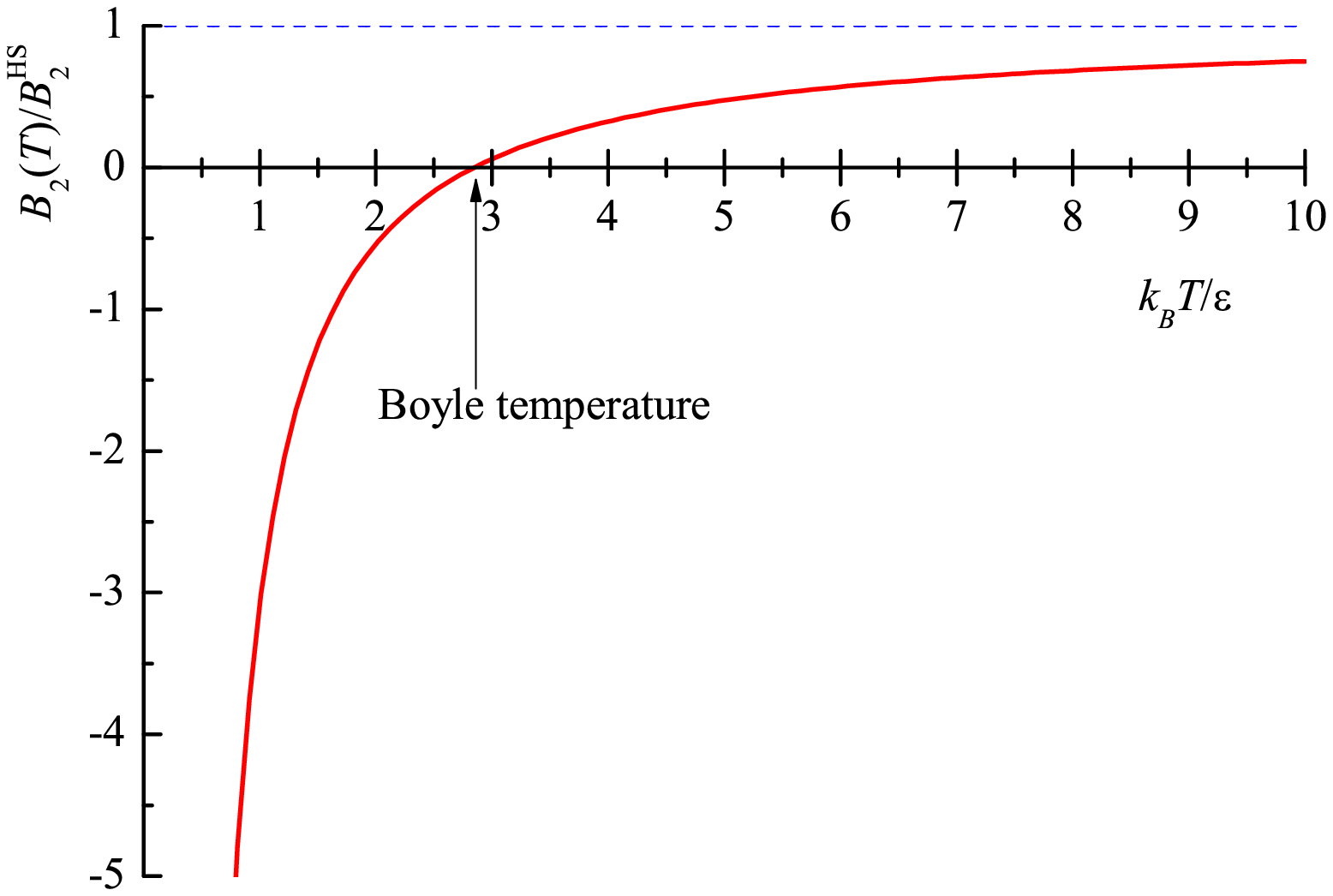}\hfill
\includegraphics[width=.5\columnwidth]{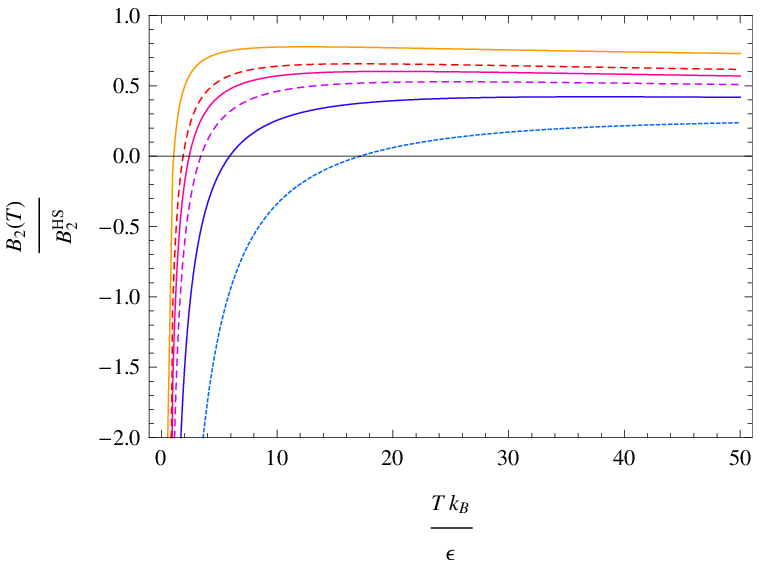}
\caption{Left panel: $B_2(T)$ for a square-well fluid with $\sigma'/\sigma=1.5$. Right panel: $B_2(T)$ for a Lennard-Jones ($2s$-$s$) fluid with $s=4$, $5$, $6$, $7$, $8$, and $12$, from bottom to top.\label{fig7.3}}
\end{figure}

Figure \ref{fig7.3} shows the temperature-dependence of $B_2$, relative to the hard-sphere value with the same $\sigma$, for (three-dimensional) square-well and Lennard-Jones fluids \cite{note_13_09}. For low temperatures the attractive part of the potential dominates and thus $B_2<0$, meaning that in the low-density regime the pressure is smaller than that of an ideal gas at the same density. Reciprocally, $B_2>0$ for high temperatures, in which case the repulsive part of the potential prevails. The transition between both situations takes place at the so-called Boyle temperature $T_B$, where $B_2=0$. Note that, while the square-well second virial coefficient monotonically grows with temperature and asymptotically tends to the hard-sphere value, the Lennard-Jones coefficient reaches a maximum (smaller than the hard-sphere value corresponding to a diameter $\sigma$) and then decreases very slowly. This reflects the fact that for very high temperatures the system behaves practically as a hard-sphere system but with an \emph{effective} diameter smaller than the nominal value $\sigma$.

\subsubsection{Higher-Order Virial Coefficients for Hard Spheres}
The evaluation of virial coefficients beyond $B_2$ becomes a formidable task as the order increases and it is necessary to resort to numerical Monte Carlo methods to perform the multiple integrals involved. Needless to say, the task is much more manageable in the case of hard spheres. In the one-component case, the third and fourth virial coefficients are analytically known \cite{CM04b,L05} and $B_5$--$B_{12}$ have been numerically evaluated \cite{CM04a,LKM05,CM05,CM06,W13}.

  \begin{figure}[t]
    \sidecaption[t]
\includegraphics[width=.64\columnwidth]{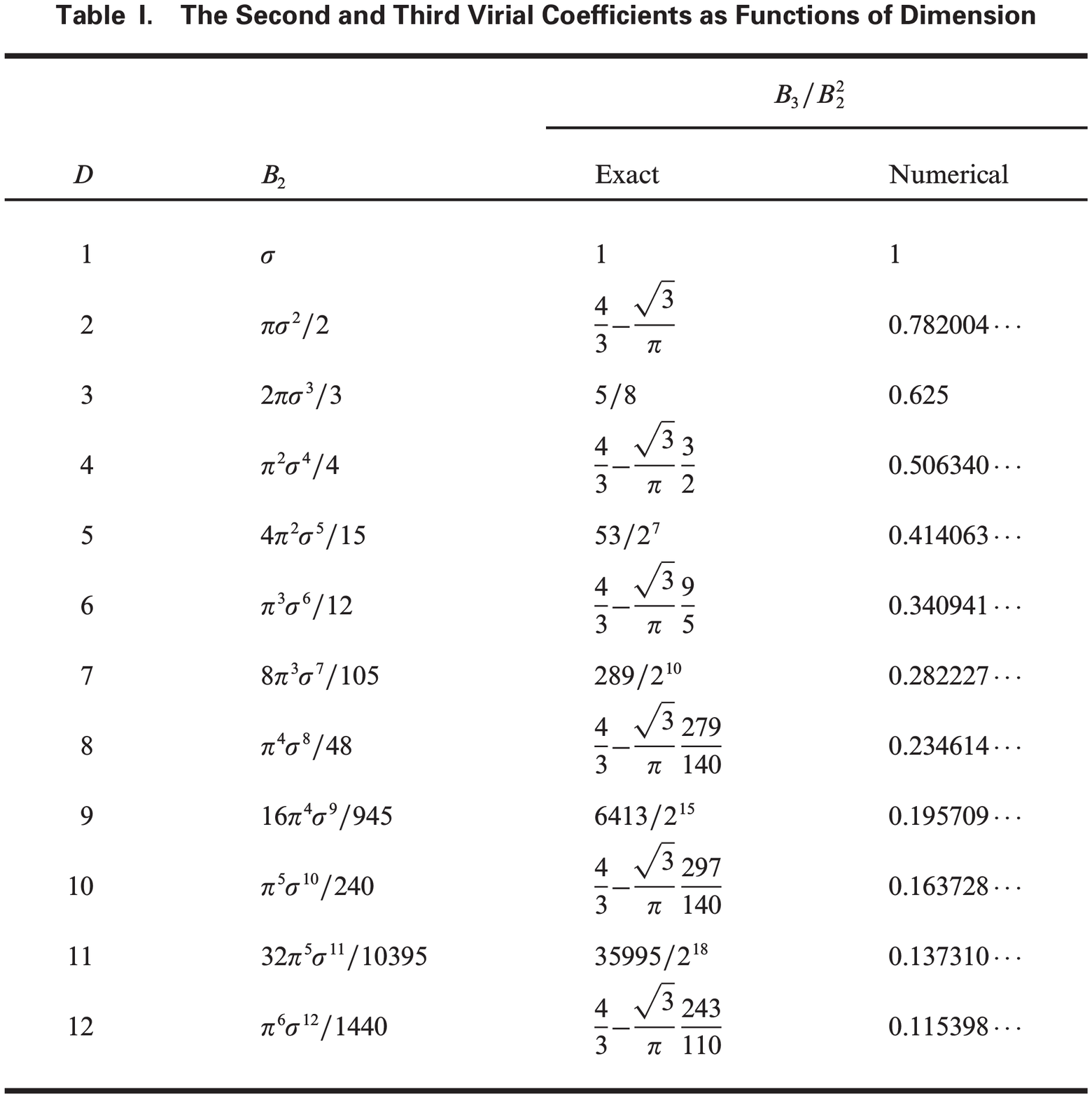}
\caption{Second and third virial coefficients for $d$-dimensional hard spheres. Source:  \protect\cite{CM04a}. \label{fig7.4}}
\end{figure}
  \begin{figure}[t]
  \sidecaption[t]
\includegraphics[width=.64\columnwidth]{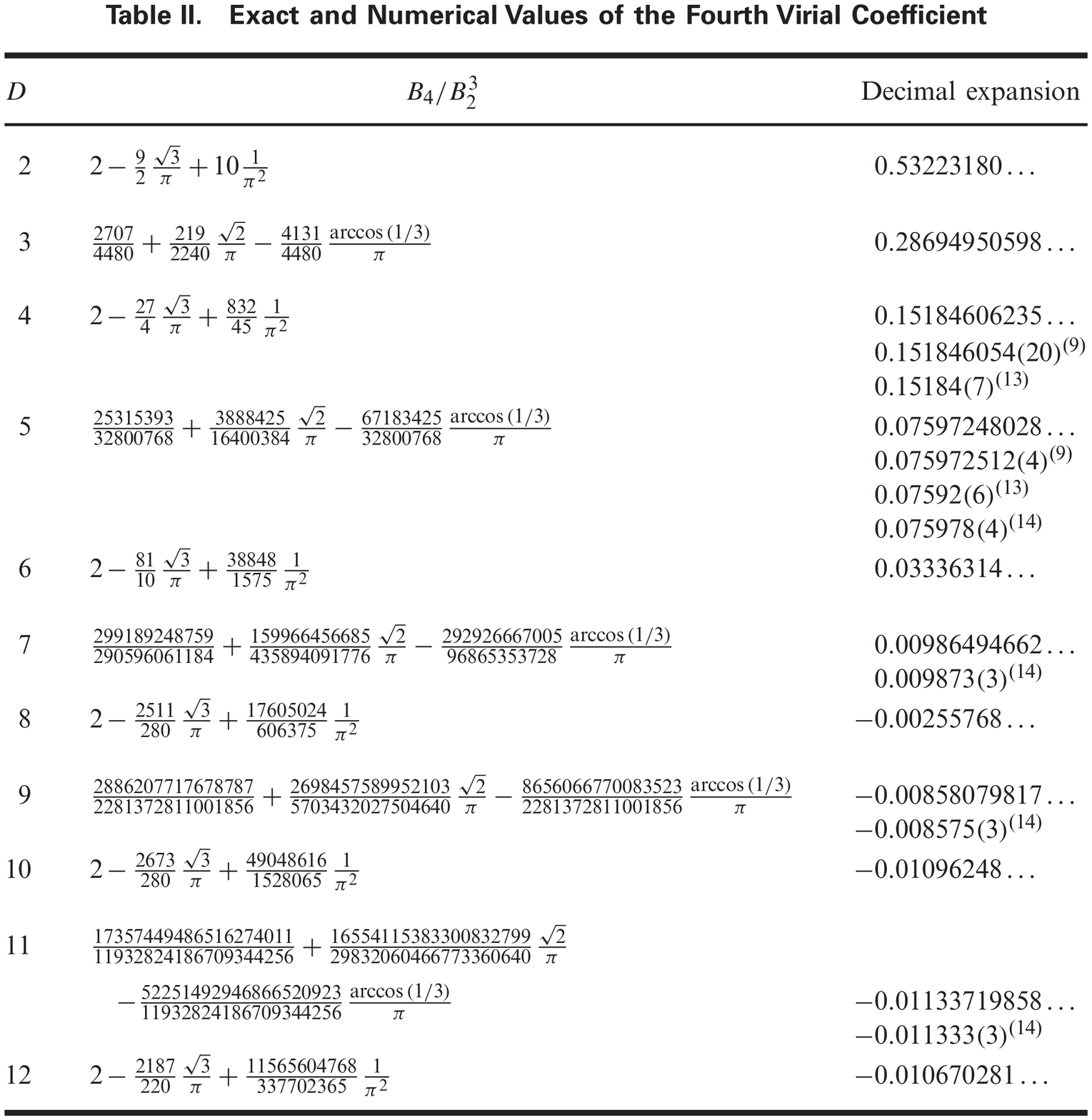}
\caption{Fourth virial coefficient for $d$-dimensional hard spheres. Source:  \protect\cite{L05}. \label{fig7.5}}
\end{figure}

The third virial coefficient is \cite{LB82}
\beq
\frac{B_3}{B_2^2}=2 \text{I}_{3/4}\left(\frac{d+1}{2},\frac{1}{2}\right)\;,
\label{7.81}
\eeq
where $\text{I}_x(a,b)=\text{B}_x(a,b)\Gamma(a+b)/\Gamma(a)\Gamma(b)$ is the \emph{regularized} incomplete beta function [see \eqref{7.58}].
The explicit expressions of $B_2$ and $B_3/B_2^2$ for $d\leq 12$ can be found in Fig.\ \ref{fig7.4}. We note that $B_3/B_2^2$ is a rational number if $d=\text{odd}$, while it is an irrational number (since it includes $\sqrt{3}/\pi$) if $d=\text{even}$. The influence of the parity of $d$ is also present in the exact evaluation of $B_4$, which has been carried out separately for $d=\text{even}$ \cite{CM04a} and $d=\text{odd}$ \cite{L05}. The results for $d\leq 12$ are shown in Fig.\ \ref{fig7.5}. We see that $B_4/B_2^3$ is always an irrational number that includes $\sqrt{3}/\pi$ and $1/\pi^2$ if $d=\text{even}$, while it includes $\sqrt{2}/\pi$  and $\cos^{-1}(1/3)/\pi$ if $d=\text{odd}$.
Interestingly, the fourth virial coefficient becomes negative for $d\geq 8$.

\begin{figure}[t]
\includegraphics[width=\columnwidth]{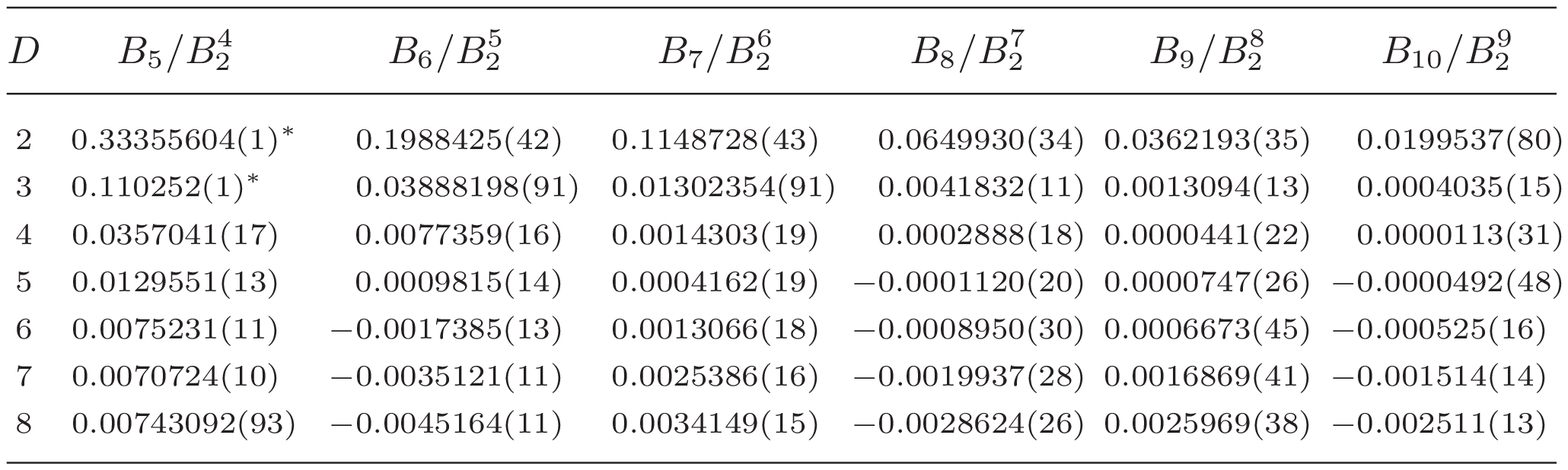}
\caption{Fifth to tenth virial coefficient for $d$-dimensional hard spheres. The numbers in parentheses indicate the
statistical error in the last significant digits. Source:  \protect\cite{CM05}. \label{fig7.6}}
\end{figure}

The Monte Carlo numerical values of the virial coefficients $B_5$--$B_{10}$ up to $d=8$ \cite{CM05,CM06} are displayed in Fig.\ \ref{fig7.6}.
While $B_5$, $B_7$, and $B_9$ remain positive (at least if $d\leq 8$), $B_6$, $B_8$, and $B_{10}$ become negative if $d\geq 6$, $d\geq 5$, and $d\geq 5$, respectively. While the known first ten and twelve virial coefficients are positive if $d=4$ and $d=3$ \cite{W13}, respectively, the behavior observed when $d\geq 5$ shows that this does not need to be necessarily the case for all the virial coefficients. It is then legitimate to speculate that, for three-dimensional hard-sphere systems, a certain high-order coefficient $B_k$ (perhaps with $k=\text{even}$) might become negative, alternating in sign thereafter. This scenario would be consistent with a singularity of the equation of state on the (density) negative real axis that would determine the radius of convergence of the virial series \cite{CM05,CM06,RRHS08}.

\subsubsection{Simple Approximations}
In terms of the packing fraction $\eta$, the virial series \eqref{7.1.2} becomes
\beq
Z=1+2^{d-1}\eta+b_3\eta^2+b_4\eta^3+\cdots\;,\quad b_k\equiv B_k/(v_d\sigma^d)^{k-1}\;.
\label{7.82}
\eeq
Although incomplete, the knowledge of the first few virial coefficients is practically the only access to exact information about the equation of state of the hard-sphere fluid. If the packing fraction $\eta$ is low enough, the virial expansion truncated after a given order is an accurate representation of the exact equation of state. However, this tool is not practical at moderate or high values of $\eta$. In those cases, instead of truncating the series, it is far more convenient to construct an \emph{approximant} which, while keeping a number of exact virial coefficients, includes all of orders in density \cite{MGPC08}. The most popular class is made by Pad\'e approximants \cite{BO87}, where the compressibility factor $Z$ is approximated by the ratio of two polynomials. Obviously, as the number of  retained exact virial coefficients increases so does the complexity of the approximant. 
Here, however, we will deal with simpler, but yet accurate, approximations.

\paragraph{\textbf{Hard disks ($d=2$)}}

\begin{figure}[t]
\sidecaption[t]
\includegraphics[width=.3\columnwidth]{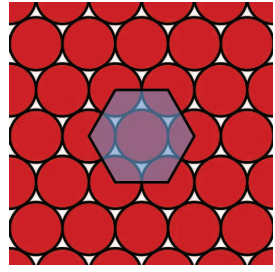}
\caption{Close-packing configuration in a system of hard disks. The fraction of the total area occupied by the disks is $\eta_{\cp}=\frac{\sqrt{3}\pi}{6}$. Source:
\url{http://en.wikipedia.org/wiki/Packing\_problem}. \label{fig7.8}}
\end{figure}

  In the two-dimensional case, the virial series truncated after the third virial coefficient is
    \beq
    Z=1+2 \eta+b_3\eta^2+\cdots,\quad \eta\equiv \frac{\pi}{4}n\sigma^2\;,
    \label{7.83}
    \eeq
    where
    \beq
    b_3=4\left(\frac{4}{3}-\sqrt{3}{\pi}\right)=3.128\cdots\simeq \frac{25}{8}\;.
    \label{7.84}
    \eeq

    Henderson's approximation \cite{H75} consists of
    \beq
    \boxed{Z=\frac{1+\eta^2/8}{(1-\eta)^2}}=1+2 \eta+\frac{25}{8}\eta^2+\cdots\;.
    \label{7.85}
    \eeq
As we see, it retains the exact  second virial coefficient and a rational-number approximation of the third virial coefficient. On the other hand, \eqref{7.85} assumes that the pressure is finite for any $\eta<1$, whereas by geometrical reasons the maximum conceivable packing fraction is the close-packing value $\eta_{\cp}=\frac{\sqrt{3}\pi}{6}\simeq 0.907$ (see Fig.\ \ref{fig7.8}).

Another simple approximation \cite{SHY95,HSY98} exploits the second virial coefficient $b_2=2$ only but imposes a pole at  $\eta_{\cp}$. Thus, the constraints are
\beq
Z=\begin{cases}
1+2\eta+\cdots\;,\quad \eta\ll 1\;,\\
\infty\;,\quad \eta\to\eta_{\cp}\;.
\end{cases}
\label{7.87}
\eeq
A simple approximation satisfying those requirements is
\beq
\boxed{Z=\frac{1}{1-2\eta+\frac{2\eta_{\cp}-1}{\eta_{\cp}^2}\eta^2}}\;.
\label{7.88}
\eeq

\begin{figure}[t]
\sidecaption[t]
\includegraphics[width=.64\columnwidth]{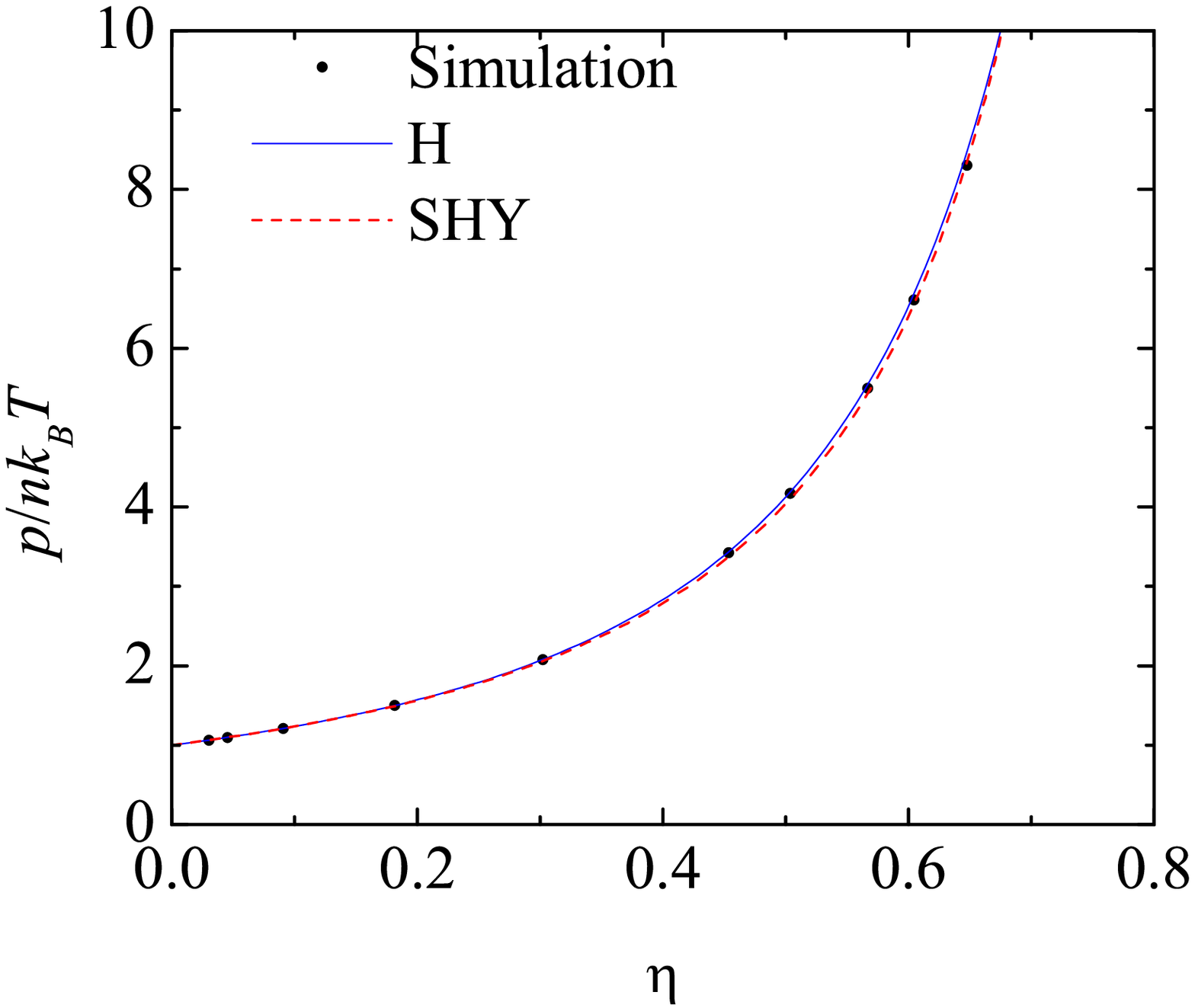}
\caption{Comparison between computer-simulation values of the equation of state of a hard-disk fluid \protect\cite{EL85} and the theoretical approximations \protect\eqref{7.85} (label H) and \protect\eqref{7.88} (label SHY). \label{fig7.9}}
\end{figure}

Figure \ref{fig7.9} compares the predictions of \eqref{7.85} and \eqref{7.88} against computer simulations \cite{EL85}. Despite their simplicity, both approximations exhibit an excellent performance, even at packing fractions where the pressure is about ten times higher than the ideal-gas one.

  \paragraph{\textbf{Hard spheres ($d=3$)}}
  In the three-dimensional case, $\eta=(\pi/6)n\sigma^3$ and the second and third reduced virial coefficients are integer numbers: $b_2=4$ and $b_3=10$. The fourth virial coefficient, however, is a transcendental number (see Fig.\ \ref{7.5}), namely $b_4=18.36476838\cdots$. If we round off this coefficient ($b_4\simeq 18$), we realize that $b_4-b_3=(b_3-b_2)+2$. Interestingly, by continuing the rounding-off process, the relationship $b_k-b_{k-1}=(b_{k-1}-b_{k-2})+2$ extends up to $k=6$, as shown in Table \ref{table7.2}.

  \begin{table}[t]
\caption{The second row shows the round-off integer of the known first twelve reduced virial coefficients $b_k$ of a three-dimensional hard-sphere fluid.
The third row gives the values obtained from the formula $k^2+k-2$. Finally, the deviation  $\varDelta b_k$ of the latter values from the true values of $b_k$ are shown
in the fourth row.}
\label{table7.2}       
\begin{tabular}{p{1.5cm}p{.8cm}p{.8cm}p{.8cm}p{.8cm}p{.8cm}p{.8cm}p{.8cm}p{.8cm}p{.8cm}p{.8cm}p{.8cm}}
\hline\noalign{\smallskip}
$k$& 2 & 3 & 4 & 5 & 6 & 7 & 8 & 9 & 10 &11&12\\
\noalign{\smallskip}\svhline\noalign{\smallskip}
       Round-off& 4 & 10 & 18 & 28 & 40 & 53 & 69 & 86& 106&128(5)& 111(30)\\
        $k^2+k-2$& 4 & 10 & $18$ & 28 & 40 & {54} & {70} & {88}& {108}&130&154 \\
        $\varDelta b_k$&0&0&$-0.36$&$-0.22$ &$0.18$ &$0.66$ &$1.5$& $2.2$& $2.2(4)$&$2(5)$&$43(30)$\\
        \noalign{\smallskip}\hline\noalign{\smallskip}
\end{tabular}
\end{table}

In the late sixties only the first six virial coefficients were accurately known and thus Carnahan and Starling \cite{CS69} proposed to extrapolate the relationship  $b_k-b_{k-1}=(b_{k-1}-b_{k-2})+2$ to any $k\geq 2$, what is equivalent to the approximation
\beq
b_k=k^2+k-2\;.
\label{7.89a}
  \eeq
By summing the virial series within that approximation, they obtained the famous Carnahan--Starling (CS) equation of state:
\beq
\boxed{Z_{\text{CS}}=\frac{1+\eta+\eta^2-\eta^3}{(1-\eta)^3}}\;.
\label{7.89}
\eeq
The corresponding isothermal susceptibility is
\beq
\chi_{\text{CS}}=\left[\frac{\partial\left(\eta Z_{\text{CS}}\right)}{\partial\eta}\right]^{-1}=\frac{(1-\eta)^4}{1+4\eta+4\eta^2-4\eta^3+\eta^4}\;.
\label{7.89b}
\eeq

    \begin{figure}[t]
\sidecaption[t]
\includegraphics[width=.64\columnwidth]{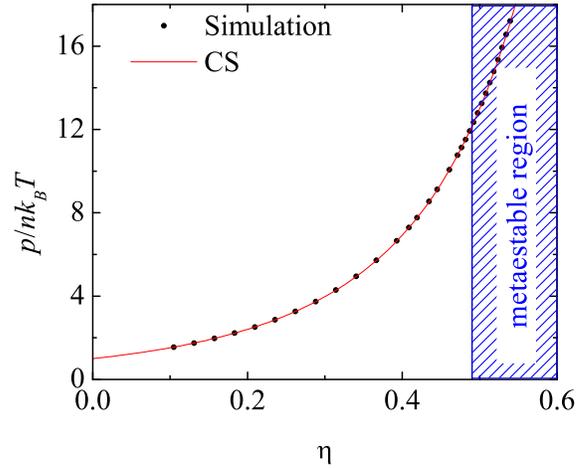}
\caption{Compressibility factor for  three-dimensional hard spheres, as obtained from computer simulations \protect\cite{KLM04} and from the Carnahan--Starling equation of state \protect\eqref{7.89}.\label{fig7.10}}
\end{figure}

Figure \ref{fig7.10} shows that, despite its simplicity, the Carnahan--Starling equation exhibits an excellent performance over the whole fluid stable region and even in the metastable fluid region ($\eta\geq 0.492$ \cite{FMSV12}), where the crystal is the stable phase. This is remarkable because, as shown in Table \ref{table7.2}, the approximation $b_k=k^2+k-2$ fails to capture the rounding-off of the virial coefficient $b_k$ for $k\geq 7$, the deviation $\varDelta b_k$ tending to increase with $k$. The explanation might partially lie in the fact that the Carnahan--Starling recipe underestimates $b_4$ and $b_5$ but this is compensated by an overestimate of the higher virial coefficients.
Apart from that, and analogously to Henderson's equation \eqref{7.85}, the Carnahan--Starling equation \eqref{7.89} provides finite values even for packing fractions higher than the close-packing value $\eta_\cp=\pi\sqrt{2}/6\simeq 0.7405$.

\section{Ornstein--Zernike Relation and Approximate Integral Equation Theories}
\label{sec8}
Similarly to what was said above in connection with the formal virial expansion \eqref{7.1.2} of the equation of state, the virial representation \eqref{7.56} of the radial distribution function is only practical in the low-density regime, in which case the expansion can be truncated after a certain low order. On the other hand, at moderate or high densities this strategy is not useful and in that case it is better to resort to approximations that include all the orders of density, in analogy to what was done in the hard-sphere equation-of-state case with \eqref{7.85}, \eqref{7.88}, and \eqref{7.89}. In order to construct those approximations, a crucial quantity is the \emph{direct correlation function} $c(r)$.

\subsection{Direct Correlation Function}
We recall that the total correlation function is defined by \eqref{4.14}. This function owes its name to the fact that it measures the degree of spatial correlation between two particles separated a distance $r$ due not only to their \emph{direct} interaction but also \emph{indirectly} through other intermediate or ``messenger'' particles. In fact, the range of $h(r)$ is usually much larger than that of the potential $\phi(r)$ itself, as illustrated by Figs.\ \ref{fig4.2} and \ref{fig6.4}. In fluids with a gas-liquid phase transition, $h(r)$ decays algebraically at the \emph{critical point}, so that the integral $\int \D \rr\, h(r)$ diverges and so does the isothermal compressibility $\kappa_T$ [see \eqref{5.1}], a phenomenon known as \emph{critical opalescence} \cite{B74b,HM06}.

    \begin{figure}[t]
    \sidecaption[t]
\includegraphics[width=.3\columnwidth]{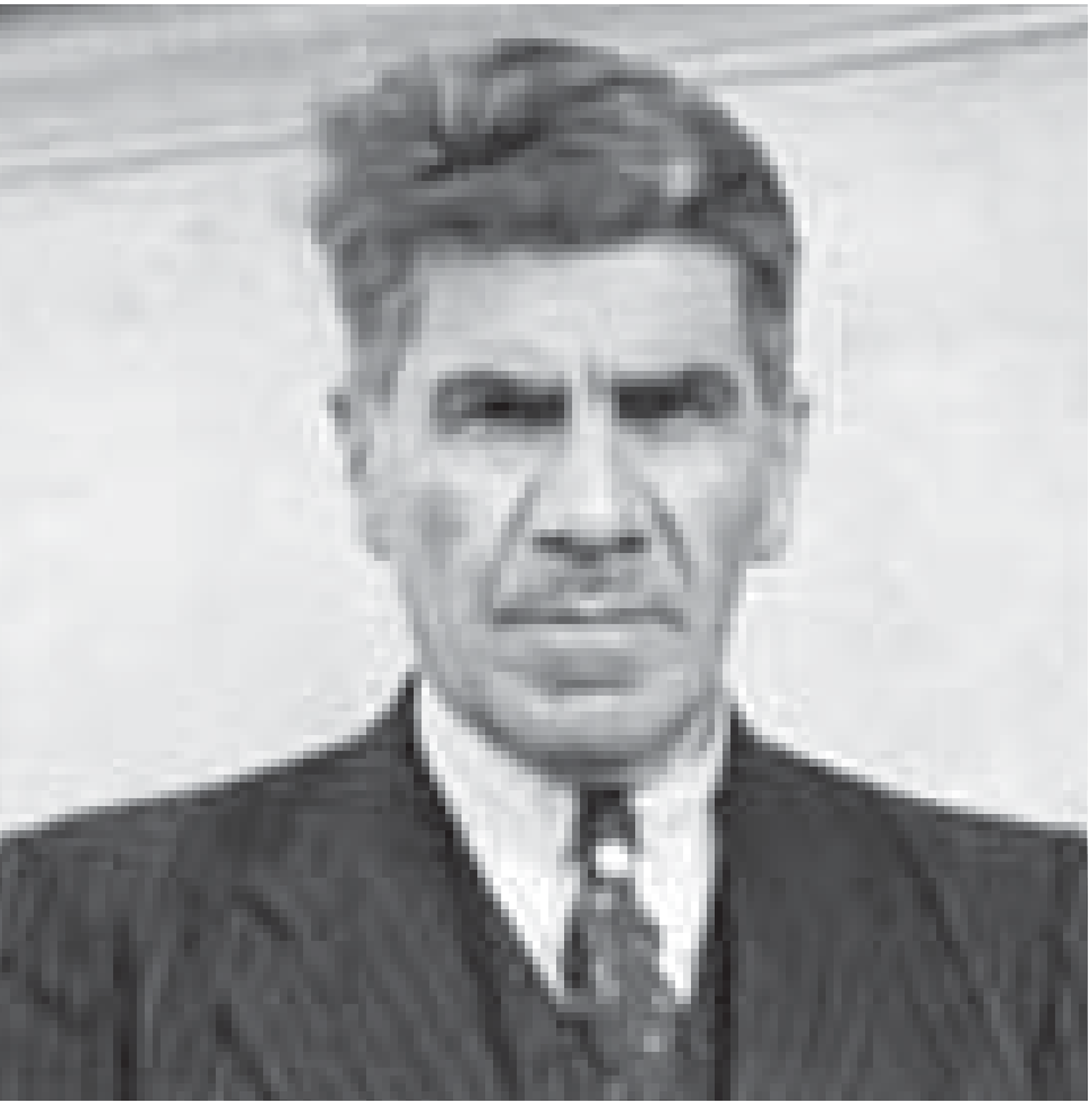}\hfill
\includegraphics[width=.22\columnwidth]{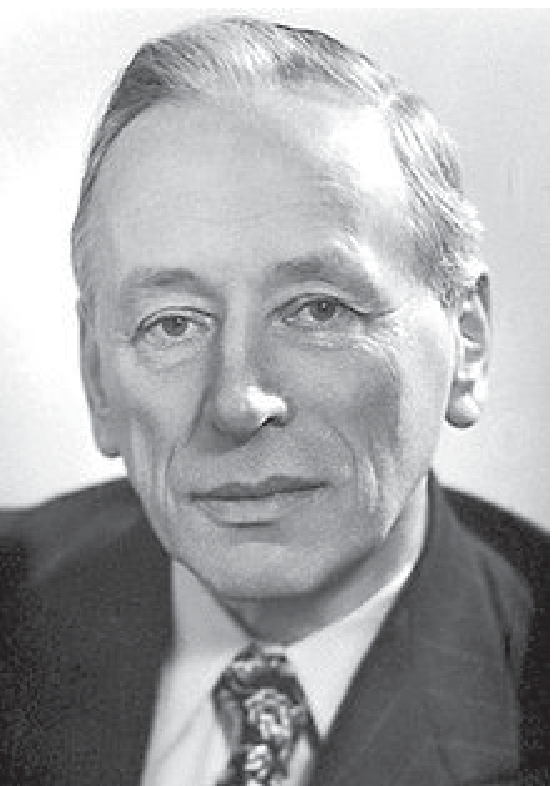}
\caption{Left panel: Leonard Salomon Ornstein (1880--1941). Right panel:  Frits Zernike (1888--1966). \label{fig8.0}}
\end{figure}

\begin{figure}[t]
    \sidecaption[t]
\includegraphics[width=.64\columnwidth]{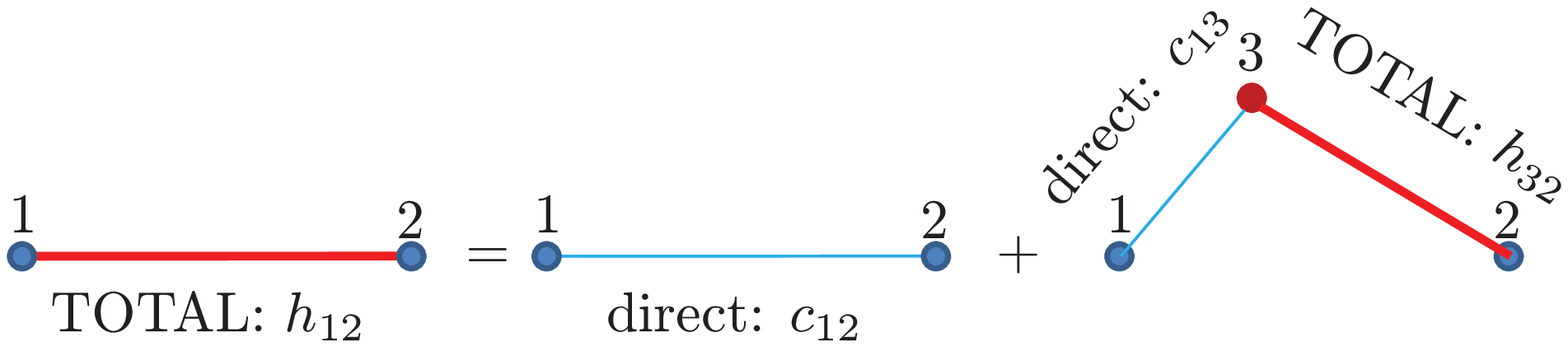}
\caption{Sketch of the meaning of the Ornstein--Zernike relation \protect\eqref{8.1}. \label{fig8.2}}
\end{figure}

It is then important to disentangle from $h(r)$ its direct and indirect contributions. This aim was addressed in 1914 by the Dutch physicists L. S. Ornstein and F. Zernike (see Fig.\ \ref{fig8.0}). They defined the \alert{\emph{direct}} correlation function \alert{$c(r)$} by the integral relation
   \beq
 \boxed{h(r_{12})=\alert{c(r_{12})}+n\int \D\mathbf{r}_3\, \alert{c(r_{13})}h(r_{32})}\;.
  \label{8.1}
  \eeq
The idea behind the Ornstein--Zernike (OZ) relation \eqref{8.1} is sketched in Fig.\ \ref{fig8.2}: the total correlation function $h_{12}$ between particles 1 and 2 can be decomposed into the direct correlation function $c_{12}$ plus the indirect part, the latter being mediated by a messenger particle 3 that is directly correlated to 1 and totally correlated to 2.

Thanks to the convolution structure of the indirect part,  the Ornstein--Zernike relation \eqref{8.1} becomes $\widetilde{h}(k)=\widetilde{c}(k)+ n \widetilde{c}(k)\widetilde{h}(k)$ in  Fourier space or, equivalently,
  \beq
   \widetilde{h}(k)=\frac{\widetilde{c}(k)}{1-n\widetilde{c}(k)}\;,\quad \widetilde{c}(k)=\frac{\widetilde{h}(k)}{1+n\widetilde{h}(k)}\;.
   \label{8.2}
  \eeq
Thus, the compressibility route to the equation of state \eqref{5.1} can be rewritten as
  \beq
\boxed{\chi\equiv n k_BT\kappa_T=\frac{1}{1-n\widetilde{c}(0)}}\;.
\label{8.3}
\eeq
Therefore, even if $\widetilde{h}(0)\to \infty$ (at the critical point), $\widetilde{c}(0)\to n^{-1}=\text{finite}$, thus showing that $c(r)$ is much shorter ranged than $h(r)$, as expected.

It is important to bear in mind that the Ornstein--Zernike relation \eqref{8.1}   \emph{defines} $c(r)$. Therefore, it is \emph{not} a closed equation.
However, if an \emph{approximate closure} of the form $c(r)=\mathcal{F}[h(r)]$ is assumed, one can obtain a \emph{closed integral equation}:
    \beq
  h(r)=\mathcal{F}[h(r)]+n\int \D\mathbf{r}'\,\mathcal{F}[h(r')]h(|\rr-\rr'|)\;.
  \label{8.4}
  \eeq
  In contrast to a truncated density expansion, a closure is applied to \emph{all orders} in density.

Before addressing the \emph{closure problem}  let us first derive \emph{formally exact} relations between $c(r)$, $h(r)$, and some other functions.

\subsection{Classification of Diagrams}

We recall from \eqref{7.56} (see also Fig.\ \ref{fig7.2}) that
\beq
y(r_{12})=1+n \rtwoRoneA+\frac{n^2}{2}\left(2\rtwoRtwoAA+4\rtwoRtwoCC+\stwoStwoAB
+\stwoStwoBC\right)+\cdots\;.
\label{8.5}
\eeq

We now introduce the following classification of \emph{open} stars:
\begin{itemize}
  \item
  \begin{svgraybox}
    \textbf{``Chains'' (or nodal diagrams), $\mathcal{C}(r)$}: Subset of \emph{open} diagrams  having at least one \emph{node}. A node is a field particle which must be \emph{necessarily} traversed when going from one root to the other one.
  \end{svgraybox}
The first few terms in the expansion of  $\mathcal{C}(r_{12})$ are
\beq
\mathcal{C}(r_{12})
=n\rtwoRoneA+\frac{n^2}{2}\left(2\rtwoRtwoAA+4\rtwoRtwoCC\right)+\cdots\;.
\label{8.7}
\eeq
\item
\begin{svgraybox}
   \textbf{Open ``parallel'' diagrams (or open ``bundles''), $\mathcal{P}(r)$}: Subset of \emph{open} diagrams with no nodes, such that there are at least \emph{two } {totally} independent (``parallel'') paths to go from one root to the other one. The existence of  parallel paths means that if the roots (together with their bonds) were removed, the resulting diagram would fall into two or more pieces.
\end{svgraybox}
The function $\mathcal{P}(r)$ is of second order in density:
\beq
\mathcal{P}(r_{12})
=\frac{n^2}{2}\stwoStwoAB+\cdots\;.
\label{8.8}
\eeq
 \item
 \begin{svgraybox}
   \textbf{``Bridge'' (or ``elementary'') diagrams, $\mathcal{B}(r)$}: Subset of \emph{open} diagrams  with no {nodes}, such that there do \emph{not} exist two  {totally} independent ways to go from one root to the other one.
\end{svgraybox}
Analogously to $\mathcal{P}(r)$, the bridge function $\mathcal{B}(r)$ is of order $n^2$:
\beq
\mathcal{B}(r_{12})
=\frac{n^2}{2}\stwoStwoBC+\cdots\;.
\label{8.9}
\eeq
  \end{itemize}

 \begin{figure}[t]
 \sidecaption[t]
\includegraphics[width=.64\columnwidth]{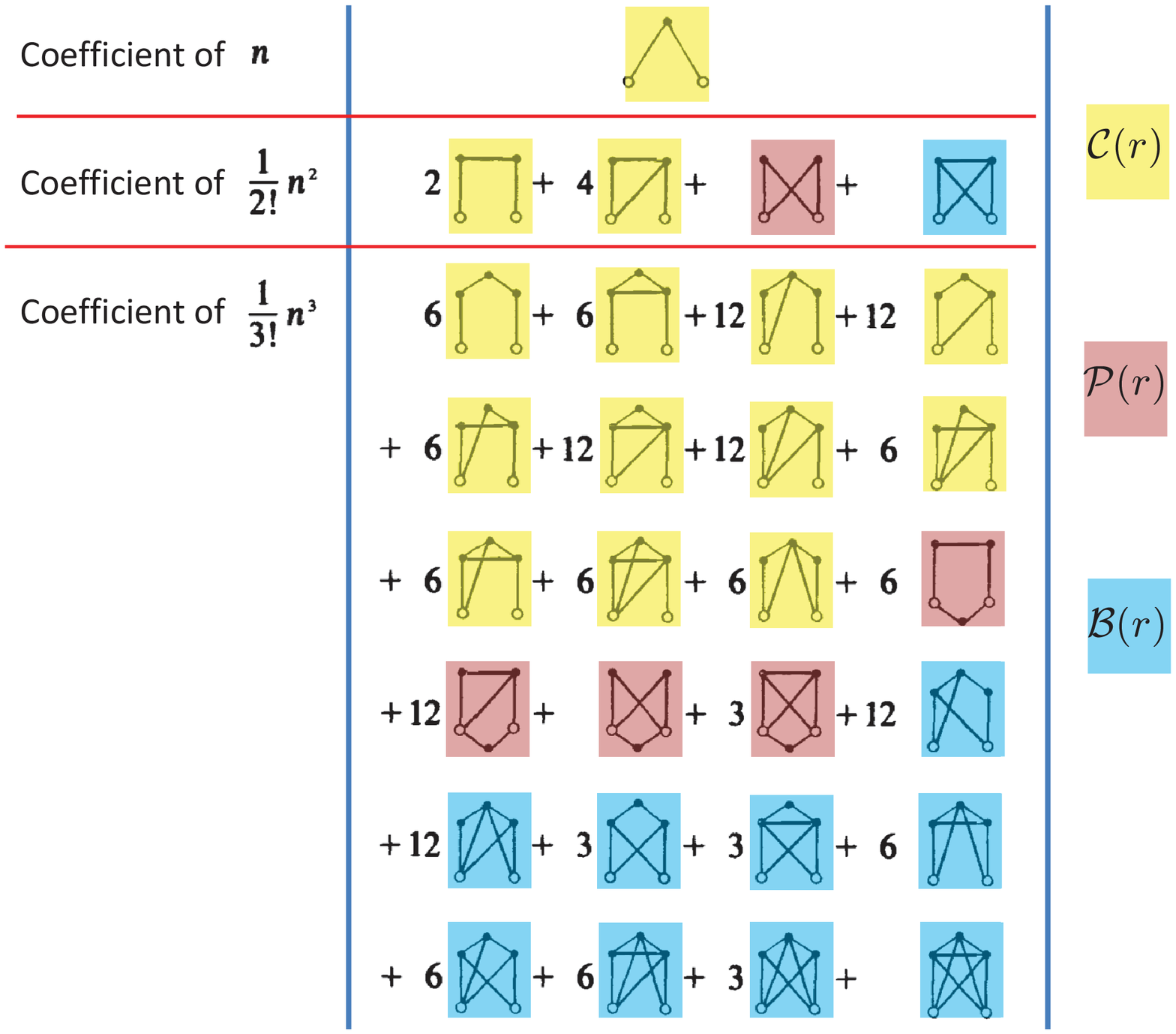}
\caption{First few chain, open parallel, and bridge diagrams. Adapted from Table 8.3.1 of  \protect\cite{B74b}.\label{fig8.1}}
\end{figure}

Figure \ref{fig8.1} shows the classification to order $n^3$. Since the three classes exhaust all the open stars, we can write
\beq
\boxed{y(r)=1+\mathcal{C}(r)+\mathcal{P}(r)+\mathcal{B}(r)}\;.
\label{8.10}
\eeq

As for the total correlation function, the diagrams contributing to it are
\beqa
h(r_{12})
&=&(1+\alert{\stwo})y(r_{12})-1\nn
&=&\alert{\stwo}+n\left(\rtwoRoneA+\alert{\stwoSone}\right)
+\frac{n^2}{2}\left(2\rtwoRtwoAA+4\rtwoRtwoCC+\stwoStwoAB\right.\nn
&&\left.+\stwoStwoBC+2\alert{\stwoStwoAA}+4\alert{\stwoStwoBA}+\alert{\stwoStwoBB}+\alert{\stwoStwoC}\right)+\cdots\nn
&=&\sum_{k=0}^\infty \frac{n^k}{k!}\sum \text{\emph{open} and \alert{\emph{closed}} stars with $2$ roots and $k$ field points.}
\label{8.6}
\eeqa
It is not worth classifying the closed diagrams. Instead, they join the open bundles to create an augmented class:
\begin{itemize}
  \item
  \begin{svgraybox}
\textbf{``Parallel'' diagrams (or ``bundles''), $\mathcal{P^+}(r)$}: All \alert{\emph{closed}} diagrams plus the \emph{open} bundles.
\end{svgraybox}
The first few ones are
\beqa
\mathcal{P^+}(r_{12})
&=&\alert{\stwo}+n\alert{\stwoSone}+\frac{n^2}{2}\left(\stwoStwoAB+2\alert{\stwoStwoAA}+4\alert{\stwoStwoBA}\right.\nn
&&\left.+\alert{\stwoStwoBB}+\alert{\stwoStwoC}\right)+\cdots\;.
\label{8.11}
\eeqa
\end{itemize}
    Obviously,
    \beq
\boxed{h(r)=\mathcal{C}(r)+\mathcal{P}^+(r)+\mathcal{B}(r)}\;.
\label{8.12}
\eeq

Why this classification? There are two main reasons.   First, open parallel diagrams ($\mathcal{P}$) factorize into products of chains ($\mathcal{C}$) and bridge diagrams ($\mathcal{B}$). For instance,
\beq
\stwoStwoAB=\left(\rtwoRoneA\right)^2\;.
\label{8.13}
\eeq
As a consequence, it can be proved that
\beqa
\mathcal{P}&=&\frac{1}{2}(\mathcal{C}+\mathcal{B})^2+\frac{1}{3!}(\mathcal{C}+\mathcal{B})^3+\cdots\nn
&=&\E^{\mathcal{C}+\mathcal{B}}-(1+\mathcal{C}+\mathcal{B})\Rightarrow \boxed{\mathcal{C}+\mathcal{B}=\ln(1+\mathcal{C}+\mathcal{P}+\mathcal{B})}\;.
\label{8.14}
\eeqa
Making use of \eqref{8.14} in \eqref{8.10}, we obtain $\ln y=\mathcal{C}+\mathcal{B}$ or, equivalently,
\beq
\boxed{\ln g(r)=-\beta\phi(r)+\mathcal{C}(r)+\mathcal{B}(r)}\;.
\label{8.15}
\eeq
The second important reason for the classification of open stars is that, as we are about to see, the chains ($\mathcal{C}$) \emph{do not} contribute to the direct correlation function $c(r)$.

Let us first rewrite \eqref{8.6} as
  \beqa
  h(r_{12})
&=&{\stwo}+n\left(\Cblu{\rtwoRoneA}+{\stwoSone}\right)+\frac{n^2}{2}\left(2\Cblu{\rtwoRtwoAA}+4\Cblu{\rtwoRtwoCC}+\stwoStwoAB\right.\nn
&&\left.+\stwoStwoBC+2{\stwoStwoAA}+4{\stwoStwoBA}+{\stwoStwoBB}+{\stwoStwoC}\right)+\cdots \;,
\label{8.16}
\eeqa
where the \Cblu{chains} are marked in blue. Next, the Ornstein--Zernike relation \eqref{8.1} or \eqref{8.2} can be iterated to yield
\beq
c=h-nh*h+n^2h*h*h-n^3h*h*h*h+\cdots\;,
\label{8.17}
\eeq
where the asterisk denotes a convolution integral. The diagrams representing those convolutions are always chains. For instance,
  \beq
h*h=\int \D\mathbf{r}_3 \,h(r_{13})h(r_{32})=\Cblu{\rtwoRoneA}+2n\left(\Cblu{\rtwoRtwoAA}+\Cblu{\rtwoRtwoCC}\right)+\cdots\;,
\label{8.18}
\eeq
\beq
h*h*h=\int \D\mathbf{r}_3\int \D\mathbf{r}_4\, h(r_{13})h(r_{34})h(r_{42})=\Cblu{\rtwoRtwoAA}+\cdots\;.
\label{8.19}
\eeq
Inserting \eqref{8.16}, \eqref{8.18}, and \eqref{8.19} into \eqref{8.17}, one obtains
\beqa
c(r_{12})
&=&\stwo+n\stwoSone+\frac{n^2}{2}\left(\stwoStwoAB+\stwoStwoBC+2\stwoStwoAA+4\stwoStwoBA\right.\nn
&&\left.+\stwoStwoBB+\stwoStwoC\right)+\cdots\;.
\label{8.20}
\eeqa
Thus, as anticipated, all chain diagrams cancel out! This is not surprising after all since the chains are the open diagrams that more easily can be ``stretched out'', thus allowing particles 1 and 2 to be be correlated via intermediate particles, even if the distance $r_{12}$ is much larger than the interaction range. Note, however, that the direct correlation function is not limited to closed diagrams but also includes the open diagrams with no nodes. Therefore,
\beq
 \boxed{c(r)=\mathcal{P}^+(r)+\mathcal{B}(r)}\;.
 \label{8.21}
 \eeq

{}From \eqref{8.10}, \eqref{8.12}, \eqref{8.15}, and \eqref{8.21} we can extract the chain function in three alternative ways:
  \begin{equation}
\mathcal{C}(r)=\E^{\beta\phi(r)}g(r)-1-\mathcal{P}(r)-\mathcal{B}(r)\;,
\label{8.22}
\end{equation}
\begin{equation}
\mathcal{C}(r)={\ln g(r)+\beta\phi(r)-\mathcal{B}(r)}\;,
\label{8.23}
\end{equation}
\begin{equation}
\mathcal{C}(r)=h(r)-c(r)\;.
\label{8.24}
\end{equation}
Combination of  \eqref{8.22} and \eqref{8.24} yields
\begin{equation}
\boxed{c(r)=g(r)\left[1-\E^{\beta\phi(r)}\right]+\mathcal{P}(r)+\mathcal{B}(r)}\;.
\label{8.25}
\end{equation}
Similarly, combining \eqref{8.23} and \eqref{8.24} one gets
\begin{equation}
\boxed{c(r)=g(r)-1-\ln g(r)-\beta \phi(r)+\mathcal{B}(r)}\;.
\label{8.26}
\end{equation}

\subsection{Approximate Closures}
  Equations \eqref{8.25} and \eqref{8.26} are formally exact, but they are not closed since they have the structure
      $c(r)=\mathcal{F}[h(r),\mathcal{P}(r)+\mathcal{B}(r)]$ and
           $c(r)=\mathcal{F}[h(r),\mathcal{B}(r)]$,
    respectively.

In most of the cases, a closure $c(r)=\mathcal{F}[h(r)]$ [see \eqref{8.4}] is an \emph{ad hoc} approximation whose usefulness must be judged \emph{a posteriori}.
  The two prototype closures are      the hypernetted-chain (HNC) closure \cite{M58,vLGB59} and
            the Percus--Yevick (PY) closure \cite{PY58}.

\subsubsection{HNC and Percus--Yevick Integral Equations}
The HNC closure consists of setting $\mathcal{B}(r)=0$ in  \eqref{8.26}:
\beq
 \boxed{c(r)=g(r)-1-\ln g(r)-\beta \phi(r)}\quad \text{(\hnc)}\;.
 \label{8.27}
 \eeq
Similarly, the Percus--Yevick closure is obtained by setting $\mathcal{P}(r)+\mathcal{B}(r)=0$ in  \eqref{8.25}, what results in
\beq
 \boxed{c(r)=g(r)\left[1-\E^{\beta\phi(r)}\right]}\quad \text{(\py)}\;.
 \label{8.28}
 \eeq
By inserting the above closures into the Ornstein--Zernike relation \eqref{8.1} we obtain the HNC and Percus--Yevick integral equations, respectively:
        \beq
    \hnc\Rightarrow\ln \left[g(r)\E^{\beta\phi(r)}\right]=-n\int\D \rr'\,\left\{\ln \left[g(r')\E^{\beta\phi(r')}\right]-h(r')\right\}h(|\rr-\rr'|)\;,
    \label{8.29}
    \eeq
            \beq
   \py\Rightarrow g(r)\E^{\beta\phi(r)}-1=-n\int\D \rr'\,\left[g(r')\E^{\beta\phi(r')}-1-h(r')\right]h(|\rr-\rr'|)\;.
    \label{8.30}
    \eeq
Interestingly, if one formally assumes that $y(r)\equiv   g(r)\E^{\beta\phi(r)}\approx 1$ and applies the linearization property     $\ln \left[g(r)\E^{\beta\phi(r)}\right]\to g(r)\E^{\beta\phi(r)}-1$, then the HNC integral equation \eqref{8.29} becomes the Percus--Yevick integral equation \eqref{8.30}. On the other hand, the Percus--Yevick theory stands by itself, even if  $y(r)$ is not close to 1.

 \begin{figure}[t]
 \sidecaption[t]
\includegraphics[width=.64\columnwidth]{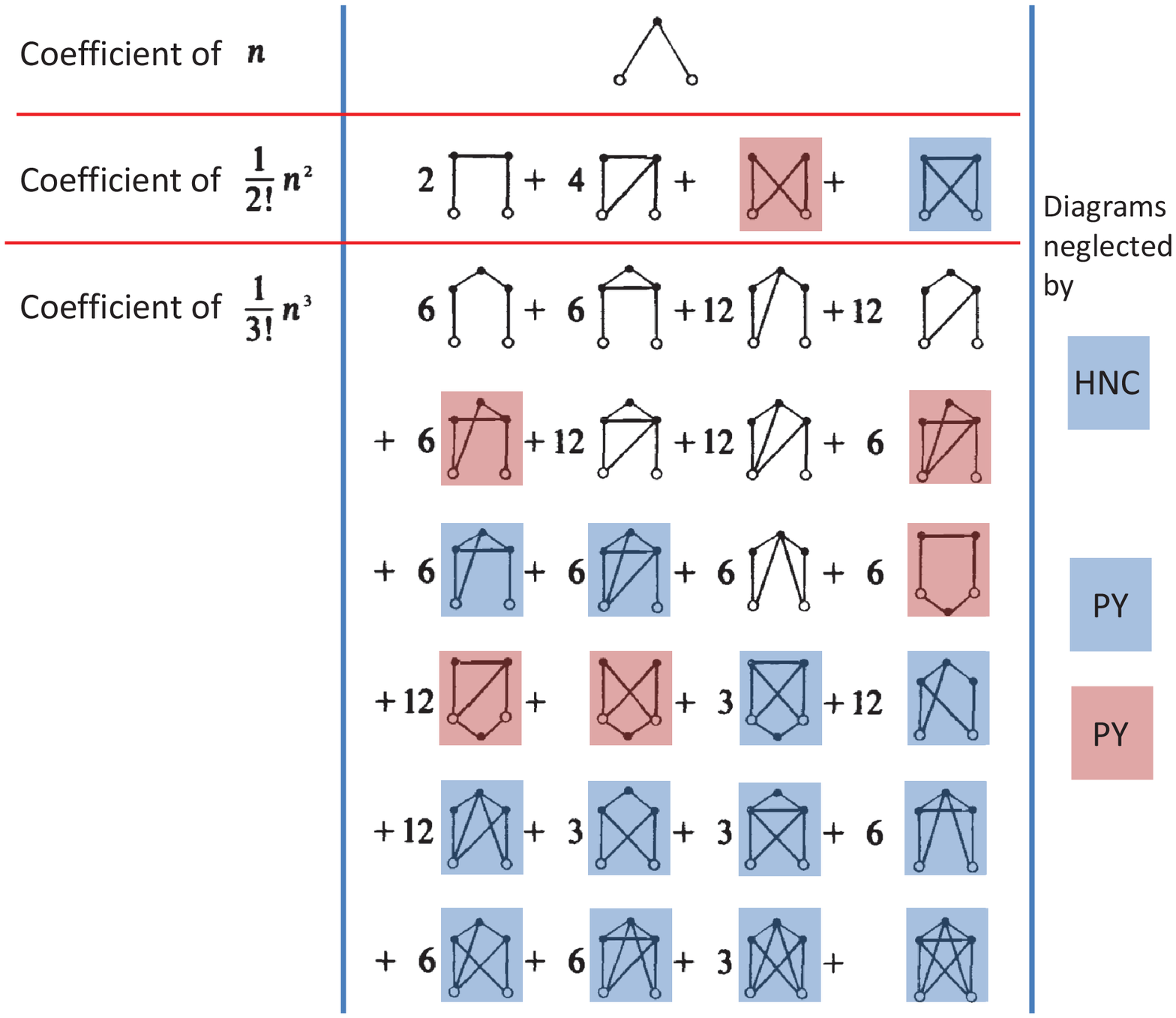}
\caption{The colored diagrams are those neglected by the HNC and the Percus--Yevick approximations. Adapted from Table 8.3.1 of  \protect\cite{B74b}.\label{fig8.2b}}
\end{figure}

A few comments are in order. First, the density expansion of $h_\hnc(r)$ and $y_\hnc(r)$ can be obtained from the closed integral equation by iteration. It turns out that  not only the  bridge diagrams disappear, but also \emph{some} chain and open parallel diagrams are not retained either.   This is because the neglect of $\mathcal{B}(r)$ at the level of \eqref{8.26} propagates to other non-bridge diagrams at the level of \eqref{8.10}. For instance, while \eqref{8.14} is an identity, we cannot neglect $\mathcal{B}(r)$ on both sides, i.e.,
$ \mathcal{C}\neq \ln(1+\mathcal{C}+\mathcal{P})$. A similar comment applies to $h_\py(r)$ and $y_\py(r)$, in which case some chain diagrams disappear along with all the bridge and open parallel diagrams. This is illustrated by comparison between Figs.\ \ref{fig8.1} and \ref{fig8.2b}.

\begin{figure}[t]
\sidecaption[t]
\includegraphics[width=.64\columnwidth]{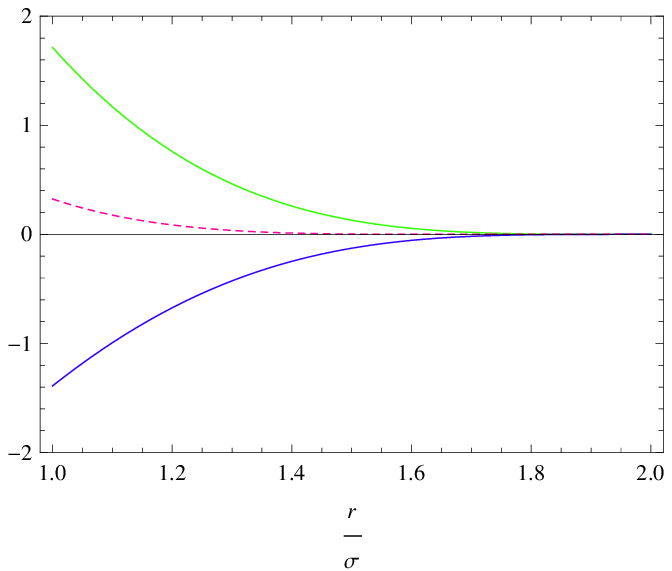}
\caption[Text]{Plot of
$\stwoStwoABtiny$ (top curve), $\stwoStwoBCtiny$ (bottom curve),
and $\stwoStwoABtiny+\stwoStwoBCtiny$
(middle curve) for  three-dimensional hard spheres.
\label{fig8.3}}
\end{figure}

Another interesting feature is that all the diagrams neglected in the density expansion of $y_\hnc(r)$ are neglected in the density expansion of $y_\py(r)$ as well. However, the latter neglects extra diagrams which are retained by $y_\hnc(r)$.
Thus, one could think that the HNC equation is \emph{always} a better approximation than the Percus--Yevick equation. On the other hand, this is not necessarily the case, especially for hard-sphere-like systems. In those cases the diagrams neglected in the Percus--Yevick equation may cancel each other to a reasonable degree, so that adding more diagrams (as HNC does) may actually worsen the result.
    For instance, the combination of the two diagrams neglected by the Percus--Yevick approximation to first order in density is
    \beq
    \stwoStwoAB+\stwoStwoBC=\stwoStwoBCd\;,
    \label{8.31}
    \eeq
where the dotted line on the right-hand side means an $e$-bond between the field particles 3 and 4, i.e., a factor $1+f(r_{34})=\E^{-\beta\phi(r_{34})}$. In the hard-sphere case the three diagrams in \eqref{8.31} vanish if $r_{12}>2\sigma$ since in that case it is impossible that either particle 3 or particle 4 can be separated from both 1 and 2 a distance smaller than $\sigma$. If $r_{12}<2\sigma$, the only configurations which contribute to the diagram on the right-hand side of \eqref{8.31} are those where $r_{13}<\sigma$, $r_{23}<\sigma$, $r_{14}<\sigma$, and $r_{24}<\sigma$ \emph{but} $r_{34}>\sigma$. It is obvious that those configurations represent a smaller volume than the ones contributing to any of the two diagrams on the left-hand side of \eqref{8.31}, especially if $r_{12}>\sigma$. In fact, as can be seen from \eqref{7.61b} and \eqref{7.62}, the right-hand side of \eqref{8.31} vanishes if $r>\sqrt{3}\sigma$ in the three-dimensional case. The three diagrams in \eqref{8.31} are plotted in Fig.\ \ref{fig8.3} in the range $1\leq r_{12}/\sigma\leq 2$.

Being approximate, the $g(r)$ obtained from either the Percus--Yevick or the HNC theory is \emph{not} thermodynamically consistent, i.e.,  virial route$\neq$chemical-potential route$\neq$compressibility route$\neq$energy route.
    However, it can be proved that the virial and energy routes are equivalent  in the HNC approximation for any interaction potential \cite{BH76,M60}.

What makes the Percus--Yevick integral equation particularly appealing is that it admits a non-trivial \emph{exact} solution for three-dimensional hard-sphere liquids \cite{W63,T63,W64}, sticky hard spheres \cite{B68}, additive hard-sphere mixtures \cite{L64}, additive sticky-hard-sphere mixtures \cite{PS75,B75}, and their generalizations to  $d=\text{odd}$ dimensions \cite{FI81,L84,RS07,RS11}.

\subsubsection{A Few Other Closures}
Apart from the classical Percus--Yevick and HNC approximations, many other ones have been proposed in the literature \cite{BH76,HM06}. Most of them as formulated as closing the formally \emph{exact} relation \eqref{8.26} with an \emph{approximation} of the form
$\mathcal{B}(r)=\mathcal{F}[\gamma(r)]$, where
\beq
\boxed{\gamma(r)\equiv h(r)-c(r)}
\label{8.32}
\eeq
is the \emph{indirect} correlation function.
In particular,
\beqa
\hnc&\Rightarrow &\mathcal{B}(r)=0,\nn
\py&\Rightarrow& \mathcal{B}(r)=\ln\left[1+\gamma(r)\right]-\gamma(r)-1\;.
\label{8.33}
\eeqa
In several cases the closure contains an \emph{adjustable} parameter fitted to guarantee the thermodynamic consistency between two routes (usually virial and compressibility).
A few examples are
  \begin{itemize}
  \item Verlet (modified) \cite{V80}
      \beq
\mathcal{B}(r)=-\frac{1}{2}\frac{\left[\gamma(r)\right]^2}{1+a_{1} \gamma(r)}\;, \quad a_1=\frac{4}{5}\;,
\eeq
\item
Martynov--Sarkisov \cite{MS83}
\beq
\mathcal{B}(r)=\sqrt{1+2\gamma(r)}-\gamma(r)-1\;,
\eeq
\item
Rogers--Young \cite{RY84}
\beq
\mathcal{B}(r)=\ln\left\{1+\frac{\exp\left[(1-\E^{-a_{2}
r})\gamma(r)\right]-1}{1-\E^{-a_{2}
r}} \right\}-\gamma(r)\;, \quad a_{2}=0.160\;,
\eeq
\item
Ballone--Pastore--Galli--Gazzillo \cite{BPGG86}
\beq
\mathcal{B}(r)=\left[1+a_{3}
\gamma(r)\right]^{1/a_{3} }-\gamma(r)-1\;,\quad a_{3}=\frac{15}{8}\;.
\eeq
  \end{itemize}

\subsubsection{Linearized Debye--H\"uckel   and Mean Spherical Approximations}
We end this section with two more simple approximate theories. First, the linearized Debye--H\"uckel (LDH) theory consists of retaining only the \emph{linear} chain diagrams in the expansion of $y(r)$ [see \eqref{8.5}]:
    \beq
    w(r)\equiv y(r)-1= n{\circ}\hspace{-0.14mm}\mbox{---}\hspace{-1mm}\bullet\hspace{-1mm}\mbox{---}\hspace{-0.14mm}{\circ}+n^2
    {\circ}\hspace{-0.14mm}\mbox{---}\hspace{-1mm}\bullet\hspace{-1mm}\mbox{---}\hspace{-1mm}\bullet\hspace{-1mm}\mbox{---}\hspace{-0.14mm}{\circ}
    +n^3
    {\circ}\hspace{-0.14mm}\mbox{---}\hspace{-1mm}\bullet\hspace{-1mm}\mbox{---}\hspace{-1mm}\bullet\hspace{-1mm}
    \mbox{---}\hspace{-1mm}\bullet\hspace{-1mm}\mbox{---}\hspace{-0.14mm}{\circ}
    +\cdots\;.
    \label{8.34}
    \eeq
This apparently crude approximation is justified in the case of \emph{long-range} interactions (like Coulomb's) since the linear chains are the most divergent diagrams but their sum gives a convergent result \cite{HM06}. The approximation \eqref{8.34} is also valid for \emph{bounded} potentials in the high-temperature limit \cite{AS04}. For those potentials $|f(r)|$ can be
made arbitrarily small by increasing the temperature and thus, at any order in density, the linear chains (having the least number of bonds) are the dominant ones.

In Fourier space, \eqref{8.34} becomes
    \beq
 \text{LDH}\Rightarrow   \widetilde{w}(k)=n\left[\widetilde{f}(k)\right]^2+n^2\left[\widetilde{f}(k)\right]^3+n^3\left[\widetilde{f}(k)\right]^4+\cdots=
    \frac{n\left[\widetilde{f}(k)\right]^2}{1-n\widetilde{f}(k)}\;.
    \label{8.35}
    \eeq
The conventional Debye--H\"uckel theory is obtained from \eqref{8.35} by assuming that (i) $\ln y(r)\approx w(r)$ and (ii) $f(r)\approx -\beta\phi(r)$. In that case \eqref{7.56c} yields $\beta\widetilde{\psi}(k)\approx\beta\widetilde{\phi}(k)-\widetilde{w}(k)\approx \beta\widetilde{\phi}(k)/[1+n\beta\widetilde{\phi}(k)]$.

Another approximation closely related to the linearized Debye--H\"uckel theory \eqref{8.35} is the mean spherical approximation (MSA). First, we start from the identity $h(r)=f(r)y(r)+y(r)-1$. Next, in the same spirit as the assumption (i) above, we assume $f(r)y(r)\approx y(r)$, so that $\widetilde{h}(k)\approx \widetilde{f}(k)+\widetilde{w}(k)$. Insertion of \eqref{8.35} yields $\widetilde{h}(k)\approx \widetilde{f}(k)/[1-n\widetilde{f}(k)]$. According to the Ornstein--Zernike relation \eqref{8.2}, the above approximation is equivalent to $\widetilde{c}(k)=\widetilde{f}(k)$. Going back to real space, $c(r)=f(r)$. Finally, repeating the assumption (ii) above, we get
\beq
\text{MSA}\Rightarrow c(r)=-\beta\phi(r)\Rightarrow \widetilde{h}(k)=\frac{-\beta \widetilde{\phi}(k)}{1+n\beta\widetilde{\phi}(k)}\;.
\label{8.36}
\eeq
It must be noted that in the mean spherical approximation the direct correlation function is independent of density but differs from its correct zero-density limit $c(r)\to f(r)$ [see \eqref{8.20}].

The mean spherical approximation \eqref{8.36} has usually been applied to bounded and soft potentials \cite{MKN06}.
For potentials with a hard core at $r=\sigma$ plus an attractive tail for $r\geq \sigma$, the mean spherical approximation \eqref{8.36} is replaced by the double condition
\beq
\begin{cases}
  g(r)=0\;,&r<\sigma\;,\\
  c(r)=-\beta\phi(r)\;,&r>\sigma\;.
\end{cases}
\label{8.37}
\eeq
This version of the mean spherical approximation is exactly solvable for Yukawa fluids \cite{W73,T03}.

\section{Some Thermodynamic Consistency Relations in Approximate Theories}
\label{sec9}
As sketched in Fig.\ \ref{5.3}, an approximate $g(r)$ does not guarantee thermodynamic consistency among the different routes. However, there are a few cases where either a partial consistency or a certain relationship may exist.

\subsection{Are $B_4^{(\hnc,v)}$ and $B_4^{(\py,c)}$ Related?}

\begin{figure}[t]
\sidecaption[t]
\includegraphics[width=.64\columnwidth]{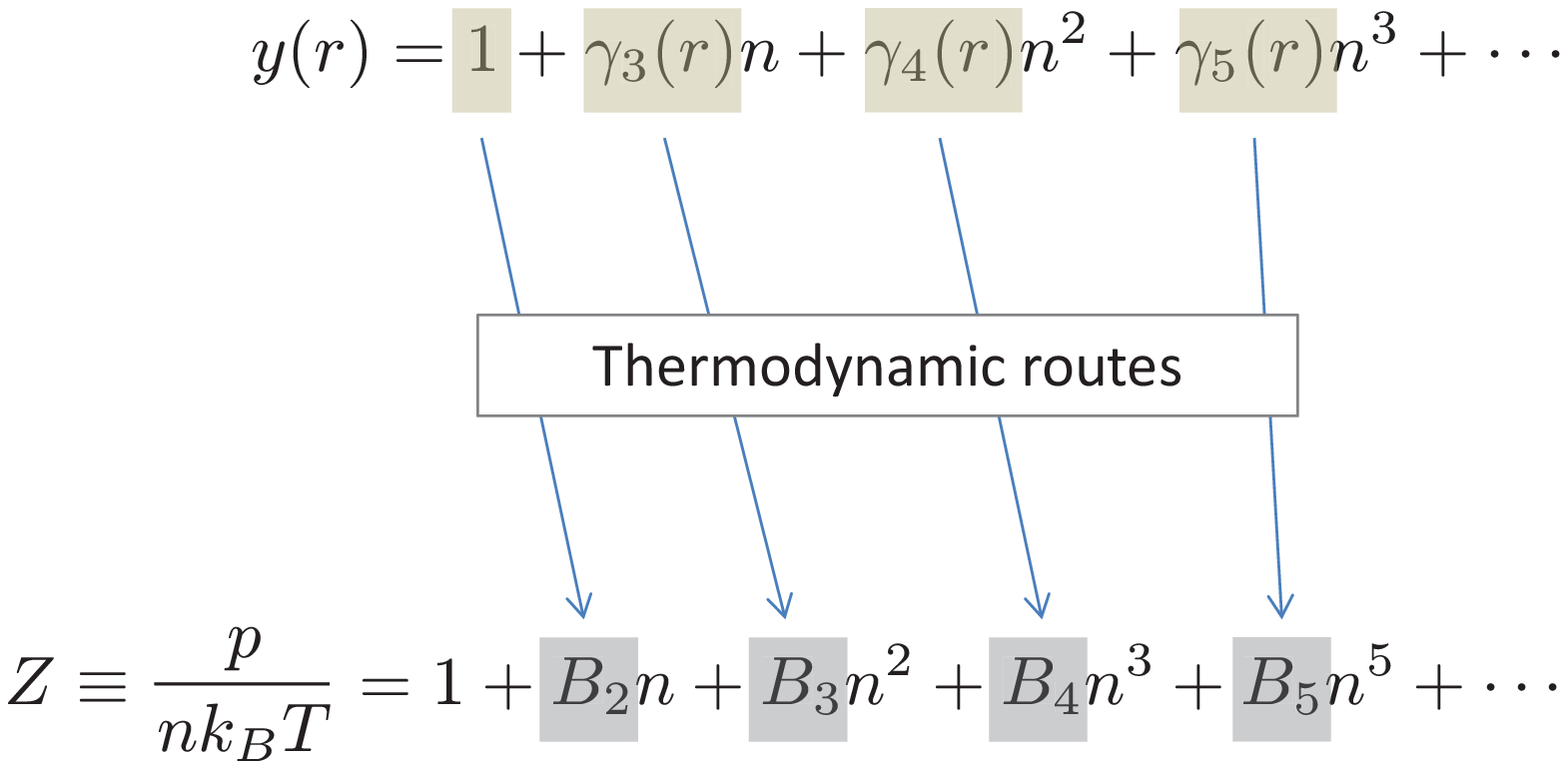}
\caption{Scheme of the relationship between the functions $\gamma_{k}(r)$ and the virial coefficients $B_k$.\label{fig9.1}}
\end{figure}}

As summarized in Fig.\ \ref{fig9.1}, the knowledge of the coefficients $\gamma_k(r)$ in the density expansion of the cavity function allows one to obtain the virial coefficients $B_k$. In general, unless the functions $\gamma_k(r)$ are exact, the virial coefficients $B_k$ will depend on the thermodynamic route followed. Here, we will focus on the  compressibility route [see \eqref{5.1}] and the virial route [see \eqref{5.13}],  denoting the corresponding virial coefficients by $B_k^{(c)}$ and $B_k^{(v)}$, respectively.

As shown before [see \eqref{7.65}], the virial route yields
\beq
    B_k^{(v)}=\frac{1}{2d}\int \D\mathbf{r}\, \gamma_k(r) r \frac{\partial f(r)}{\partial r}\;.
    \label{9.1}
        \eeq
As for the     compressibility route, from \eqref{5.1} one has
    \beqa
    \chi&=&1+n\int \D\rr\,\left\{\left[f(r)+1\right]y(r)-1\right\}\nn
    &=&1+\chi_2 n+\chi_3 n^2+\chi_4 n^3+\cdots\;,
    \label{9.2}
    \eeqa
    where
    \beq
    \chi_2=\int \D\rr\,f(r)\;,\quad \chi_k=\int \D\rr\,\left[f(r)+1\right]\gamma_k(r)\;,\quad k\geq 3\;.
    \label{9.3}
    \eeq
Then, taking into account that
    $\chi\equiv \left[{\partial\left(nZ\right)}/{\partial n}\right]^{-1}$,
    we obtain
    \beq
    B_2^{(c)}=-\frac{1}{2}\chi_2\;,\quad
    B_3^{(c)}=-\frac{1}{3}\left(\chi_3-\chi_2^2\right)\;,\quad B_4^{(c)}=-\frac{1}{4}\left(\chi_4-2\chi_2\chi_3+\chi_2^3\right)\;.
    \label{9.4}
    \eeq

\subsubsection{HNC and Percus--Yevick Theories}
Let us now particularize to the HNC and Percus--Yevick theories.
    Since $\gamma_3^{(\py)}(r)=\gamma_3^{(\hnc)}(r)=\gamma_3^{(\text{exact})}(r)$ (see Fig.\ \ref{fig8.2b}), it follows that
    \beq
    B_3^{(\py,v)}=B_3^{(\py,c)}=B_3^{(\hnc,v)}=B_3^{(\hnc,c)}=B_3^{(\text{exact})}\;.
    \label{9.5}
    \eeq
On the other hand, $\gamma_4^{(\py)}(r)\neq\gamma_4^{(\hnc)}(r)\neq\gamma_4^{(\text{exact})}(r)$ (see again Fig.\ \ref{fig8.2b}). Therefore, it can be expected that
    \beq
    B_4^{(\py,v)}\neq B_4^{(\py,c)}\neq B_4^{(\hnc,v)}\neq B_4^{(\hnc,c)}\neq B_4^{(\text{exact})}\;.
    \label{9.6}
    \eeq

However, interestingly enough,   $B_4^{(\py,c)}$ and $B_4^{(\hnc,v)}$ turn out to be closely related. More specifically, our aim is to prove that \cite{SM10}
    \beq
    \boxed{B_4^{(\hnc,v)}=\frac{3}{2}B_4^{(\py,c)}}
    \label{9.7}
    \eeq
    for any potential $\phi(r)$ and dimensionality $d$.

\subsubsection{A ``Flexible'' Function $\gamma_4(r)$}
The exact function $\gamma_4(r)$ is given by \eqref{7.55}. As shown by Fig.\ \ref{fig8.2b}, the HNC approximation neglects the last diagram and the Percus--Yevick approximation neglects the two last diagrams. In order to account for all of these possibilities, let us construct the function
\beq
 \gamma_{4}=\frac{1}{2}\left(2\rtwoRtwoAA+4\rtwoRtwoCC+\lambda_1\stwoStwoAB+\lambda_2\stwoStwoBC\right)\;.
  \label{9.8}
  \eeq
The cases $(\lambda_1,\lambda_2)=(1,1)$, $(1,0)$, and $(0,0)$ correspond to the exact, HNC, and Percus--Yevick functions, respectively.

  Inserting \eqref{9.8} into \eqref{9.1}, one has
  \beq
B_{4}^{(v)}
=\frac{1}{4d}\Bigg(2\ytwoAd+4\ytwoBd+\lambda_1\ytwoCd+\lambda_2\ytwoDd\Bigg)\;,
\eeq
where a dashed line denotes a factor $r\partial f(r)/\partial r$. By integrating by parts, the following properties can be proved \cite{SM10}:
\beq
\ytwoAd=-\frac{3d}{4}\soneSthreeA\;,
\eeq
\beq
\ytwoBd+\frac{1}{4}\ytwoCd=-\frac{3d}{4}\soneSthreeBA\;,
\eeq
\beq
\ytwoDd=-\frac{d}{2}\soneSthreeC\;.
\eeq
Consequently,
\beq
{B_{4}^{(v)}=-\frac{3}{8}\soneSthreeA-\frac{3}{4}\soneSthreeBA-\frac{\lambda_2}{8}\soneSthreeC+\frac{\lambda_1-1}{4d}\ytwoCd}\;.
\label{9.9}
\eeq

In the case of the compressibility route, \eqref{9.3} yields
  \beq
\chi_2=\soneSone\;, \quad
\chi_3=\roneRtwoB+\soneStwo\;,
\label{9.10}
\eeq
\beq
\chi_{4}=\roneRthreeAA+\frac{2+\lambda_1}{2}\soneSthreeA+2\roneRthreeCC+\frac{4+\lambda_1+\lambda_2}{2}\soneSthreeBA+\frac{\lambda_2}{2}\soneSthreeC\;,
\label{9.11}
\eeq
where in \eqref{9.11} use has been made of the property
\beq
\soneSthreeBB=\soneSthreeBA= V^{-1}\SfourB\;.
\eeq
Noting that
\beq
\chi_2\chi_3=\roneRthreeAA+\roneRthreeCC\;,\quad \chi_2^3=\roneRthreeAA\;,
\eeq
and using \eqref{9.4}, we finally obtain
    \beq
B_{4}^{(c)}
={-\frac{2+\lambda_1}{8}\soneSthreeA-\frac{4+\lambda_1+\lambda_2}{8}\soneSthreeBA
-\frac{\lambda_2}{8}\soneSthreeC}\;.
\label{9.12}
\eeq

Comparison between \eqref{9.9} and \eqref{9.12} shows that
\beq
 \boxed{B_4^{(v)}\;\left(\text{with $\lambda_1=1$ and $\lambda_2=\frac{3\lambda}{2+\lambda}$}\right)
= \frac{3}{2+\lambda}
B_4^{(c)}\;\left(\text{with $\lambda_1=\lambda_2=\lambda$}\right)}\;.
\label{9.13}
\eeq
In the case of the exact $\gamma_4(r)$ we have $\lambda=1$ in both sides of \eqref{9.13} and therefore $B_4^{(\text{exact},v)}=B_4^{(\text{exact},c)}$, as expected. On the other hand, the choice $\lambda=0$ makes the left- and right-hand sides correspond to the HNC and Percus--Yevick approximations, respectively, and then \eqref{9.13} reduces to the sought result  \eqref{9.7}.

\begin{figure}[t]
\sidecaption[t]
\includegraphics[width=.64\columnwidth]{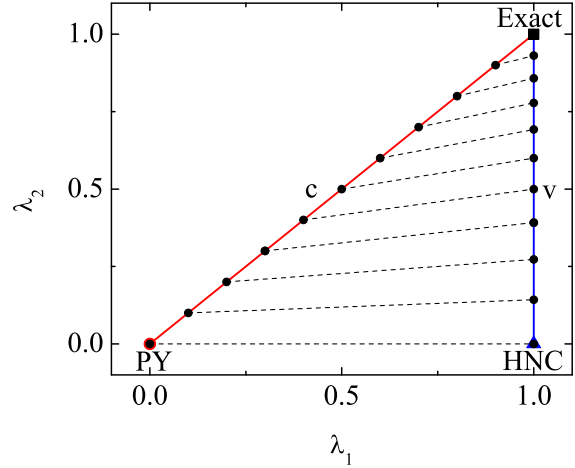}
\caption{
The diagonal (labeled c) and vertical (labeled v) lines represent the classes of approximations $\lambda_1=\lambda_2$ and $\lambda_1=1$, respectively. The dashed tie lines connect the pairs of approximations whose respective values of $B_4^{(c)}$ and $B_4^{(v)}$ are related by \protect\eqref{9.13}.
\label{fig9.2}}
\end{figure}

More in general, \eqref{9.13} implies that for
any  approximation of the class $\lambda_1=\lambda_2$ there exists a specific
 approximation of the class $\lambda_1=1$, such that the compressibility
and virial values, respectively, of $B_4$ are proportional to each
other. The connection between both classes is schematically
illustrated in Fig.\ \ref{fig9.2}. Interestingly, the largest deviation of the proportionality factor from 1 occurs  in the case of the Percus--Yevick and HNC pair.
The proof of \eqref{9.13} can be easily extended to mixtures \cite{SM10}.

\subsection{Energy and Virial Routes in the Linearized Debye–-H\"uckel Theory and in the Mean Spherical Approximation}
As said in Sect.\ \ref{sec8}, the energy and virial routes are equivalent in the HNC approximation. Now we will see that the same property holds in the linearized Debye--H\"uckel approximation \eqref{8.35} \cite{SFG09} and in the mean spherical approximation \eqref{8.36} \cite{S07a}.

\subsubsection{Linearized Debye--H\"uckel Theory}

We start by recalling the energy and pressure routes \eqref{5.6} and \eqref{5.13}, respectively. In terms of $w(r)=y(r)-1$, they are given by
  \beq
u^\ex\equiv \frac{\llangle E\rrangle^\ex}{N}=-\frac{n}{2}\int \D\mathbf{r}\,  \left[1+w(r)\right]\frac{\partial f(r)}{\partial \beta}\;,
\label{9.14}
\eeq
\beq
Z\equiv \frac{\beta p}{n}=1+\frac{n}{2d}\int \D\mathbf{r}\,\left[1+w(r)\right] \rr\cdot \nabla f(r)\;.
\label{9.15}
\eeq

The consistency condition between both routes is provided by the Maxwell relation
\beq
n\frac{\partial u^\ex}{\partial n}=\frac{\partial Z}{\partial
\beta}\;.
\label{9.16}
\eeq
Given the mathematical identity
\beq
 -\int \D\mathbf{r}\,  \frac{\partial f(r)}{\partial \beta}=\frac{1}{d}\frac{\partial}{\partial \beta}\left[\int \D\mathbf{r}\, \rr\cdot \nabla f(r)\right]\;,
 \label{9.16b}
\eeq
the consistency condition \eqref{9.16} becomes
 \beq
 -\frac{\partial}{\partial n}\left[n\int \D\mathbf{r}\,  w(r)\frac{\partial f(r)}{\partial \beta}\right]=\frac{1}{d}\frac{\partial}{\partial \beta}\left[\int \D\mathbf{r}\,w(r) \rr\cdot \nabla f(r)\right]\;.
 \label{9.17}
\eeq

Since the linearized Debye--H\"uckel approximation \eqref{8.35} is formulated in Fourier space, it is convenient to express the spatial integrals in \eqref{9.17} as wavevector integrals:
\beq
\frac{\partial}{\partial n} \left[ n \int \D \mathbf{k}\,\widetilde{w}(k) \frac{\partial \widetilde{f}\left(k\right)}
{\partial \beta} \right] =\frac{1}{d} \frac{\partial}{\partial \beta} \left\{ \int \D \mathbf{k}
\,
\widetilde{w}(k) \nabla_{\mathbf{k}} \cdot \left[ \mathbf{k} \widetilde{f}\left(k\right) \right]
\right\}\;.
\label{9.18}
\eeq
We now make use of the mathematical identity
\beqa
\frac{\partial}{\partial \beta} \left\{ \widetilde{w}\left({k}\right) \nabla_{\mathbf{k}} \cdot \left[\mathbf{k}
\widetilde{f}\left({k}\right)\right] \right\}&=&
d \frac{\partial \widetilde{w}\left({k}\right)}{\partial \beta} \widetilde{f}\left({k}\right) + \nabla_{\mathbf{k}} \cdot
\left[\mathbf{k}
\widetilde{w}\left({k}\right) \frac{\partial \widetilde{f} \left({k}\right)}{\partial \beta} \right]\nonumber\\
&&+
\mathbf{k}\cdot \left[\frac{\partial \widetilde{w}\left({k}\right)}{\partial \beta} \nabla_\mathbf{k}
\widetilde{f}\left({k}\right)
-\frac{\partial \widetilde{f}\left({k}\right)}{\partial \beta} \nabla_\mathbf{k}\widetilde{w}\left({k}\right)\right]
\label{9.19}
\eeqa
to rewrite \eqref{9.18} as
 \beqa
\frac{\partial}{\partial n} \left[ n \int {\D \mathbf{k}}\,\widetilde{w}
\left({k}\right) \frac{\partial \widetilde{f}\left({k}\right)}
{\partial \beta} \right] &=&
\frac{1}{d} \int {\D \mathbf{k}}\,
\mathbf{k}\cdot \left[\frac{\partial \widetilde{w}\left({k}\right)}{\partial \beta} \nabla_\mathbf{k} \widetilde{f}\left({k}\right)-
\frac{\partial \widetilde{f}\left({k}\right)}{\partial \beta} \nabla_\mathbf{k}\widetilde{w}\left({k}\right) \right]
\nonumber\\
&&
+\int {\D \mathbf{k}}\,\frac{\partial \widetilde{w}
\left({k}\right)}{\partial \beta}
\widetilde{f} \left({k}\right)\;.
\label{9.20}
\eeqa
It must be emphasized that no approximations have been carried out so far. Therefore,  \textit{any} $\widetilde{w}({k})$ satisfying the condition \eqref{9.20} gives thermodynamically
consistent results
via the energy and virial routes.

Let us suppose a \emph{closure} relation of the form
    \beq
    \widetilde{w}(k)=n^{-1}\mathcal{F}\left(n\widetilde{f}(k)\right)\;,\quad \mathcal{F}(z)=\text{arbitrary}\;.
    \label{9.21}
    \eeq
     This implies the relations
    \beq
\frac{\partial}{\partial n} \left[n \widetilde{w}\left({k}\right)\right]= \mathcal{F}'\left(n \widetilde{f}\left({k}\right)\right)
\widetilde{f}\left({k}\right)\;,
\eeq
\beq
\frac{\partial \widetilde{w}\left({k}\right)}{\partial \beta} =\mathcal{F}'\left(n \widetilde{f}\left({k}\right)\right)
\frac{\partial \widetilde{f}\left({k}\right)}{\partial \beta}\;,
\eeq
\beq
\nabla_\mathbf{k} \widetilde{w}\left({k}\right)=\mathcal{F}'\left(n \widetilde{f}\left({k}\right)\right)
\nabla_\mathbf{k} \widetilde{f}\left({k}\right)\;.
\eeq
It is then straightforward to check that the energy-virial consistency condition \eqref{9.20} is identically satisfied.

As a corollary, the linearized Debye--H\"uckel approximation \eqref{8.35} belongs to the scaling class \eqref{9.21} with the particular choice $\mathcal{F}(z)=z^2/(1-z)$, what closes the proof.

\subsubsection{Mean Spherical Approximation}
The proof in the case on the mean spherical approximation \eqref{8.36} follows along similar lines \cite{S07a}. Now, instead of \eqref{9.14} and \eqref{9.15}, we start from the energy and virial routes written in the forms \eqref{5.4} and \eqref{5.12}, namely
 \beq
u^\ex=\frac{n}{2}\int \D\mathbf{r}\,  \left[1+h(r)\right]\frac{\partial \left[\beta\phi(r)\right]}{\partial \beta}\;,
\label{9.21b}
\eeq
\beq
Z=1-\frac{n}{2d}\int \D\mathbf{r}\,\left[1+h(r)\right] \rr\cdot \nabla \left[\beta\phi(r)\right]\;.
\label{9.22}
\eeq
We observe that \eqref{9.14} and \eqref{9.15} become \eqref{9.21b} and \eqref{9.22}, respectively, with the formal changes $w(r)\to h(r)$ and $f(r)\to -\beta\phi(r)$. Since all the steps leading from \eqref{9.16} to \eqref{9.20} are purely technical, it is clear that we obtain a consistency condition analogous to \eqref{9.20}, except for the formal changes $\widetilde{w}(k)\to \widetilde{h}(k)$ and $\widetilde{f}(k)\to -\beta\widetilde{\phi}(k)$.
Consequently, that consistency condition is automatically satisfied by closures of the form
    \beq
    \widetilde{h}(k)=n^{-1}\mathcal{F}\left(-n\beta\widetilde{\phi}(k)\right)\;,\quad \mathcal{F}(z)=\text{arbitrary}\;.
    \label{9.23a}
    \eeq
As shown in \eqref{8.36}, the mean spherical approximation belongs to that class of closures with the particular choice $\mathcal{F}(z)=z/(1-z)$.

\subsection{Energy Route in Hard-Sphere Liquids}
We saw in \eqref{5.33} that the energy route is \emph{useless} for hard spheres. In fact,
the consistency condition \eqref{9.16} is \emph{trivially} satisfied since
    \beq
    n\frac{\partial u^\ex_\hs}{\partial n}=0\;,\quad \frac{\partial Z_\hs}{\partial
\beta}=0\;.
\eeq
The last equality expresses the fact that the hard-sphere compressibility factor
\beq
    Z_\hs(\eta)=1+2^{d-1}\eta y_\hs(\sigma;\eta)
    \label{9.23}
    \eeq
is independent of temperature. Thus, there is no possibility of extracting thermodynamic information from $u^\ex_\hs$.

However, a physical meaning can be allocated to the energy route for hard spheres if \emph{first} it is applied to a non-hard-sphere system that includes the hard-sphere system as a special case and \emph{then} the hard-sphere limit is taken.

  \subsubsection{A ``Core-Softened'' Potential. The Square-Shoulder Interaction}

  \begin{figure}[t]
\includegraphics[width=0.5\columnwidth]{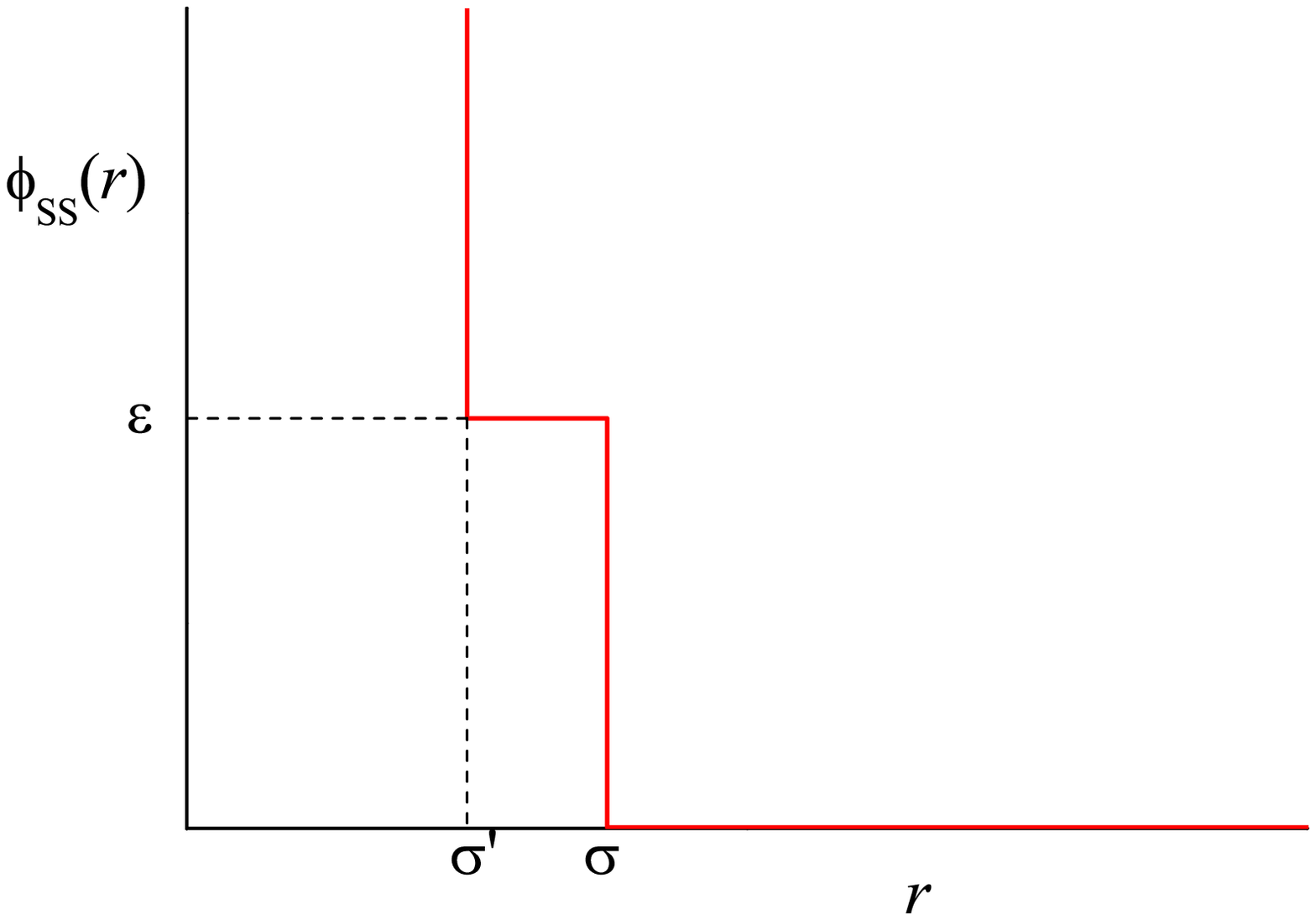}
\includegraphics[width=0.5\columnwidth]{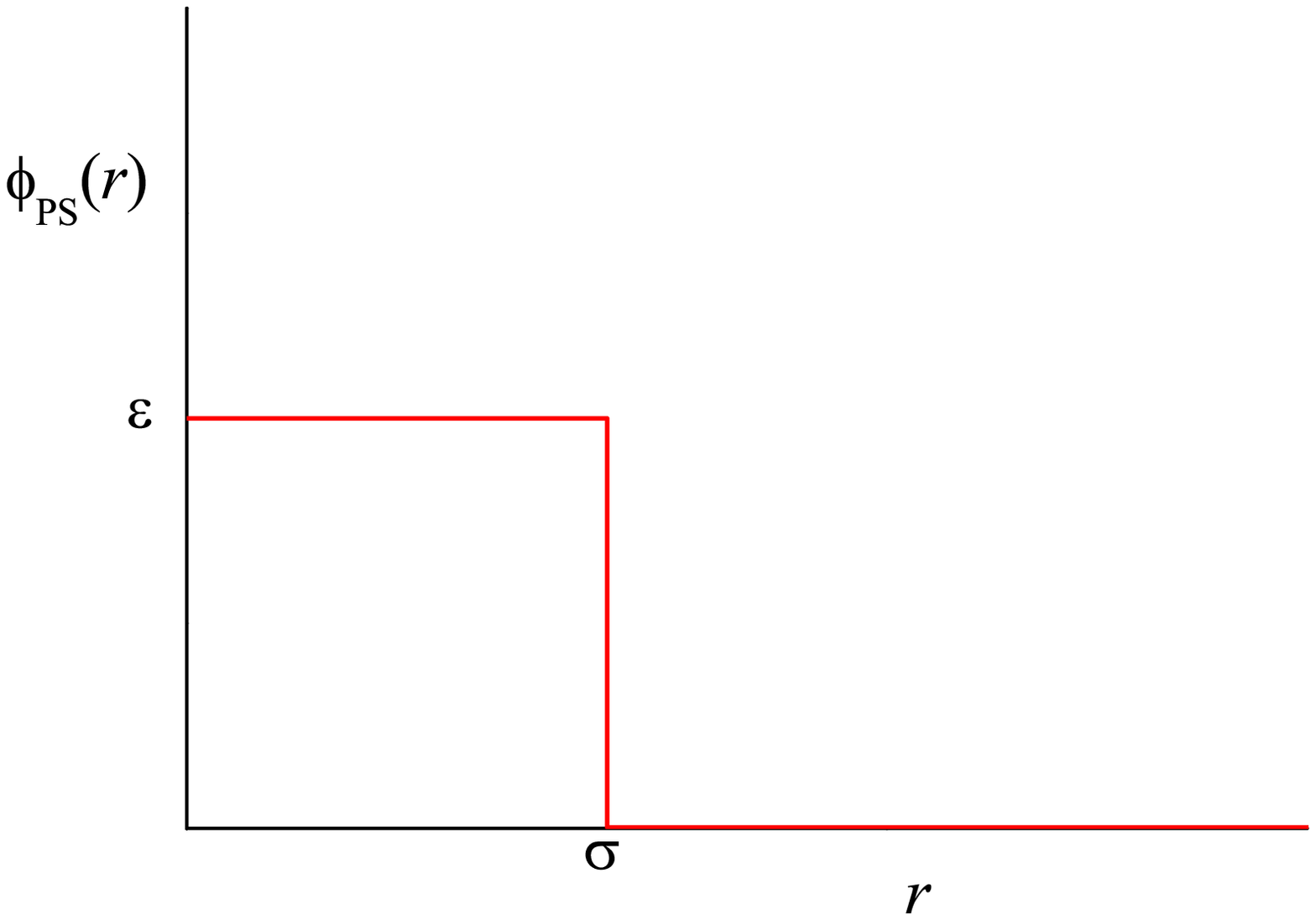}
\caption{Left panel: Square-shoulder potential. Right panel: Penetrable-sphere  potential\label{fig9.3}}
\end{figure}

Let us take the square-shoulder (SS) potential
\beq
\phi_\Ss(r)=\begin{cases}
\infty\;,&r<\sigma'\;,\\
\Cred{\varepsilon}\;,&\sigma'<r<\sigma\;,\\
0\;,&r>\sigma
\end{cases}
\label{9.24}
\eeq
as a convenient choice of a non-hard-sphere potential (see Fig.\ \ref{fig9.3}, left panel).
The square-shoulder potential is the simplest example of a core-softened potential, i.e., a potential with a two-length scale repulsive part exhibiting a softening region where the slope changes dramatically \cite{YBGS05}.

The square-shoulder potential has the interesting property of reducing to the hard-sphere potential in three independent limits:
   \beq
    \lim_{\beta\varepsilon\to 0}\phi_\Ss(r)=\phi_\hs(r)\text{ (diameter $\sigma'$)}\;,
    \label{9.25}
    \eeq
    \beq
    \lim_{\beta\varepsilon\to \infty}\phi_\Ss(r)=\phi_\hs(r)\text{ (diameter $\sigma$)}\;,
    \label{9.26}
    \eeq
   \beq
    \lim_{\sigma'\to \sigma}\phi_\Ss(r)=\phi_\hs(r)\text{ (diameter $\sigma'=\sigma$)}\;.
    \label{9.27}
    \eeq
It also reduces to the so-called penetrable-sphere (PS) potential (see Fig.\ \ref{fig9.3}, right panel) in the limit $\sigma'\to 0$:
\beq
    \lim_{\sigma'\to 0}\phi_\Ss(r)=\phi_\text{PS}(r)\;.
    \label{9.27b}
    \eeq

\subsubsection{Equation of State from the Energy Route}
Suppose an \emph{approximate} cavity function $y_\Ss(r;n,\beta)$ is known (for instance, as the solution to an integral equation) for the square-shoulder fluid. Then, the {energy route} \eqref{5.6} gives
   \beq
u_\Ss^\ex(n,\beta)=d2^{d-1}v_dn\varepsilon
\E^{-\beta\varepsilon}\int_{\sigma'}^{\sigma} \D r\, r^{d-1}
y_\Ss(r;n,\beta)\;.
\label{9.28}
\eeq
Then, the energy-route equation of state is obtained from \eqref{9.16} as
\beqa
Z_\Ss(n,\beta)&=&Z_\hs(n{\sigma'}^d)+n\frac{\partial}{\partial n}\int_0^\beta\D\beta'\, u^\ex_\Ss(n,\beta')\nn
&=&Z_\hs(n{\sigma'}^d)+
d2^{d-1}v_dn \varepsilon\frac{\partial}{\partial n}n\int_0^\beta\D\beta'\, \E^{-\beta'\varepsilon}
\int_{\sigma'}^{\sigma} \D r\, r^{d-1}y_\Ss(r;n,\beta')\;,\nn
\label{9.29}
\eeqa
where in the first step we have fixed the integration constant by the physical condition \eqref{9.25}, while in the second step we have used \eqref{9.28}.

As a second step, we now, take the limit $\beta\epsilon\to\infty$ on both sides of \eqref{9.29}, apply \eqref{9.26}, and divide both sides by $n{\sigma}^d-n{\sigma'}^d$. The result is
\beq
\frac{Z_\hs(n{\sigma}^d)-Z_\hs(n{\sigma'}^d)}{n{\sigma}^d-n{\sigma'}^d}=
\frac{d2^{d-1}v_d\varepsilon}{{\sigma}^d-{\sigma'}^d}\frac{\partial}{\partial n}n
\int_0^\infty\D\beta\, \E^{-\beta\varepsilon}
\int_{\sigma'}^{\sigma} \D r\, r^{d-1}y_\Ss(r;n,\beta)\;.
\label{9.30}
\eeq

Finally, we take the limit $\sigma'\to\sigma$. The left-hand side of \eqref{9.30} becomes
\beq
\lim_{\sigma'\to\sigma}\frac{Z_\hs(n{\sigma}^d)-Z_\hs(n{\sigma'}^d)}{n{\sigma}^d-n{\sigma'}^d}
=\sigma^{-d}\frac{\partial}{\partial n}Z_\hs(n\sigma^d)\;.
\label{9.31}
\eeq
Moreover, the spatial integral on the right-hand side of \eqref{9.30} reduces to
\beq
\lim_{\sigma'\to\sigma}\frac{1}{{\sigma}^d-{\sigma'}^d}\int_{\sigma'}^{\sigma}
\D r\,
r^{d-1}y_\Ss(r;n,\beta)=\frac{1}{d}y_\hs(\sigma;n\sigma^d)\;,
\label{9.32}
\eeq
where the third limit \eqref{9.27} has been used.
Taking into account \eqref{9.31} and \eqref{9.32} in \eqref{9.30}, one gets
\beq
\frac{\partial}{\partial n}Z_\hs(n\sigma^d)=2^{d-1}v_d
\frac{\partial}{\partial n}n\sigma^d
y_\hs(\sigma;n\sigma^d)\;.
\label{9.33}
\eeq
Integration over density and application of the ideal-gas boundary condition $Z_\hs(0)=1$ yields \eqref{9.23}, which is not but the virial equation of state! The generalization to mixtures follows essentially the same steps \cite{S05}.

\begin{figure}[t]
\sidecaption[t]
\includegraphics[width=.64\columnwidth]{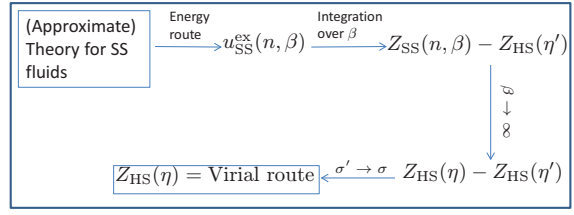}
\caption{Scheme of the steps followed to derive \protect\eqref{9.23} starting from \protect\eqref{9.28}.\label{fig9.4}}
\end{figure}

In summary, the ill definition
of the energy route to the equation of state of hard spheres can be
avoided by first considering a square-shoulder fluid and then taking the limit of a vanishing shoulder width. The resulting equation of state
coincides exactly with the one obtained through the virial
route. \emph{From that point of view}, the energy and virial routes to
the equation of state of hard-sphere fluids can be considered as equivalent. Figure \ref{fig9.4} presents a scheme of the $\text{energy route}\to\text{virial route}$ path.

It must be emphasized that the application of the three limits \eqref{9.25}--\eqref{9.27} is \emph{essential} to derive \eqref{9.23} from \eqref{9.28} \cite{S06}. For instance, if the limit $\sigma'\to 0$ (instead of $\sigma'\to\sigma$) is taken in \eqref{9.30}, the result is
\beq
Z_\hs(n{\sigma}^d)=1+
{d2^{d-1}nv_d\varepsilon}\frac{\partial}{\partial n}n
\int_0^\infty\D\beta\, \E^{-\beta\varepsilon}
\int_{0}^{\sigma} \D r\, r^{d-1}y_\text{PS}(r;n,\beta)\;,
\label{9.34}
\eeq
where the change $y_\Ss\to y_\text{PS}$ is a consequence of \eqref{9.27b}. Equation \eqref{9.34} is an alternative recipe to obtain the hard-sphere equation of state from the energy route applied to penetrable spheres. In general, it gives a result different from \eqref{9.23} when an approximate $y_\text{PS}$ is used. For instance, in the Percus--Yevick approximation for three dimensional systems, \eqref{9.23} gives a (reduced) fourth virial coefficient $b_4=16$, while \eqref{9.34} gives $b_4=1814/175\simeq 10.37$ \cite{S06}.

\section{Exact Solution of the Percus--Yevick Equation for Hard Spheres \ldots and Beyond}
\label{sec10}

As said in Sect.\ \ref{sec8}, one of the milestones of the statistical-mechanical theory of liquids in equilibrium was the exact analytical solution of the Percus--Yevick integral equation \eqref{8.30} for three-dimensional hard spheres \cite{B74b,W63,T63,W64}.

The statement of the problem is as follows. Particularized to $d=3$, the Ornstein--Zernike relation \eqref{8.1} can be written as
    \beq
  h(r)=c(r)+\frac{2\pi n}{r}\int_0^\infty \D r'\,r' c(r')\int_{|r-r'|}^{r+r'}\D r''\, r''h(r'')\;,
  \label{10.1}
  \eeq
  where bipolar coordinates have been used. In the hard-sphere case, one necessarily has $g(r)=0$ for $r<\sigma$. Moreover, the Percus--Yevick closure \eqref{8.28} implies that $c(r)=0$ for $r>\sigma$.
Thus, the mathematical problem consists in solving \eqref{10.1}   subject to the boundary conditions
  \beq
  \begin{cases}
    g(r)=0\;,& r<\sigma \text{ (exact hard-core condition)}\;,\\
    c(r)=0\;, &r>\sigma \text{ (Percus--Yevick approximation for hard spheres)}\;.
      \end{cases}
      \label{10.2}
  \eeq
The solution relies on the use of Laplace transforms, as suggested by the structure of \eqref{10.1}, and stringent analytical properties of \emph{entire functions} of complex variable.

Here, however, we will follow an alternative method \cite{S13,YS91,YHS96,HYS08} that does not make explicit use of \eqref{10.2} and lends itself to extensions and generalizations.

\subsection{An Alternative Approach. The Rational-Function Approximation}
The main steps we will follow are the following ones:
\begin{svgraybox}

    \begin{enumerate}[(I)]
      \item
      Introduce the Laplace transform $G(s)$ of $r g(r)$.

      \item
      Define an auxiliary function $F(s)$ directly related to $G(s)$.

      \item
      Find the exact properties of $F(s)$ for small $s$ and for large $s$.

      \item
      Propose a rational-function \emph{approximation} (RFA) for $F(s)$ satisfying the previous exact properties.
    \end{enumerate}
   \end{svgraybox}
As will be seen, the \emph{simplest} approximation (i.e., the one with the least number of parameters) yields the Percus--Yevick solution. Furthermore, the next-order approximation contains two free parameters which can be determined by prescribing a given equation of state and thermodynamic consistency between the virial and compressibility routes.

The same approach can be extended to    mixtures, to other related systems with  piece-wise constant potentials, and to higher dimensionalities with $d=\text{odd}$.

We now proceed with the four steps described above.

\begin{description}
  \item
  \paragraph{\textbf{(I) Introduction of $G(s)$}}

Let us introduce the Laplace transform of $rg(r)$:
    \beq
    G(s)=\mathcal{L}\left[rg(r)\right](s)=\int_0^\infty \D r\; \E^{-s r} r g(r)\;.
    \label{10.3}
    \eeq
The choice of $rg(r)$ instead of $g(r)$ as the function to be Laplace transformed is suggested by the structure of \eqref{10.1} and also by the link of $G(s)$ to the \emph{Fourier}  transform $\widetilde{h}(k)$ of $h(r)=g(r)-1$ and hence to the structure function $S(k)=1+n \widetilde{h}(k)$:
   \beq
    \widetilde{h}(k)=-2\pi\left[\frac{H(s)-H(-s)}{s}\right]_{s=\I k}=-2\pi\left[\frac{G(s)-G(-s)}{s}\right]_{s=\I k}\;,
    \label{10.4}
    \eeq
    where $H(s)=G(s)-s^{-2}$ is the Laplace transform of $rh(r)$. Had we defined $G(s)$ as the Laplace transform of $g(r)$, \eqref{10.4} would have involved the derivative $G'(s)$, what would be far less convenient.

    In the more general case of $d=\text{odd}\geq 3$, it can be seen that the right choice for $G(s)$ is \cite{RS07}
    \beq
     G(s)=\int_0^\infty \D r\; \E^{-s r}  \theta_{(d-3)/2}(s r)r g(r)\;,
    \label{10.5}
    \eeq
where
\beq
\label{theta}
\theta_k(x)= \sum_{j=0}^{k} \frac{(2k-j)!}{2^{k-j}(k-j)!j!} x^j
\eeq
are the so-called {\em reverse Bessel polynomials} \cite{C57}. In this more general case, \eqref{10.4} becomes
    \beq
    \widetilde{h}(k)=(-2\pi)^{(d-1)/2}\left[\frac{H(s)-H(-s)}{s^{d-2}}\right]_{s=\I k}=(-2\pi)^{(d-1)/2}\left[\frac{G(s)-G(-s)}{s^{d-2}}\right]_{s=\I k}\;,
    \label{10.7}
    \eeq
    where $H(s)=G(s)-(d-2)!! s^{-2}$ is defined as in \eqref{10.5}, except for the replacement $g(r)\to h(r)$.

\item

\paragraph{\textbf{(II) Definition of $F(s)$}}

Henceforth we return to the three-dimensional case ($d=3$) and, for simplicity, we take $\sigma=1$ as the length unit.
Taking \eqref{7.56}    and \eqref{7.59} into account, the hard-sphere radial distribution function to first order in density is
\beq
    g(r)=\Theta(r-1)\left[1+\Theta(2-r)\left(r-2\right)^2\left(\frac{r}{2}+2\right)\eta+\cdots\right]\;.
    \label{10.8}
    \eeq
To that order, the Laplace transform of $rg(r)$  is given by
      \beq
     s^{-1}{G(s)}=\left[\alert{F_0(s)}+F_1(s)\eta\right]\E^{-s}-12\eta \alert{\left[F_0(s)\right]^2} \E^{-2s}+\cdots\;,
     \label{10.9}
      \eeq
     where
      \beq
      F_0(s)=s^{-2}+s^{-3}\;,\quad F_1(s)=\frac{5}{2}s^{-2}-2s^{-3}-6s^{-4}+12s^{-5}+12s^{-6}\;.
      \label{10.10}
      \eeq

The exact form \eqref{10.9} of $G(s)$ to order $\eta$ \emph{suggests} the \emph{definition} of an auxiliary function $F(s)$ through
      \beqa
    s^{-1}G(s)&=&F(s) \E^{-s}-12\eta \left[F(s)\right]^2\E^{-2s}+(12\eta)^2 \left[F(s)\right]^3 \E^{-3s}-\cdots\nn
   &=& \frac{F(s)\E^{-s}}{1+12\eta F(s)\E^{-s}}\;.
   \label{10.11}
    \eeqa
    Equivalently,
    \beq
    \boxed{F(s)\equiv \E^s\frac{s^{-1}G(s)}{1-12\eta s^{-1}G(s)}}\;.
    \label{10.12}
    \eeq
      Of course, $F(s)$ depends on $\eta$. To first order,
      \beq
      F(s)=F_0(s)+F_1(s)\eta+\cdots\;.
      \label{10.13}
      \eeq

    In analogy with the one-dimensional case [see \eqref{6.43}], the introduction of $F(s)$ allows one to express $g(r)$ as a succession of \emph{shells} ($1<r<2$, $2<r<3$, \ldots) in a natural way. First, according to \eqref{10.11},
    \beq
    G(s)=\sum_{\ell=1}^\infty \left(-12\eta\right)^{\ell-1}s\left[F(s)\right]^\ell \E^{-\ell s}\;.
    \label{10.14}
    \eeq
    Then, Laplace inversion term by term gives
    \beq
      g(r)=\frac{1}{r}\sum_{\ell=1}^\infty \left(-12\eta\right)^{\ell-1}\Psi_{\ell}(r-\ell)\alert{\Theta(r-\ell)}\;,
      \label{10.15}
    \eeq
    where
    \beq
    \Psi_\ell(r)=\mathcal{L}^{-1}\left[s\left[F(s)\right]^\ell\right](r)\;.
    \label{10.16}
    \eeq

\item

\paragraph{\textbf{(III) Exact Properties of $F(s)$ for Small $s$ and Large $s$}}

In order to derive the exact behavior of $G(s)$ for large $s$, and in view of \eqref{10.15}, we need to start from the behavior of $g(r)$ for $r\gtrsim 1$:
    \beq
    g(r)=\Theta(r-1)\left[g(1^+)+g'(1^+)(r-1)+\frac{1}{2}g''(1^+)(r-1)^2+\cdots\right]\;.
    \eeq
    In Laplace space,
    \beq
    s \E^s G(s)=g(1^+)+\left[g(1^+)+g'(1^+)\right]s^{-1}+\mathcal{O}(s^{-2})\;.
    \eeq
   Therefore, according to \eqref{10.12},
    \beq
   \boxed{ \lim_{s\to\infty} s^2 F(s)=g(1^+)=\text{finite}}\;.
   \label{10.17}
   \eeq
Thus, we see that $F(s)$ must necessarily behave as $s^{-2}$ for large $s$.

Now we turn to the small-$s$ behavior.
Let us expand the
    Laplace transform of $r h(r)$ in powers of $s$:
     \beq
    H(s)=H^{(0)}+H^{(1)} s+\cdots\;,
    \eeq
    where
    \beq
    H^{(0)}\equiv \int_0^\infty \D r\,  r h(r)\;,\quad H^{(1)}\equiv -\int_0^\infty \D r\,  r^2 h(r)\;.
    \eeq
    In particular, $H^{(1)}$ is directly related to the isothermal compressibility [see \eqref{5.1}]:
    \beq
    \chi=1+n\widetilde{h}(0)=1-24\eta H^{(1)}\;.
    \label{10.17b}
    \eeq
  Since $\chi$ must be finite, and recalling that $H(s)=G(s)-s^{-2}$, we find
  \beq
  s^2 G(s)=\alert{1}+\alert{0}\times s+H^{(0)}s^2+H^{(1)} s^3+\mathcal{O}(s^4)\;.
  \eeq
Therefore, from \eqref{10.12} the small-$s$ behavior of $F(s)$ is found to be
  \beqa
  \frac{\E^s}{F(s)}&=&-12\eta+\frac{s}{G(s)}\nn
  &=&\alert{-12\eta}+\alert{0}\times s +\alert{0}\times s^2+\alert{1}\times s^3+\alert{0}\times s^4-H^{(0)}s^5-H^{(1)} s^6+\mathcal{O}(s^7)\;.\nn
  \label{10.17c}
  \eeqa
Thus, just the condition $\chi=\text{finite}$ univocally fixes the first \emph{five} coefficients in the power series expansion of $F(s)$. More specifically,
   \beq
   \boxed{F(s)=-\frac{1}{12\eta}\left[1+s+\frac{s^2}{2}+\frac{1+2\eta}{12\eta}s^3+\frac{1+\eta/2}{12\eta}s^4
   \right]+\mathcal{O}(s^5)}\;.
   \label{10.18}
   \eeq

\item

\paragraph{\textbf{(IV) Construction of the Approximation}}

Thus far, all the results are formally exact. To summarize, we have defined the Laplace transform $G(s)$ in \eqref{10.3} and the auxiliary function $F(s)$ in \eqref{10.12}. This latter function  must comply with the two basic requirements \eqref{10.17} and \eqref{10.18}.

    A simple way of satisfying both conditions is by means of a \emph{rational-function} form:
    \beq
    F(s)=\frac{\text{Polynomial in $s$ of degree $k$}}{\text{Polynomial in $s$ of degree $k+2$}}
    \label{10.19a}
    \eeq
    with $2k+3\geq 5\Rightarrow k\geq 1$. The
   \emph{ simplest} rational-function approximation  corresponds to $k=1$:
   \beq
   \boxed{F(s)=-\frac{1}{12\eta}\frac{1+L^\one s}{1+S^\one s+S^\two s^2+S^\three s^3}}\;,
   \label{10.19}
   \eeq
where the coefficients are determined from \eqref{10.18}. They are
    \beq
    L^\one=\frac{1+\eta/2}{1+2\eta}\;,
    \label{10.20}
     \eeq
         \beq
    S^\one=-\frac{3}{2}\frac{\eta}{1+2\eta},\quad  S^\two=-\frac{1}{2}\frac{1-\eta}{1+2\eta},\quad
     S^\three=-\frac{1}{12\eta}\frac{(1-\eta)^2}{1+2\eta}\;.
     \label{10.21}
     \eeq

\end{description}

\subsection{Structural Properties}
Once $F(s)$ and hence $G(s)$ have been completely determined by the approximation \eqref{10.19}, it is easy to go back to real space and obtain the corresponding $g(r)$.
     Three alternative ways are possible. First, one can invert numerically the Laplace transform $G(s)$ by means of efficient algorithms \cite{AW92}. A second method consists of obtaining $\widetilde{h}(k)$ from \eqref{10.4} and then performing a numerical Fourier inversion.
The third method is purely analytical and is based on \eqref{10.15} and \eqref{10.16}. {}From a practical point of view, one is interested in determining $g(r)$ up to a certain distance $r_\text{max}$ since $g(r)\to 1$ for large distances. In that case, the summation in \eqref{10.15} can be truncated for $\ell> r_\text{max}$. In obtaining $\Psi_\ell(r)$ from \eqref{10.16} and \eqref{10.19} one only needs the roots of the cubic equation $1+S^\one s+S^\two s^2+S^\three s^3$ and to apply the residue theorem. This latter method is the one employed in  \cite{note_13_10}

       As for the structure function, application of \eqref{4.15} and \eqref{10.4} yields the explicit expression
  \beqa
    \frac{1}{S(k)}&=&1+\frac{72\eta^2(2+\eta)^2}{(1-\eta)^4}k^{-4}+\frac{288\eta^2(1+2\eta)^2}{(1-\eta)^4}k^{-6}-\cos k\left[\frac{12\eta(2+\eta)}{(1-\eta)^2}k^{-2}\right.\nn
    &&\left.+\frac{72\eta^2(2-4\eta-7\eta^2)}{(1-\eta)^4}k^{-4}+\frac{288\eta^2(1+2\eta)^2}{(1-\eta)^4}k^{-6}\right]\nn
    &&+\sin k\left[\frac{24\eta(1-5\eta-5\eta^2)}{(1-\eta)^3}k^{-3}-\frac{288\eta^2(1+2\eta)^2}{(1-\eta)^4}k^{-5}\right]\;.
    \label{10.22}
  \eeqa

To complete the description of the structural properties stemming from the approximation \eqref{10.19}, let us consider the direct correlation function. Its Fourier transform can be obtained from $\widetilde{h}(k)$ via the Ornstein--Zernike relation \eqref{8.2}. The inverse Fourier transform can be performed analytically with the result
     \beq
     c(r)=
     \begin{cases}
  -\frac{(1+2\eta)^2}{(1-\eta)^4}+\frac{6\eta(1+\eta/2)^2}{(1-\eta)^4}{r}-
  \frac{\eta(1+2\eta)^2}{2(1-\eta)^4}{r}^3\;,&r<1\;,\\
  0\;,&r>1\;.
  \end{cases}
   \label{10.23}
  \eeq
We observe that $c(r)=0$ for $r>1$. But this is the \emph{signature} of the Percus--Yevick approximation for hard spheres [see \eqref{10.2}].
  This shows that the \emph{simplest} realization \eqref{10.19} of the rational-function approximation \eqref{10.19a} turns out to coincide with the exact Percus--Yevick solution.

  \begin{figure}[t]
\includegraphics[width=.5\columnwidth]{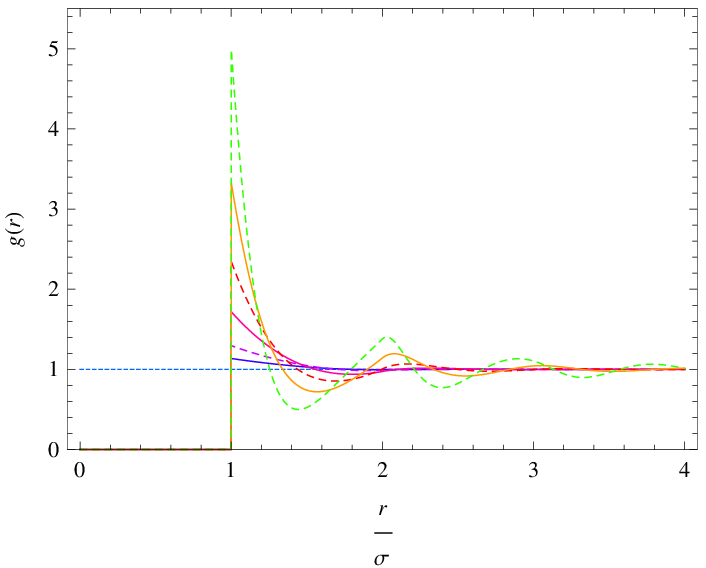}
\includegraphics[width=.5\columnwidth]{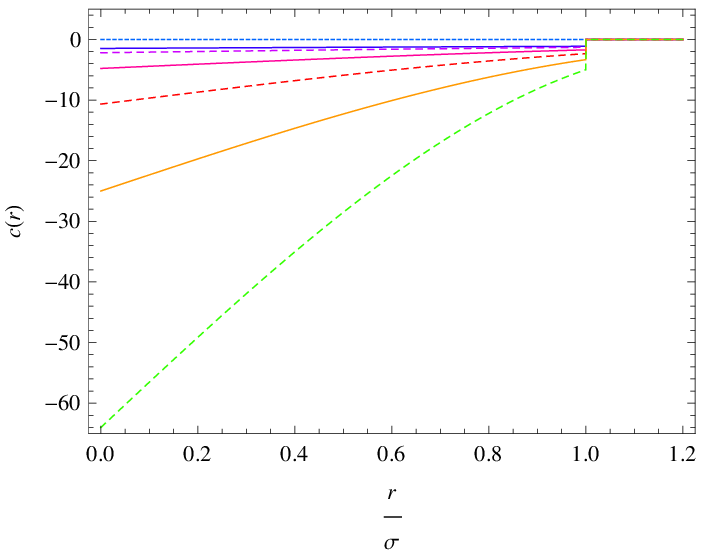}
\caption{Radial distribution function (left panel) and direct correlation function (right panel) of a three-dimensional hard-sphere fluid, as obtained from the Percus--Yevick approximation. at several values of the packing fraction $\eta\equiv (\pi/6)n\sigma^3=0.05$, $0.1$, $0.2$, $0.3$, $0.4$, and $0.5$, in increasing order of complexity.\label{fig10.1}}
\end{figure}

Figure \ref{fig10.1} displays the Percus--Yevick functions $g(r)$ and $c(r)$ at several densities. The corresponding structure factor curves were plotted in Fig.\ \ref{fig4.3}.

\subsection{Equation of State}
Once $G(s)$ is fully determined, one can obtain the equation of state. As expected, the result depends on the thermodynamic route employed.
Let us start with the virial route. According to \eqref{9.23}, the virial route in the three-dimensional case is
    \beq
    Z^{(v)}=1+4\eta g(1^+)\;.
    \label{10.24}
    \eeq
The contact value is obtained from \eqref{10.17} as
    \beq
     g(1^+)=-\frac{1}{12\eta}\frac{L^\one}{S^\three}=\frac{1+\eta/2}{(1-\eta)^2}\;.
     \label{10.25}
     \eeq
     Thus,
     \beq
     \boxed{Z^{(v)}_\py=\frac{1+2\eta+3\eta^2}{(1-\eta)^2}}\;.
     \label{10.26}
     \eeq

In the case of the compressibility route, \eqref{10.17b} shows that we need $H^\one$. This quantity is evaluated from the coefficient of $s^6$ in the Taylor expansion of $\E^s/F(s)$, as shown in \eqref{10.17c}. The result is
\beq
  H^\one=\frac{8-2\eta+4\eta^2-\eta^3}{24(1+2\eta)^2} \;.
  \eeq
Insertion into \eqref{10.17b}  yields
    \beq
  \boxed{\chi_\py=\frac{(1-\eta)^4}{(1+2\eta)^2}}\;.
  \label{10.27}
  \eeq
The associated compressibility factor is obtained upon integration as
  \beq
  \boxed{Z_\py^{(c)}=\frac{1}{\eta}\int_0^\eta \frac{\D\eta'}{\chi_\py(\eta')}=\frac{1+\eta+\eta^2}{(1-\eta)^3}}\;.
  \label{10.28}
  \eeq

Finally, we consider the chemical-potential equation of state. In the three-dimensional one-component case, \eqref{5.41} gives
  \beq
  \beta \mu^\ex=-\ln(1-\eta)+24\eta\int_{\frac{1}{2}}^1\D\sigma_{01}\sigma_{01}^2g_{01}(\sigma_{01}^+)\;.
  \label{10.29}
  \eeq
We see that the contact value \eqref{10.25} is not enough to compute $\mu^\ex$.
  We need to ``borrow'' the solute-solvent contact value $g_{01}(\sigma_{01}^+)$ from the Percus--Yevick solution for mixtures \cite{L64}:
  \beq
  g_{01}(\sigma_{01}^+)=\frac{1}{1-\eta}+\frac{3}{2}\frac{\eta}{(1-\eta)^2}\left(2-\frac{1}{\sigma_{01}}\right)\;.
  \label{10.30}
  \eeq
This expression is exact if $\sigma_{01}=\frac{1}{2}$ \cite{S12b} and reduces to \eqref{10.25} if $\sigma_{01}=1$.
Performing the integration in \eqref{10.29} one finds
\beq
\boxed{\beta\mu^\ex_\py=-\ln(1-\eta)+\eta\frac{7+\eta/2}{(1-\eta)^2}}\;.
\label{10.31}
\eeq
The excess free energy $F^\ex$ consistent with \eqref{10.31} is obtained taking into account the thermodynamic relation \eqref{2.9}, i.e., $\mu^\ex=\partial (F^\ex/V)/\partial n$, as
\beq
\frac{\beta F_\py^\ex}{N}=\frac{1}{\eta}\int_0^\eta {\D\eta'}\,\beta\mu_\py^\ex(\eta')=\frac{9-\eta}{\eta}\ln(1-\eta)+\frac{3}{2}\frac{6-\eta}{1-\eta}\;.
\label{10.31b}
\eeq
Then, the equation of state is derived from the thermodynamic relation  \eqref{2.9}, i.e., $Z=1+n\partial(\beta F^\ex/N)/\partial n$. The result is
 \beq
  \boxed{Z_\py^{(\mu)}=-9\frac{\ln(1-\eta)}{\eta}-8\frac{1-31\eta/16}{(1-\eta)^2}}\;.
  \label{10.32}
  \eeq
Surprisingly, while the virial and compressibility equations of state \eqref{10.26} and \eqref{10.28}, respectively, are known since 1963 \cite{T63}, the chemical-potential equation of state \eqref{10.32} has remained hidden until recently \cite{S12b}.

\begin{table}[t]
\caption{First eleven (reduced) virial coefficients $b_k$ as obtained exactly and from several equations of state related to the Percus--Yevick theory.}
\label{table10.1}
\begin{tabular}{p{.5cm}p{1.5cm}p{1.cm}p{1.cm}p{1.8cm}p{1.cm}p{2cm}p{2cm}}
\hline\noalign{\smallskip}
$k$ &exact&$Z_\py^{(v)}$&$Z_\py^{(c)}$&$Z_\py^{(\mu)}$&$Z_{\text{CS}}$&$Z^{(\mu c,1)}$&$Z^{(\mu c,2)}$\\
\hline\noalign{\smallskip}
${2} $ &$4$&$4$&$4$&$4$&$4$&$4$&$4$\\
$ {3} $ & $10$&$10$&$10$&$10$&$10$&$10$&$10$\\
$ {4} $ & $18.36476\cdots$&$16$&$19$&$\frac{67}{4}=16.75$&$18$&$\frac{181}{10}=18.1$&$\frac{145}{8}=18.125$\\
$ {5} $ & $28.2245(3)$&$22$&$31$&$\frac{119}{5}=23.8$&$28$&$\frac{703}{25}=28.12$&$\frac{141}{5}=28.2$\\
${6} $ & $39.8151(9)$&$28$&$46$&$31$&$40$&$40$&$\frac{241}{6}\simeq40.2$\\
$ {7} $ &$53.344(4)$&$34$&$64$&$\frac{268}{7}\simeq38.3$&$54$&$\frac{376}{7}\simeq 53.7$&$54$\\
$ {8} $ & $68.54(2)$&$40$&$85$&$\frac{365}{8}=45.6$&$70$&$\frac{277}{4}=69.25$&$\frac{1115}{16}\simeq69.7$\\
$ {9} $ & $85.81(9)$&$46$&$109$&$53$&$88$&$\frac{433}{5}=86.6$&$\frac{785}{9}\simeq87.2$\\
$ {10} $ & $105.8(4)$&$52$&$136$&$\frac{302}{5}=60.4$&$108$&$\frac{2644}{25}=105.76$&$\frac{533}{5}=106.6$\\
$ {11} $ & $128(5)$&$58$&$166$&$\frac{746}{11}\simeq 67.8$&$130$&$\frac{1394}{11}\simeq126.7$&$\frac{1406}{11}\simeq127.8$\\
\noalign{\smallskip}\hline\noalign{\smallskip}
\end{tabular}
\end{table}

The reduced virial coefficients $b_k$ [see \eqref{7.82}] predicted by the three equations of state \eqref{10.26}, \eqref{10.28}, and \eqref{10.32} are
\beq
b_k^{(\py, v)}=2(3k-4)\;,\quad b_k^{(\py, c)}=\frac{3k^2-3k+2}{2}\;,\quad b_k^{(\py,\mu)}=\frac{15k^2-31k+18}{2k}\;.
\label{10.33}
\eeq
Those virial coefficients are compared with the exact values \cite{LKM05,CM06,W13} in Table \ref{table10.1}. We observe that \eqref{10.28} overestimates the known coefficients, while \eqref{10.26} and \eqref{10.32} underestimate them, the chemical-potential route being slightly more accurate than the virial one.

    Interestingly, the Carnahan--Starling equation of state [see \eqref{7.89a} and \eqref{7.89}] can be recovered as an \emph{interpolation} between the Percus--Yevick virial and compressibility equations:
    \beq
    Z_\text{CS}=\frac{1}{3}Z_\py^{(v)}+\frac{2}{3} Z_\py^{(c)}\;.
    \label{10.33b}
    \eeq
As shown by Fig.\ \ref{fig7.10}, $Z_\text{CS}$ is an excellent equation of state.
On the other hand, since   $Z_\py^{(\mu)}$ is more reliable than $Z_\py^{(v)}$, one may wonder whether a similar interpolation formula, this time between $Z_\py^{(\mu)}$ and $Z_\py^{(c)}$, i.e.,
    \beq
    Z^{(\mu c)}=\lambda Z_\py^{(\mu)}+(1-\lambda) Z_\py^{(c)}\;,
    \label{10.34}
    \eeq
might be even more accurate. {}From an analysis of the virial coefficients one can check that the optimal value of the interpolation parameter is $\lambda\approx 0.4$. In particular, the two choices
\beq
 \lambda=\frac{2}{5}\Rightarrow Z^{(\mu c,1)}\; ,\quad \lambda=\frac{7}{18}\Rightarrow Z^{(\mu c,2)}
 \label{10.35}
\eeq
are analyzed in Table \ref{table10.1} at the level of the virial coefficients, where
\beq
b_k^{(\mu c,1)}=\frac{9k^3+21k^2-56k+36}{10k}\;,\quad b_k^{(\mu c,2)}=\frac{11k^3+24k^2-65k+42}{12k}\;.
\eeq
A better performance than that of the Carnahan--Starling coefficients is clearly observed (except in the cases of $b_6$ and $b_7$).

    \begin{figure}[t]
    \sidecaption[t]
  \includegraphics[width=.64\columnwidth]{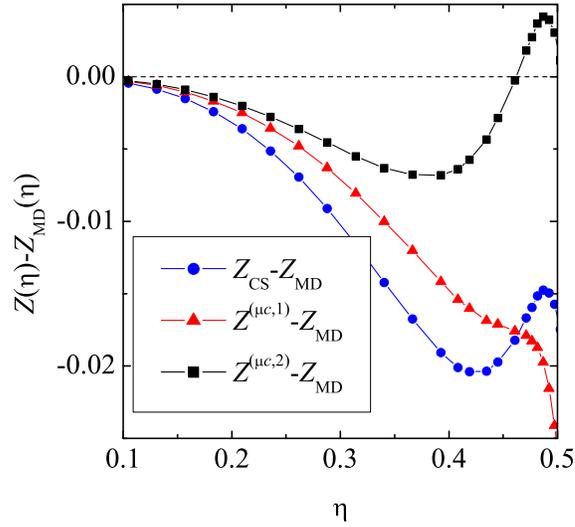}
  \caption{Plot of $Z_{\text{CS}}(\eta)-Z_{\text{MD}}(\eta)$ (circles), $Z^{(\mu c, 1)}(\eta)-Z_{\text{MD}}(\eta)$ (triangles), and  $Z^{(\mu c, 2)}(\eta)-Z_{\text{MD}}(\eta)$ (squares).\label{fig10.2}}
  \end{figure}

The good quality of $Z^{(\mu c,1)}$ and $Z^{(\mu c,2)}$, even better than that of $Z_{\text{CS}}$, is confirmed by Fig.\ \ref{fig10.2}, where the deviations of those three compressibility factors from molecular dynamics simulation values ($Z_{\text{MD}}$) \cite{KLM04} are plotted as functions of the packing fraction.

It is worth mentioning that  \eqref{10.25}, \eqref{10.26}, \eqref{10.27}, \eqref{10.28}, \eqref{10.31}, \eqref{10.31b}, and \eqref{10.32} are extended to additive hard-sphere mixtures as \cite{SR13,L64}
\beq
g_{\alpha\gamma}(\sigma_{\alpha\gamma}^+)=\frac{1}{1-\eta
}+\frac{3}{2}\frac{\eta }{(1-\eta )^{2}}\frac{\sigma_\alpha\sigma_\gamma
}{\sigma_{\alpha\gamma}}\frac{M_2}{M_3}\;,
\label{10.36}
\eeq
\beq
Z_\py^{(v)}=\frac{1}{1-\eta}+\frac{3\eta}{(1-\eta)^2}\frac{M_1M_2}{M_3}
+\frac{3\eta^2}{(1-\eta)^2}\frac{M_2^3}{M_3^2}\;,
\label{10.37}
\eeq
\beq
\chi_\py^{-1}=\frac{1}{(1-\eta)^2}+\frac{6\eta}{(1-\eta)^3}\frac{M_1M_2}{M_3}
+\frac{9\eta^2}{(1-\eta)^4}\frac{M_2^3}{M_3^2}\;,
\label{10.38a}
\eeq
\beq
Z_\py^{(c)}=\frac{1}{1-\eta}+\frac{3\eta}{(1-\eta)^2}\frac{M_1M_2}{M_3}
+\frac{3\eta^2}{(1-\eta)^3}\frac{M_2^3}{M_3^2}\;,
\label{10.38}
\eeq
\beqa
\beta\mu_{\py,\nu}^\ex&=&-\ln(1-\eta)+\frac{3\eta}{1-\eta}\frac{M_2}{M_3}\sigma_\nu+\frac{3\eta}{1-\eta}\left(
\frac{M_1M_2}{M_3}+\frac{3}{2}\frac{\eta}{1-\eta}\frac{M_2^3}{M_3^2}\right)\frac{\sigma_\nu^2}{M_2}\nn
&&+\frac{\eta}{1-\eta}\left(1
+3\frac{\eta}{1-\eta}\frac{M_1M_2}{M_3}
\right)\frac{\sigma_\nu^3}{M_3}\;,
\label{10.38.1}
\eeqa
\beqa
\frac{\beta F_\py^\ex}{N}&=&-\ln(1-\eta)+\frac{3\eta}{1-\eta}\frac{M_1M_2}{M_3}
+\frac{3\eta^2}{2(1-\eta)^2}\frac{M_2^3}{M_3^2}\nn
&&+
\frac{3M_2^3}{2M_3^2}\left[\frac{6-9\eta+2\eta^2}{(1-\eta)^2}
+6\frac{\ln(1-\eta)}{\eta}\right]\;,
\label{10.38.2}
\eeqa
\beq
Z_\py^{(\mu)}=\frac{1}{1-\eta}+\frac{3\eta}{(1-\eta)^2}\frac{M_1M_2}{M_3}
+\frac{3\eta^2}{(1-\eta)^3}\frac{M_2^3}{M_3^2}-
\frac{9M_2^3}{M_3^2}\left[\frac{1-\frac{3}{2}\eta}{(1-\eta)^2}
+\frac{\ln(1-\eta)}{\eta}\right]\;,
\label{10.39}
\eeq
where
\beq
M_q\equiv \sum_{\alpha} x_\alpha\sigma_\alpha^q\;.
\eeq

\subsection{Beyond the Percus--Yevick Solution}
Once we have obtained the exact solution of the Percus--Yevick integral equation for hard spheres as the simplest application of the rational-function approximation methodology, let us go beyond it either by improving the approximation or by considering other interaction models.
\subsubsection{Next-Order Approximation for Hard-Sphere Fluids}
   In the spirit of the rational-function approximation \eqref{10.19a}, the next-order approximation is obtained with $k=2$, i.e.,
   \beq
   F(s)=-\frac{1}{12\eta}\frac{1+L^\one s\alert{+L^\two s^2}}{1+S^\one s+S^\two s^2+S^\three s^3\alert{+S^\four s^4}}\;.
   \label{10.40}
   \eeq
From the exact series expansion \eqref{10.18}  one can obtain
      \beq
        L^\one=L^\one_\py+\frac{12\eta}{1+2\eta}\left[\frac{1}{2}\alert{L^\two}-\alert{S^\four}\right]\;,
        \label{10.41}
        \eeq
        \beq
    S^\one=S^\one_\py+\frac{12\eta}{1+2\eta}\left[\frac{1}{2}\alert{L^\two}-\alert{S^\four}\right]\;,
     \eeq
         \beq
    S^\two=S^\two_\py+\frac{12\eta}{1+2\eta}\left[\frac{1-4\eta}{12\eta}\alert{L^\two}+\alert{S^\four}\right]\;,
    \eeq
    \beq
      S^\three=S^\three_\py-\frac{12\eta}{1+2\eta}\left[\frac{1-\eta}{12\eta}\alert{L^\two}+\frac{1}{2}\alert{S^\four}\right]\;,
      \label{10.42}
     \eeq
where $L^\one_\py$, $S^\one_\py$, $S^\two_\py$, and $S^\three_\py$ are given by \eqref{10.20} and \eqref{10.21}.

So far, the two coefficients $L^\two$ and $S^\four$ remain free.
They can be fixed by imposing any desired contact value $g(1^+)$ (or compressibility factor $Z$) and the corresponding consistent isothermal susceptibility $\chi=[\partial(\eta Z)/\partial \eta]^{-1}$.
First, the exact condition \eqref{10.17} fixes the ratio $L^\two/S^\four$, so that
  \beq
    L^\two=-3(Z-1)S^\four\;.
    \label{10.43}
  \eeq
Next, the expansion \eqref{10.17c} allows us to identify $H^\one$ and, by means of \eqref{10.17b}, relate $\chi$, $L^\two$, and $S^\four$. Using \eqref{10.43}, one gets a quadratic equation for $S^\four$ \cite{YHS96}, whose physical solution is
  \beq
      S^\four=\frac{1-\eta}{36\eta(Z-\frac{1}{3})}\left[1-\sqrt{1+
\frac{Z-\frac{1}{3}}{Z-Z_{\text{PY}}^{(v)}}\left(\frac{\chi}{\chi_{\text{PY}}}
-1\right)}\right]\;,
\eeq
where $Z_{\text{PY}}^{(v)}$ and $\chi_\py$ are given by \eqref{10.26} and \eqref{10.27}, respectively.

  \begin{figure}[t]
\includegraphics[width=.8\columnwidth]{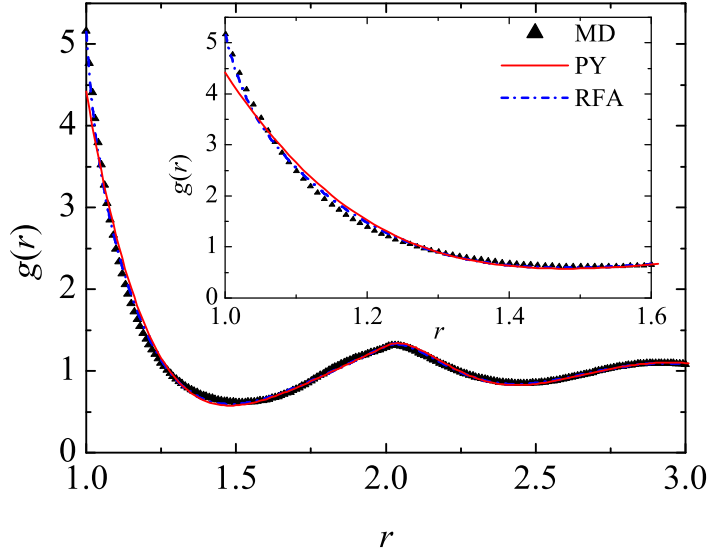}
\caption{Radial distribution function $g(r)$ of a hard-sphere fluid at a packing fraction $\eta=0.471$ as obtained by molecular dynamics simulations \protect\cite{KLM04} and from the Percus--Yevick and rational-function approximation approaches. \label{fig10.3}}
\end{figure}

Figure \ref{fig10.3} compares computer simulations results of $g(r)$ at $\eta=0.471$ \cite{KLM04} with the predictions obtained from the Percus--Yevick solution \eqref{10.19} and from the next-order rational-function approximation \eqref{10.40}. In the latter, $Z$ and $\chi$ have been chosen as given by the Carnahan--Starling equation of state [see \eqref{7.89} and \eqref{7.89b}]. We observe that both theories describe quite well the behavior of $g(r)$ but the Percus--Yevick approximation underestimates the contact value and then decays by crossing the simulation data. Both features are satisfactorily corrected by the rational-function approximation.

It is interesting to note that the rational-function approximation \eqref{10.40} coincides with the solution of the so-called generalized mean-spherical approximation (GMSA) \cite{W73,HB76,HB77}, where the direct correlation function $c (r)$ outside
the hard core ($r > 1$), which vanishes in the Percus--Yevick theory, is assumed to be given by a Yukawa form. The rational-function approximation method, however, is mathematically much more economical and open to applications to other systems.

\subsection{Non-Hard-Sphere Systems}
    The rational-function approximation methodology has been applied to systems different from one-component three-dimensional  hard spheres.
Those systems can be  classified into two categories: (i) systems amenable to an exact solution of the Percus--Yevick equation and (ii) systems non-amenable to an exact solution of the Percus--Yevick equation. The first class includes  sticky hard spheres (see Fig.\ \ref{fig6.3}, right panel) \cite{B68}, additive hard-sphere mixtures \cite{L64},  additive sticky-hard-sphere mixtures \cite{PS75,B75}, and hard hyperspheres \cite{FI81,L84}.
In that class of systems, the rational-function approximation method recovers the Percus--Yevick solution as the \emph{simplest} possible approach, just as in the hard-sphere case [see \eqref{10.19}]. The next-order approach allows one to make contact with empirical equations of state, thus improving the predictions. The interested reader can consult the references \cite{RS07,RS11,HYS08,YSH98,SYH98,MMYSH02,MYSH07,YSH08,RS11b} for further details.

The application of the rational-function approximation to systems of the second class includes the penetrable-sphere model (see Fig.\ \ref{9.3}, right panel) \cite{MS06,MYS07}, the penetrable-square-well model \cite{FGMS09}, the square-well potential  (see Fig.\ \ref{fig6.3}, left panel) \cite{YS94,AS01,LSYS05}, the square-shoulder potential (see Fig.\ \ref{fig9.3}, left panel) \cite{YSH11}, piece-wise constant potentials with more than one step \cite{SYH12,SYHBO13}, nonadditive hard-sphere mixtures \cite{FS11,FS13}, and Janus particles with constrained orientations \cite{MFGS13}. In those cases,  the \emph{simplest} rational-function approximation is already quite accurate, generally improving on the (numerical) solution of the Percus--Yevick equation.

\section{Concluding Remark}
These lecture notes are already too long, so let this author conclude just by saying that he will feel fully satisfied if the notes are useful to some of the students who attended the 5th Warsaw School of Statistical Physics, to some of the readers who have had the patience to read them, or to some instructors who might find something profitable for their own courses.

\begin{acknowledgement}
I am very grateful to the organizers of the 5th Warsaw School of Statistical Physics (especially Jaros{\l}aw Piasecki) for their kind invitation to be one of the speakers.
I also want to express my deep gratitude to Mariano L\'opez de Haro (who  critically read the manuscript) and Santos B. Yuste for having shared with me so many joyful hours ``playing with marbles.'' Finally, financial
support from the Spanish Government through Grant No.\ FIS2010-16587 and from
the Junta de Extremadura (Spain) through Grant No.\ GR10158, partially
financed by Fondo Europeo de Desarrollo Regional (FEDER) funds, is gratefully acknowledged.
\end{acknowledgement}

\bibliographystyle{spphys}

\bibliography{D:/Dropbox/Public/bib_files/liquid}

\end{document}